\documentclass[preprint,onecolumn,12pt]{elsarticle}

\usepackage{amsmath}
\usepackage{amssymb}
\usepackage{mathtools}
\usepackage{tikz}
\usepackage[a4paper]{geometry}
\usepackage{lipsum}
\usetikzlibrary{trees}

\newcommand{\pin}{\par\noindent}
\usepackage[hidelinks,colorlinks=true,linkcolor=blue, citecolor=blue]{hyperref}

\biboptions{numbers,sort&compress}

\journal{Nuclear Physics B}

\begin{document}

\begin{frontmatter}




\title{Holographic unitary renormalization group for correlated\\ electrons~-~II: insights on fermionic criticality}

\author{Anirban Mukherjee}
\ead{am14rs016@iiserkol.ac.in}
\author{Siddhartha Lal \corref{1}}
\cortext[1]{Corresponding author}
\ead{slal@iiserkol.ac.in}

\address{Department of Physical Sciences, Indian Institute of Science Education and Research-Kolkata, W.B. 741246, India\\
This article is registered under preprint number: arXiv:2004.06900v1}

\begin{abstract}
Capturing the interplay between electronic correlations and many-particle entanglement requires a unified framework for Hamiltonian and eigenbasis renormalization. In this work, we apply the unitary renormalization group (URG) scheme developed in a companion work~\cite{anirbanurg1} to the study of two archetypal models of strongly correlated lattice electrons, one with translation invariance and one without. We obtain detailed insight into the emergence of various gapless and gapped phases of quantum electronic matter by computing effective Hamiltonians from numerical evaluation of the various RG equations, as well as their entanglement signatures through their respective tensor network descriptions. For the translationally invariant model of a single-band of interacting electrons, this includes results on gapless metallic phases such as the Fermi liquid and Marginal Fermi liquid, as well as gapped phases such as the reduced Bardeen-Cooper-Schrieffer, pair density-wave and Mott liquid phases. Additionally, a study of a generalised Sachdev-Ye model with disordered four-fermion interactions offers detailed results on many-body localised phases, as well as thermalised phase. We emphasise the distinctions between the various phases based on a combined analysis of their dynamical (obtained from the effective Hamiltonian) and entanglement properties. Importantly, the RG flow of the Hamiltonian vertex tensor network is shown to lead to emergent gauge theories for the gapped phases. Taken together with results on the holographic spacetime generated from the RG of the many-particle eigenstate (seen through, for instance, the holographic upper bound of the one-particle entanglement entropy), our analysis offer an ab-initio perspective of the gauge-gravity duality for quantum liquids that are emergent in systems of correlated electrons.
\end{abstract}


\end{frontmatter}
\tableofcontents

\section{Introduction}
\pin
Strongly correlated electron systems on a lattice in two spatial dimensions or higher are known to host several exotic emergent quantum phenomena that are yet to be understood clearly, e.g., high-T$_{c}$ superconductivity, non-Fermi liquids, topological order, many body localization~\cite{anderson1972more,imada1998metal,
dagotto2005complexity,balents2010spin,castelnovo2012spin,
banerjee2016proximate,abanin2019}. Considerable effort has been made to the development of renormalisation group (RG) approaches to the understanding of these phenomena. Indeed, tremendous advancements of Wilson's original (RG) scheme have been made in the last few decades in the form of the functional RG~(FRG)~\cite{metzner2012, platt2013}. FRG deals with the RG flow of the Grassmanian many-body action~\cite{Wetterich1993,metzner2012functional,
salmhofer2001fermionic,kopietz2001exact}, incorporating all orders of quantum fluctuations by accounting for the entire hierarchy of $2n$-point vertex RG flow equations~\cite{kugler2018multiloop,tagliavini2019}. This has provided deeper insights into various emergent phases of quantum matter via effective descriptions derived from parent models of strongly correlated electrons. The latest developments in FRG involve the inclusion of self-energy~\cite{uebelacker2012,vilardi2017} and frequency dependence of vertices~\cite{husemann2009,giering2012,wang2013}, allowing the RG flows to reach theories closer to stable fixed points. Another important development has been the resummation of multiloop diagrams in FRG equations, yielding the parquet approximation~\cite{kugler2018}. Such careful computations has led to a better quantification of the effective theories in terms of measurables like susceptibility, spectral function etc~\cite{metzner2012,tagliavini2019,hille2020}. However, one crucial difficulty in the FRG program is its inability to attain stable fixed points, such that effective Hamiltonians can be obtained in the IR. One of the reasons for a lack of a controlled description of scale dependent effective Hamiltonians is the interplay between many-particle entanglement and four-fermionic interactions~\cite{orus2014advances,laflorencie2016quantum,
kaplis2017,orus2019tensor}. Upon the availability of such effective Hamiltonians, we aim to be able to track the phases emergent from fermionic criticality, as well as study their entanglement content.
\pin
In this work, we apply a novel unitary RG (URG) method developed in Refs.\cite{anirbanmotti,anirbanmott2,mukherjee2020}, and extended substantially in a companion work~\cite{anirbanurg1}, to two paradigmatic models of strongly correlated electrons. One of these is a model of a single band of translationally invariant electrons with a very general form of four-fermionic interactions. The other is one in which we consider the interplay of inter-particle interactions and disorder: the generalized Sachdev-Ye model~\cite{sachdev1993}, comprised of hopping, on-site and four-fermionic interactions, all of whose amplitudes are drawn randomly from separate Gaussian distributions. The goal is to obtain effective low-energy descriptions of the varied phases of electronic quantum matter that are emergent within these models. Towards this, we will demonstrate the primary advantage of the URG method: the RG flows of this method help obtain effective Hamiltonians, as well as insights into the many-particle entanglement content of its eigenstates, at stable IR fixed points. We briefly present the essential features of URG here.
\pin
The URG method is carried out via a sequence of unitary disentanglement operations on a graph, each of whose nodes corresponds to one electronic state. Each unitary operation on the graph disentangles an electronic state from the rest (the coupled subspace), leading simultaneously to block diagonalisation of the Hamiltonian in the occupation number (Fock) basis. The unitary operations are themselves determined from the form of the Hamiltonian, and their successive application thus generates a flow of the Hamiltonian into an iteratively block diagonalized form. 
The method yields a hierarchy of $2n$-point vertex flow equations, where each RG equation contains the summation of loop contributions into a closed form expression. Furthermore, the RG procedure reveals a family of energy scales for quantum fluctuations~($\omega$), arising from the non-commutativity between various parts of the Hamiltonian. These features lead to a non-perturbative contributions from frequency and correlation/self-energy, as observed in the structure of the denominator of the RG equations. Importantly, this structure allows the RG flows to attain stable fixed points in the IR where the effective Hamiltonian (and sometimes even the low-energy eigenstates) can be obtained. In a recent work on the 2D Hubbard modelat$1/2$-filling~\cite{anirbanmotti} as well as with hole-doping~\cite{anirbanmott2}, we used the effective Hamiltonian and ground state wavefunction obtained for the Mott insulating state at low energies to benchmark the ground state energy and double occupancy obtained from the URG against the numbers available from several other state-of-art numerical methods~\cite{leblanc2015}. Furthermore, we have also provided in Refs.\cite{anirbanmotti,anirbanmott2} a comparative study of the flow equations obtained from URG and the weak coupling FRG flow equations obtained for the same model.
\pin
We now offer some justification for our choice of the two models we study in the present work. Most importantly, both the translationally invariant four-fermi interacting model and the Sachdev-Ye model are very general in their scope, i.e., they have a wide parameter space, such that several well known phenomenological models (e.g., Fermi liquids and non-Fermi liquids, the reduced BCS Hamiltonian, Anderson's model for disordered non-interacting electrons etc.) can well lie within the sub-parameter regimes of these two models. Indeed, we will demonstrate that this is the case, and that the  
URG approach is an efficient method for the discovery of these phases. It is important to recall that problems of interacting electrons with translational invariance and an extended Fermi surface is known to be challenging, and have been studied using FRG methods over the last three decades~\cite{feldman1990,feldman1991,benfatto1990a,benfatto1990b,
benfatto1994,shankar1991,shankar1994,salmhofer1998,
benfatto2006,metzner2012,salmhofer2019}. FRG approaches have reported signatures of several novel states of electronic matter, including the the Mott insulator, non-Fermi liquid, pseudogap, d-wave superconductivity etc. phases within the realms of the four-fermi interacting model~\cite{katanin2004,rohe2005,giering2012,eberlein2014,hille2020pg}. We have earlier studied the effects of an extended and nested Fermi surface in the case of the 2D Hubbard model on the square lattice at $1/2$-filling in Refs.\cite{anirbanmotti,anirbanmott2}, as well as the case of a (Dirac) point-like spinon Fermi surface of a XXZ Kagome antiferromagnet in a finite magnetic field in Ref.\cite{pal2019}.
Here, we will present here a detailed study of the effects of electronic correlations for extended Fermi surfaces that are both nested as well as non-nested in spatial dimensions $D\geq 2$.
Further, as mentioned above, the electronic Sachdev Ye~($SY_{4}$)~\cite{sachdev1993} model described above, upon being embedded on a lattice, allows an investigation of the interplay between disorder and strong correlation. Studies on this model shows the fascinating phenomena of many-body localization and thermalization, with a novel transition between these two phases~\cite{jian2017solvable,abanin2019}. However, an ab-initio derivation of the effective Hamiltonians of these phases is a challenge that we aim to meet in this work.  
\pin
Specifically, by using the URG method, we obtain the $2$-point, $4$-point and $6$-point vertex RG flow equations for both the models. By numerically solving these RG equations for the translationally invariant model, we explore the phases that arise from the destabilization of the extended Fermi surface, whether nested or non-nested. We exlore the stable fixed point theories obtained both at high energies (of the order of the bandwidth) and low quantum fluctuation energyscale~($\omega$). At low $\omega$, the effective Hamiltonians obtained describe the Fermi liquid and the reduced BCS theory. At higher $\omega$,  we find a non-Fermi liquid phase with linear-in-temperature resistivity. In this particular case, the role of 6-point scattering vertices are found to be important, identifying the 2-electron 1-hole composite entity which replaces the Landau quasiparticle as the low-energy excitation proximate to the Fermi surface.  For the nested Fermi surfaces at $1/2$-filling, we find that the spin-exchange backscattering and the Umklapp scattering processes lead to emergent Mott liquid phases, described by the condensation of pseudospin degrees of freedom comprised of oppositely spin-paired electron-electron or electron-hole composites. We rewrite the effective Hamiltonian of such gapped phases in terms of non-local Wilson loop degrees of freedom. This allows the formulation of a Hamiltonian gauge theory for such topologically ordered gapped states of quantum matter. In such phases, we argue that the corresponding gauge theory supports Wilson loops with non-trivial anticommutation relations and describe fractionally charged excitations that interpolate between topologically degenerate ground states on the torus
~\cite{yamanaka1997nonperturbative,hastings2004lieb}. In this way, the present work shows that the vertex tensor network for gapped phases generated in the RG direction encodes an emergent gauge theory.
\pin
For the electronic Sachdev-Ye~($SY_{4}$) model, we perform a URG study by disentangling electronic states that are ordered in terms of their on-site energy (from higher to lower). By placing the model on a $D$ spatial-dimensional volume describing a specified geometry, we obtain a variety of phases from numerical evaluations of the RG equations obtained from the URG procedure. Some of these are described by effective Hamiltonians that possess translational invariant, while some other that do not. The former category includes the phases observed for the single band four-fermi interacting model discussed above. Among the phases that lack translation invariance, our analysis reveals glassy variants of the Fermi liquid as well as non-Fermi liquid phases, and display features of the phenomenon of many body localization~(MBL). On the other hand, we also find regimes describing thermalized phases, where the effective stable fixed point theory is related to the parent $SY_{4}$ model via marginal deformations, as well as a phase corresponding to the Anderson model of disordered noninteracting electrons. 
\pin
Importantly, in keeping with our presentation in Ref.\cite{anirbanurg1} for the tensor network of wavefunction coefficients that is generated holographically under RG flow by the vertex tensor network, we offer some results here for the case of gapless (e.g., the Fermi liquid and Marginal Fermi liquid phases) as well as gapped (e.g., the reduced BCS and Mott liquid phases) quantum liquids. We derive scaling relations for the single-electron entanglement entropy of these phases, and use them to obtain relations for the (holographic) upper bound of the entanglement entropy. This is also in agreement with our recent finding that the URG flow respects the holographic principle~\cite{mukherjee2020}. The rest of the work is organized as follows. We first recapitulate the important results of Ref.\cite{anirbanurg1} in Section \ref{prelims}, as well as present some new ones for the scaling relation of the single-electron entanglement entropy and its holographic upper bound. In Section \ref{SFIM}, we perform a URG treatment of the single band four-fermi interacting model, revealing the various IR fixed points as well as obtaining the tensor network representation of the various gapped/gapless phases. Section \ref{emergentGaugeTheories} describes the gauge theoretic description for the gapped theories reached under RG. In Section \ref{SY}, we perform the URG analysis of a generalized $SY_{4}$ model for electrons, revealing various translation invariant and non-invariant fixed points. We conclude in Section \ref{conclusions}. Finally, the details of certain calculations are presented in appendices.

\section{Preliminaries}\label{prelims}
\pin
In a companion manuscript~\cite{anirbanurg1}, we have presented the URG method in detail for a system of strongly coupled electrons, leading to a heirarchy of $2n$-point vertex RG flow equations. We can interpret the $2n$-point vertices as $2n$-legged tensors, thus allowing a realization of the URG as a vertex-tensor network RG. We have also shown in Ref.\cite{anirbanurg1} that, when applied to the eigenbasis of the Hamiltonian, the URG leads to the renormalization of the coefficient tensors, i.e., superposition weights of the separable states comprising the many-particle eigenstate. From the renormalisation of the entanglement tensors, a entanglement holographic mapping~(EHM)~\cite{qi2013,lee2016} network is generated along the RG direction. In Ref.\cite{mukherjee2020}, the EHM networks  for the normal metallic state and the insulating ground state of the 2D Hubbard model at half-filling has also been explicitly constructed by us. Prior to applying the URG method to some archetypal models of correlated electrons, we first lay out some of the important results from the above works.\\
\pin{\bf \textit{Hamiltonian RG flow via iterative block diagonalization}}\\
We represent a general fermionic Hamiltonian $H$ as a $2\times 2$ block matrix in the number-occupation basis of an electronic state. By performing a Gauss-Jordan elimination of one of the blocks via a rotation of the many-particle eigenbasis, we obtain a block-diagonal representation of the matrix. Such a procedure can be realized as a unitary transformation $U$ of the Hamiltonian, $H'=UHU^{\dagger}$. The unitary transformation $U_{(j)}$ is identified as a disentangler that separate an electronic state $j$ from the rest in the renormalization group step $j$. Below, we present the form of the $U$-operation
\begin{equation}
U_{(j)}=\frac{1}{\sqrt{2}}[1+\eta_{(j)}-\eta^{\dagger}_{(j)}]~,\label{Unitary operator}~~~~
\end{equation}
where $\eta^{\dagger}_{(j)}$ and $\eta_{(j)}$ are electron-hole transition operators fulfilling the algebra $\lbrace\eta^{\dagger}_{(j)},\eta_{(j)}\rbrace =1$ and $[\eta^{\dagger}_{(j)},\eta_{(j)}] = 2\hat{n}_{j}-1$.
Importantly, note that $U_{(j)}$ can also be represented as the exponential of a phase operator
\begin{eqnarray}
U_{(j)}=\exp(i\theta_{(j)})~,~ \theta_{(j)}=-i\frac{\pi}{4}(\eta_{(j)}-\eta^{\dagger}_{(j)})~,\label{phase_operator}
\end{eqnarray}
corresponding to a rotation of $\pi/2$ in the many-particle state space gathered via the generator $i(\eta_{(j)}-\eta^{\dagger}_{(j)})$. The operator $\eta_{(j)}$ is written in terms of $2n$-point off- diagonal scattering vertices (with respect to a given electronic state $j$) in the numerator and diagonal $2n$-point vertices in the denominator
\begin{equation}
\eta_{(j)}  = Tr_{j}(c^{\dagger}_{j}H_{(j)})c_{j}\frac{1}{\hat{\omega}_{(j)} -Tr_{j}(H^{D}_{(j)}\hat{n}_{j})}~
.\label{e-h transition operator}
\end{equation} 
Here, $\hat{\omega}_{(j)}$ represents the quantum fluctuation operator, and accounts for the non-commutativity between different off-diagonal $2n$-point vertices.  It is mathematically defined as
\begin{eqnarray}
\hat{\omega}_{(j)}=H_{(j-1)}^{D}+H_{(j-1)}^{X,\bar{j}}-H_{(j)}^{X,\bar{j}}~,\label{fluctuation_Op}
\end{eqnarray}
where the number diagonal part of the Hamiltonian ($H^{D}_{(j)}$) is associated with $n$-particle self/correlation energies, and the term $H_{(j)}^{X,\bar{j}}$ represents coupling only among the other degrees of freedom $\lbrace 1,\ldots,j-1\rbrace$.
$\hat{\omega}_{(j)}$ can be given a spectral decomposition as follows
\begin{eqnarray}
\hat{\omega}_{(j)}=\sum_{i}\omega^{i}_{(j)}\hat{O}_{(j)}(\omega^{i})~,~\hat{O}_{(j)}(\omega^{i})=|\Phi^{i}_{(j)}\rangle\langle\Phi^{i}_{(j)}|~,\label{fluctuation_Subspace}
\end{eqnarray} 
where $|\Phi^{i}_{(j)}\rangle$ are eigenstates of $\hat{\omega}_{(j)}$, and $\omega^{i}_{(j)}$ are the quantum fluctuation eigenvalues. At each RG step, the $\omega_{(j)}$ attains a block-diagonal form. We note that if in a number-occupation subspace $P$, the off-diagonal vertices attain an RG fixed point, then the fluctuation operator attains a number diagonal form in that subspace: $P\hat{\omega}_{(j*)}P=H^{D}_{(j^{*})}$. Thus, stable fixed points are identified by the fact that $|\Phi^{i}_{(j)}\rangle$ become simultaneous eigenstates of $H^{D}_{(j^{*})}$ and $\omega_{(j^{*})}$.
\pin 
The RG flow equation for the Hamiltonian is given by
\begin{eqnarray}
H_{(j-1)}=U_{(j)}H_{(j)}U^{\dagger}_{(j)}~,\label{HamRGflow}
\end{eqnarray}    
and with the above form for the unitary map eq.\eqref{Unitary operator}, we obtain the iterative equation for the rotated Hamiltonian
\begin{eqnarray}
\hat{H}_{(j-1)} &=&\frac{1}{2}Tr_{j}(\hat{H}_{(j)})+Tr_{j}(\hat{H}_{(j)}\tau_{j})+\tau_{j}\lbrace c^{\dagger}_{j}Tr_{j}(H_{(j)}c_{j}),\eta_{(j)}\rbrace~.~~~
\end{eqnarray}
The first and second terms represent $H^{D}_{(j-1)}$ and $H^{X,\bar{j}}_{(j-1)}$ mentioned above, while the third represents the off-diagonal processes ($H_{(j)}^{X,j}$) that are responsible for quantum fluctuations in the occupation number of state $j$.
Note that the above Hamiltonian $H_{(j-1)}$ commutes with $N-j$ Pauli-Z gates $\tau_{i}=\hat{n}_{i}-1/2$~($i=j,\ldots, N$).\\
\pin{\bf \textit{Vertex tensor network representation of the Hamiltonian}}\\ 
The Hamiltonian $H$ can be interpreted as a tensor network formed from the $2n$-point vertex tensors
\begin{eqnarray}
H_{(j)}&=& H^{D}_{(j)}+H^{X}_{(j)}~,\label{D-X decomposition}\\
H^{D}_{(j)}&=& \sum_{i=1}^{2^{j-1}}\sum_{n=1}^{N}\sum_{\alpha,\alpha'}\lbrace\tilde{c}^{\dagger}_{\alpha}\frac{\Gamma^{2n}_{\alpha\alpha'}(\omega^{i})}{2^{n}}\tilde{c}_{\alpha'}\hat{O}(\omega^{i})\rbrace_{(j)}~,\nonumber\\
H^{X}_{(j)} &=& \sum_{i=1}^{2^{j-1}}\sum_{n=1}^{a_{j}}\sum_{\alpha,\beta}\lbrace\tilde{c}^{\dagger}_{\alpha}\Gamma^{2n}_{\alpha\beta}(\omega^{i})\tilde{c}_{\beta}\hat{O}(\omega^{i})\rbrace_{(j)}~.\label{cluster decomposition}
\end{eqnarray}
We now explain the various terms and notations in the above equation. Eq.\ref{D-X decomposition} shows the decomposition of the Hamiltonian into number-diagonal $H^{D}_{(j)}$ and off-diagonal $H^{X}_{(j)}$ parts. The index $i$ ranges from $1$ to $2^{j-1}$, and labels the eigenbasis element $|\Phi^{i}_{(j)}\rangle$ of $\hat{\omega}_{(j)}$ in the entangled subspace of $j$ electronic states. The index $\alpha:=\lbrace (l,\mu)\rbrace$ is a set of paired labels: $l$ labels the electronic states participating in the entangled subspace and $\mu=(0,1)$ represents an electron occupied/unoccupied state. Therefore, $\tilde{c}^{\dagger}_{\alpha}$ represents a string of electron creation and annihilation operators. The index $\alpha':=\{(l,\bar{\mu})\}$ is similar to $\alpha$: $l$ represents the same collection of indices, but where $\bar{\mu}$ is the complement of $\mu$, i.e., $\bar{\mu}=1,0$ refers to unoccupied ($1$) and occupied ($0$). The symbol $\Gamma^{2n,(j)}_{\alpha\beta}$ represents the collection of $2n$-point off-diagonal vertex tensors, with $\beta$ being an index defined similarly to $\alpha$. Finally, $a_{j}$ represents the  maximum order of the off-diagonal vertex tensor.
\pin
The iterative unitary mapping of the Hamiltonian generates an RG flow for the vertex tensor network
\begin{eqnarray}
\Delta \Gamma^{2n,(j)}_{\alpha\beta}(\omega^{i})&=&\sum_{p_{1},p_{3}}^{2a_{j}}\sum_{\gamma}\{\Gamma^{p_{1}}_{\alpha\gamma}G^{2p_{2}}_{\gamma\gamma'}\Gamma^{p_{3}}_{\gamma'\beta}\}^{(j)}(\omega^{i})~,\label{vertexRGflows}
\end{eqnarray} 
where $2n=p_{1}+p_{3}-2p_{2}$.\\
\pin{\bf \textit{Eigenbasis RG flow via iterative block diagonalization}}\\
The RG flow equation for the eigenstate of the Hamiltonian is given by
\begin{eqnarray}
|\Psi^{i}_{(j-1)}\rangle = U_{(j)}|\Psi^{i}_{(j)}\rangle~,\label{unitaryMap} 
\end{eqnarray}
where $H_{(j)}|\Psi^{i}_{(j)}\rangle = E^{i}|\Psi^{i}_{(j)}\rangle$. There are $N-j$ good quantum numbers at RG step $j$, such that the state $|\Psi^{i}_{(j)}\rangle$ satisfies the following eigenvalue relation
\begin{eqnarray}
\hat{n}_{l}|\Psi^{i}_{(j)}\rangle=|\Psi^{i}_{(j)}\rangle~,~ l=\lbrace N,\ldots ,N-j+1\rbrace~. 
\end{eqnarray}
As a result, the many-body state $|\Psi^{i}_{(j-1)}\rangle$ can be represented as a coefficient tensor network
 \begin{eqnarray}
|\Psi^{i}_{(j)}\rangle = \sum_{\alpha}C^{i,(j)}_{\alpha}|\alpha\rangle |Q_{j}\rangle~.\label{configuration_space_expansion}
\end{eqnarray}
The coefficient $C^{i,(j)}_{\alpha}$ is a tensor with $m$ legs representing the superposition weight of the configurations with $m$ occupied electronic states. $\alpha$ represents a set of electronic labels for the occupied electronic states. $|Q_{j}\rangle$ represents the occupation-number configuration of the disentangled states. In another recent work~\cite{mukherjee2020}, we have presented the quantum circuit/tensor network representation of a specific many-body state. Altogether, we show that URG (see eq.\ref{unitaryMap}) generates a fermionic tensor network renormalization
\begin{eqnarray}
 \Delta C^{i,(j)}_{\beta}&=& (\sqrt{N^{(j)}}-1)C^{i,(j)}_{\beta}+\sqrt{N^{(j)}}\sum^{a_{j}}_{\bar{k}=1}\sum_{\alpha,\alpha',\beta'}sgn(\alpha,\alpha',\beta')\lbrace\Delta\Gamma^{2n}_{\beta^{'}\alpha}
G^{2p}_{\alpha\alpha^{'}}C^{i}_{\alpha^{'}}\rbrace^{(j)}~,~~~~
\label{Tensor_flow_eqn}
\end{eqnarray}
where $N^{(j)}$ is the normalization. Here, $\alpha':=\lbrace (l,\mu)\rbrace$ is a ordered set of $m$ pairs of indices with $\mu=1$ throughout and $1\leq l<j$. $\alpha:=\lbrace (a,\mu)\rbrace$ is an ordered set of $p$ pairs of indices ($p<m$) and $\mu=0$ throughout. Note that the electronic state labels that comprise $\alpha$ is a subset of those within $\alpha^{'}$. Finally, $\beta':=\lbrace(b,\mu)\rbrace$ is an ordered set of $2n-p$ pairs of indices with $\mu=1$ throughout. The set $\beta$ is an ordered set of $m-2p+2n$ pairs of indices that emerges from the convolution of the sets above, $\beta:=(\beta'\cup\gamma')-\gamma$. The sign $sgn(\alpha,\alpha',\beta)$ is the net phase gathered via counting the number of electrons exchanged in the scattering process $\Delta\Gamma^{2n}_{\beta^{'}\alpha}$ involving a string of $2n-p$ electron creation and $p$ annhilation operators   
\begin{eqnarray}
sgn(\alpha,\alpha',\beta')=\prod_{k=1}^{2n-p}Q_{k}\prod_{k=1}^{p}P_{k}~.\label{fermion_exchange_sign}
\end{eqnarray}
Here $\prod_{k=1}^{p}P_{k}$ and $\prod_{k=1}^{2n-p}Q_{k}$ are the net phases that arises from the number of electron exchanged via the string of electron annihilation and creation operators respectively. Below, we quantify the phases $P_{k}$ and $Q_{k}$
\begin{eqnarray}
P_{k}&=&\exp\left(i\pi\sum_{i\in \alpha'-\rho}^{a_{k}}n_{i}\right)~,~Q_{k}=\exp(i(k-1)\pi)\exp\left(i\pi\sum_{\substack{i=1,\\ i\notin \gamma\cup\alpha}}^{b_{k}}n_{i}\right)~,
\end{eqnarray}
where $\rho=\lbrace a_{1},\ldots,a_{k-1}\rbrace$ and $\gamma=\lbrace b_{1},\ldots, b_{k}\rbrace$ are ordered sets of electronic state labels where electrons are annihilated and created respectively. Note that $\rho$ is a subset of the electronic state labels contained in the set $\alpha'$. In the definition of $P_{k}$, the electron number count $n_{i}=1$ for $i\in \alpha'$ and $0$ otherwise. Upon acting the annihilation operators of the scattering vertex $\Delta\Gamma^{2n,(j)}_{\beta'\alpha}$ on $|\alpha'\rangle$, the state reached in eq.\eqref{Tensor_flow_eqn} is given by
\begin{equation}
\Delta\Gamma^{2n,(j)}_{\beta'\alpha}c^{\dagger}_{b_{2n-p}}\ldots c^{\dagger}_{b_{1}}c_{a_{p}}\ldots c_{a_{1}}|\alpha'\rangle=\prod_{k=1}^{p}P_{k}\Delta\Gamma^{2n,(j)}_{\beta'\alpha}c^{\dagger}_{b_{2n-p}}\ldots c^{\dagger}_{b_{1}}|\alpha''\rangle~. 
\end{equation}
Finally, in the definition of $Q_{k}$, the number count $n_{i}=1$ if $i\in\alpha''$ and $0$ otherwise. 
\pin
Importantly, we note that in cases when pairs of electronic states condense into bounds states, the fermion exchange sign trivializes to $sgn(\alpha,\alpha',\beta)=1$ in the RG equation \eqref{Tensor_flow_eqn}. This results in an emergent Hamiltonian theory and associated eigenbbasis at the IR fixed point that is free from fermion signs. In what follows, we apply the tensor renormalization group theory to certain strongly coupled electronic systems. The analysis reveals a class of stable IR fixed points corresponding to gapped as well as critical theories. We also show that for a certain class of IR fixed point theories, the fermion sign issues are altogether mitigated.\\    
\pin
{\bf\textit{Relation between thermal ($k_{B}T$) and quantum ($\omega$) fluctuation energy scales}}\\
In the URG formalism, the renormalized Hamiltonian is partitioned in various eigen-subspaces ($|\Phi^{i}_{(j)}\rangle$ in eq.\eqref{fluctuation_Subspace}) of the quantum fluctuation operator $\hat{\omega}$.  Naturally, the Hamiltonians in the subspaces are associated with the eigenvalues $\omega^{(j)}_{i}$ of the renormalized fluctuation operator $\hat{\omega}_{(j)}$ eq.\ref{fluctuation_Op}. The nature of RG flow equations for various $2n$-point vertices are dictated by the quantum fluctuation scales $\omega^{i}_{(j)}$, deciding ultimately whether the low-energy spectrum $H^{*}(\omega)$ at the IR fixed point is either gapped or gapless. In Ref.\cite{anirbanmotti}, $\omega_{(j)}$ was shown to be equivalent to a thermal scale upto which dominant quantum fluctuations leading to $H^{*}(\omega)$ persist
\begin{eqnarray}
T = \frac{1}{k_{B}\pi^{2}}\frac{\hbar}{\tau},~\frac{\hbar}{\tau}=\Sigma^{Im}(\omega)=\mathcal{P}\int^{\infty}_{-\infty}d\omega'\frac{\Sigma^{*}(\omega)}{\omega-\omega'}~.\label{Thermal scale}
\end{eqnarray}
The above relation shows that the finite lifetime ($\tau$) of the single-particle states with self-energy $\Sigma$ can be viewed as an effective temperature scale arising out of the unitary disentanglement: it is the highest temperature upto which the one-particle excitations can survive, and are replaced by 2e-1h composite excitation beyond.  We will see in later sections that the RG transformations lead generically to either a gapped or a gapless phase. For the first case, the above equation quantifies the thermal upper bound for the validity of the emergent condensate. On the other hand, for the second scenario, it indicates the lifetime of the gapless excitations in the neighbourhood of the Fermi surface.\\
\pin
{\bf\textit{URG scaling of the Ryu-Takayanagi entanglement entropy bound}}\\
As any nonlocal unitary rotation can be decomposed as a tensor product of $2$-local and local qubit rotations, the unitary operators of the URG framework form a entanglement holographic mapping network~\cite{lee2016,qi2013}. As a consequence, the renormalized states in the bulk of the EHM network respect the Ryu-Takayanagi entanglement entropy bound formula~\cite{ryu2006aspects}: the entanglement entropy of a region $R$ is bounded from above by the number of linkages between it and its complement. We have given an explicit demonstration of this entropy bound for the parent metallic state and the insulating ground state of the 2D Hubbard model at half-fillingin Ref.\cite{mukherjee2020}. This shows that the entanglement renormalization obtained via URG generates a holographic dual space-time along the RG direction. We aim here to reveal the URG scaling features of the Ryu-Takayanagi entanglement entropy bound for various metallic and insulating states obtained in the IR starting from generic strongly correlated models. In this section, we will obtain the expression for one-electron entanglement entropy in terms of the coefficient tensors. This is important as the maximum one-electron entanglement entropy among the electrons in a region $R$, when  multiplied by the number of entangled links at a given RG step $j$, leads to the scaling of the holographic entropy bound.
\pin  
The many-body eigenstate $|\Psi_{(j)}\rangle$ at an RG step $j$ can be written in a Schmidt-decomposed form with respect to 1-electron state $k$ and the rest of the system 
\begin{eqnarray}
|\Psi_{(j)}\rangle &=& a_{0}|\phi_{0,l}\rangle|\Psi_{0,(j)}\rangle + a_{1}|\phi_{1,l}\rangle|\Psi_{1,(j)}\rangle~.
\end{eqnarray}
Here $\langle\phi_{1,l}|\phi_{0,l}\rangle=0=\langle\Psi_{1,(j)}|\Psi_{0,(j)}\rangle$. Note that for $l<j$, the electronic state is a part of the entangled subspace, ensuring that the Schmidt coefficients $a_{1}$ and $a_{0}$ take values between $0$ and $1$. The states $|\Psi_{0,(j)}\rangle$ and $|\Psi_{1,(j)}\rangle$ can be written in terms of the coefficient tensors as follows
\begin{eqnarray}
|\Psi_{1,(j)}\rangle &=&\frac{1}{\sqrt{\sum_{\beta_{1}}|C^{1,(j)}_{\beta_{1}}|^{2}}}\sum_{\beta_{1}}C^{1,(j)}_{\beta_{1}}|\beta'_{1}\rangle~,~|\Psi_{0}\rangle =\frac{1}{\sqrt{\sum_{\alpha_{1}}|C^{0,(j)}_{\alpha_{1}}|^{2}}}\sum_{\alpha_{1}}C^{0,(j)}_{\alpha_{1}}|\alpha'_{1}\rangle~.~~~~
\end{eqnarray}
Here the labels $\alpha_{1}'$ and $\beta_{1}'$ represent the collection of electronic states that are occupied. Given the orthogonality condition $\langle\Psi_{0,(j)}|\Psi_{1,(j)}\rangle=0$, the Schmidt coefficients have the following expression
\begin{eqnarray}
a_{1,(j)}=\sqrt{\sum_{\beta_{1}}|C^{1}_{\beta_{1}}|^{2}}, a_{0,(j)}=\sqrt{\sum_{\alpha_{1}}|C^{0}_{\alpha_{1}}|^{2}}~,
\end{eqnarray}
with the constraint $a_{1,(j)}^{2}+a_{0,(j)}^{2}=1$. The one-electron entanglement entropy is obtained in terms of Schmidt coefficients
\begin{eqnarray}
S_{(j)}(k)=\log 2 -\frac{1}{2}(1+x)\log(1+x)-\frac{1}{2}(1-x)\log(1-x)~,
\end{eqnarray}
where $x=\sqrt{1-4a_{1,(j)}^{2}a_{0,(j)}^{2}}$. 
\pin
We now obtain the leading terms in $S_{(j)}$ for two extreme cases: (i) when the URG flow leads to IR fixed points where the ground state is completely separable, and (ii) when the URG flow generates a highly entangled subspace in the IR. For case (i), and with $x=1-\epsilon$ ($\epsilon\to 0$), we have
\begin{eqnarray}
S^{1}_{(j)}(k)&=&-(1-\frac{\epsilon}{2})\log
\left(1-\frac{\epsilon}{2}\right)-\frac{\epsilon}{2}\log\frac{\epsilon}{2}\approx \epsilon/2 = \sum_{\beta_{1},\alpha_{1}}|C^{1,(j)}_{\beta_{1}}C^{0,(j)}_{\alpha_{1}}|^{2}~.
\end{eqnarray}
For case (ii0, with $x\to 0$, we find
\begin{eqnarray}
S^{2}_{(j)}(k)&\approx &\log 2-\frac{x^{2}}{2}=\log 2 -\frac{1}{2}+2\sum_{\beta_{1},\alpha_{1}}|C^{1,(j)}_{\beta_{1}}C^{0,(j)}_{\alpha_{1}}|^{2}~.
\end{eqnarray}
Next, we will obtain the renormalization of entanglement entropy $S_{(j)}^{1}$ for case (i) in the lowest order of $\Delta C^{(j)}_{\rho}$
\begin{eqnarray}
\Delta S^{1}_{(j)}(k)&=&\sum_{\beta_{1},\alpha_{1}}|(C^{1,(j)}_{\beta_{1}}+\Delta C^{1,(j)}_{\beta_{1}})(C^{0,(j)}_{\alpha_{1}}+\Delta C^{0,(j)}_{\alpha_{1}})|^{2}-\sum_{\beta_{1},\alpha_{1}}|C^{1,(j)}_{\beta_{1}}C^{0,(j)}_{\alpha_{1}}|^{2}\nonumber\\
&\approx  &\sum_{\beta_{1}}2Re(C^{1,(j)}_{\beta_{1}}\Delta C^{1,(j)}_{\beta_{1}})a^{2}_{0,(j)}+\sum_{\alpha_{1}}2Re(C^{0,(j)}_{\alpha_{1}}\Delta C^{0,(j)}_{\alpha_{1}})a^{2}_{1,(j)}~.
\end{eqnarray}
Note that for a separable state, either $a_{1,(j)}\to 1$ or $a_{0,(j)}\to 1$, resulting in
\begin{eqnarray}
\Delta S^{1}_{(j)}(k)&=& \sum_{\beta_{1}}2Re(C^{1,(j)}_{\beta_{1}}\Delta C^{1,(j)}_{\beta_{1}})~.\label{entropyScaling}
\end{eqnarray}
Similarly, for highly entangled states in case (ii), $\Delta S^{2}_{(j)}(k)=\sum_{\beta_{1}}4Re(C^{1,(j)}_{\beta_{1}}\Delta C^{1,(j)}_{\beta_{1}})$. Finally, note that, following the Ryu-Takyanagi formula~\cite{ryu2006aspects}, the entanglement entropy of a region $R$ is bounded as follows
\begin{eqnarray}
\Delta S_{(j)}(R)\leq N_{(j)}(R)\Delta \max_{k\in R}S_{(j)}^{i}(k)~,\label{entropy_bound}
\end{eqnarray}
where $N_{(j)}(R)$ is the number of electrons in the region $R$ that belong to the entangled subspace at RG step $j$. In a later section, we will obtain the entropy bound scaling relation for various gapless and gapped IR fixed points obtained from a generic strongly correlated model.

\section{Tensor RG theory for the single band four-fermi interacting model}\label{SFIM}
\par\noindent
The URG formalism, introduced in a companion work~\cite{anirbanurg1}, leads to the iteratie block diagonalization of the Hamiltonian in Fock space. In Sec.\ref{prelims}, we have laid out the major results from the URG formalism. In the companion work~\cite{anirbanurg1}, we have also investigated the leading effects of such unitary transformations on a generic model of interacting fermions on a lattice. These investigations pointed towards the emergence of six-point (or three-particle) vertices that can either lead to the modification of the Fermi liquid self-energy or its complete destabilization, outcomes of logarithmic divergences in the 1-particle self-energy and 2-particle correlation energies respectively. To further investigate these log-divergences, we implement the Hamiltonian Tensor RG scheme.
\pin
We begin the analysis by representing the single-band translational invariant four- fermion interacting model (SFIM) as follows
\begin{eqnarray}
\hat{H}_{SFIM} &=& \sum_{\mathbf{k}}(\epsilon_{\mathbf{k}}-\mu)\hat{n}_{\mathbf{k}\sigma}+\sum_{\mathbf{k}\mathbf{k}'\mathbf{p}}V_{\alpha\beta}c^{\dagger}_{\mathbf{k}\sigma}c^{\dagger}_{\mathbf{p}-\mathbf{k}\sigma'}c_{\mathbf{p}-\mathbf{k}'\sigma'}c_{\mathbf{k}'\sigma}~,\label{HSFIM}
\end{eqnarray} 
where $\mathbf{p}$ is the net pair-momentum and $\mathbf{k}-\mathbf{k}'$ is the momentum transfer. The four-fermion interaction vertex can be compactly represented as $V_{\alpha\beta}=V^{\sigma\sigma'}_{\mathbf{k}\mathbf{k'}\mathbf{p}}$, where $\alpha:=\lbrace(\mathbf{k},\sigma,1) ;(\mathbf{p}-\mathbf{k},\sigma',1)\rbrace$, $\beta:=\lbrace (\mathbf{p}-\mathbf{k}',\sigma',0) ; (\mathbf{k}',\sigma,0)\rbrace$, and the indices $1$ and $0$ represent the $c^{\dagger}$ and $c$~operators respectively. The zero momentum transfer vertices are denoted as $V_{\alpha\alpha'}$ (where $\alpha'=\lbrace(\mathbf{k},\sigma,0) ;(\mathbf{p}-\mathbf{k},\sigma',0)\rbrace$). For an overall understanding of the various zero temperature phases, we develop below the Hamiltonian renormalization group scheme eq.\eqref{HamRGflow} for $H_{SFIM}$. We will, thereby, display the tensor network representation of the Hamiltonian RG flows towards various fixed point theories. 
\par
\begin{figure}
\centering
(a)\includegraphics[scale=0.35]{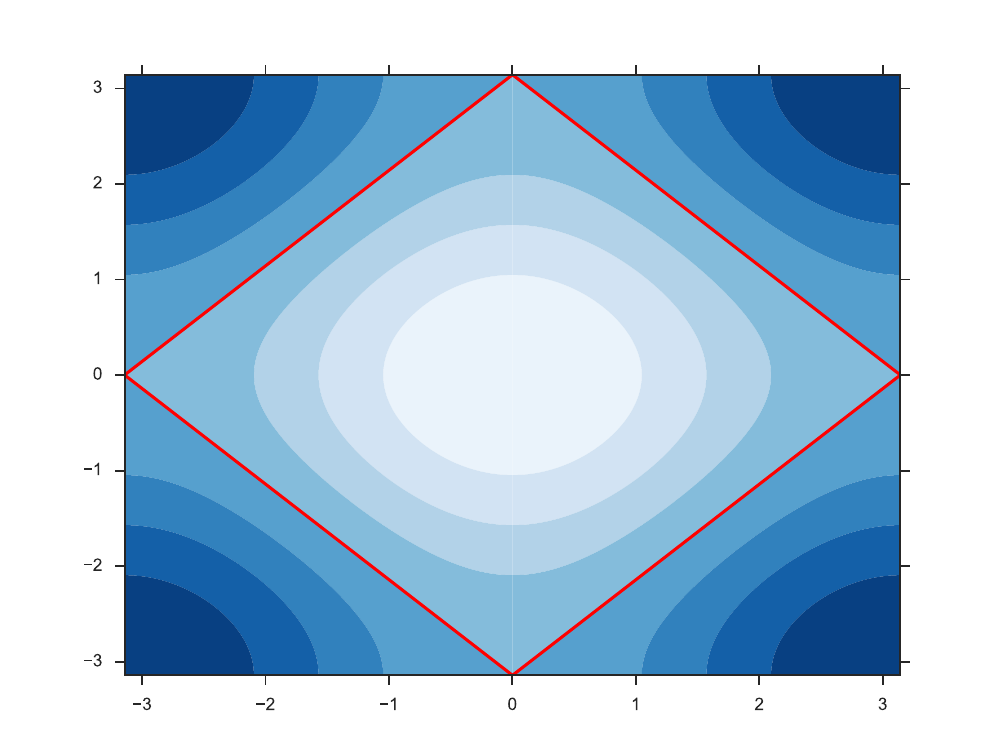} 
(b)\includegraphics[scale=0.35]{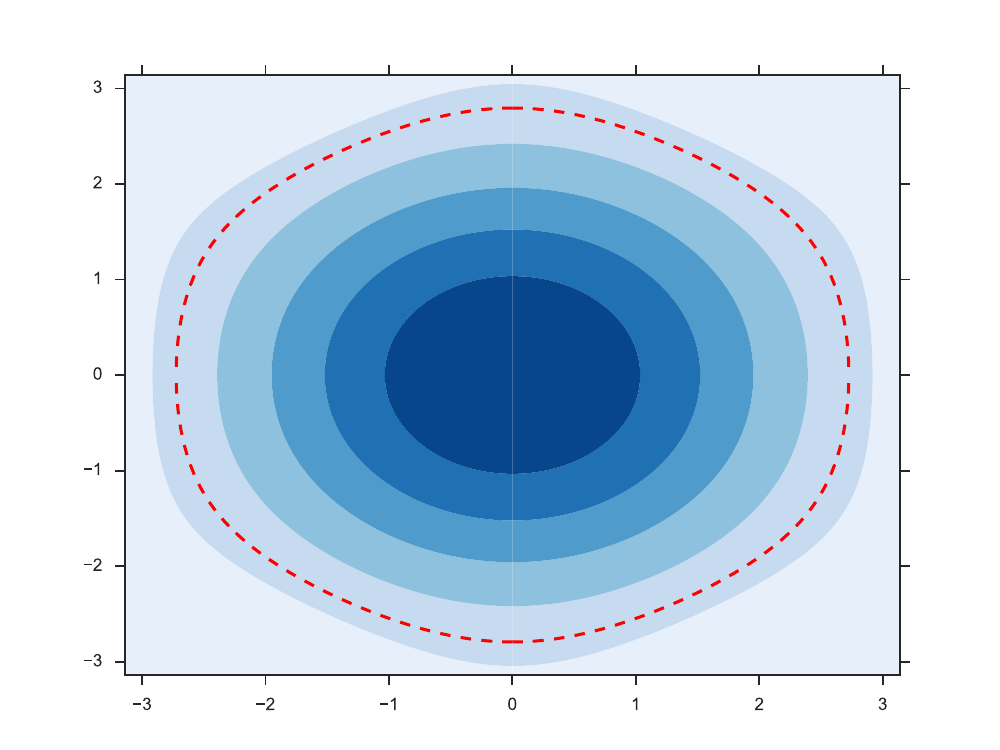}
(c)\includegraphics[scale=0.35]{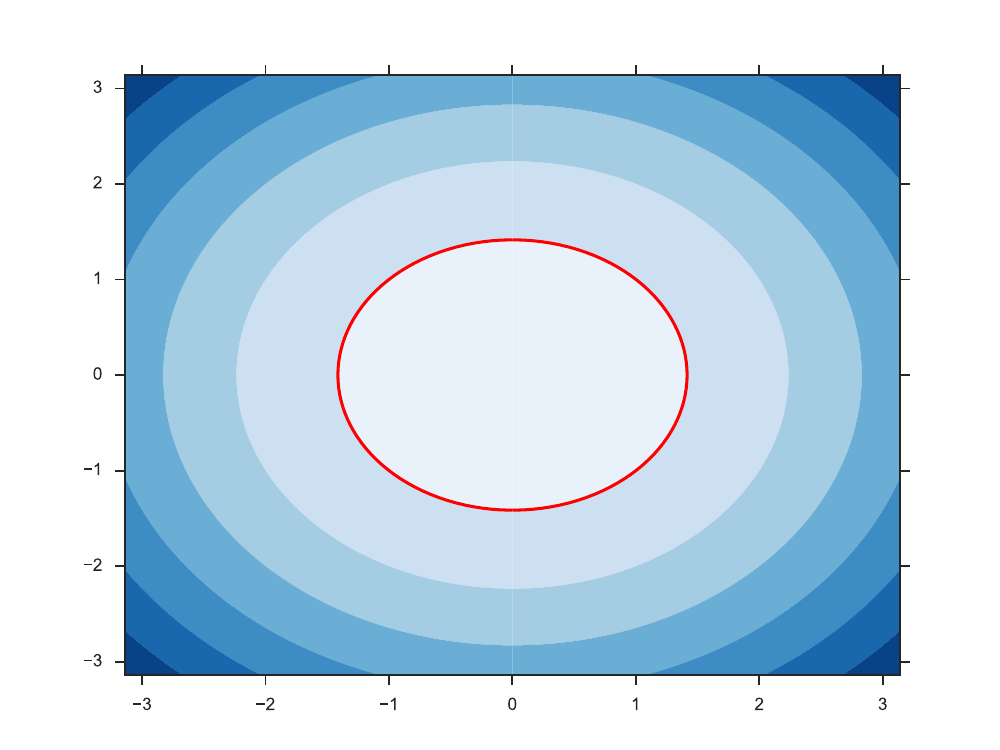}\\
\caption{Figure represents iso-geometric curves that are displaced parallely from the Fermi surface (red curve), and cover the Brillouin zone in the limit of $(L_{x},L_{y})\to \infty$ limit for (a) the square lattice at chemical potential $\mu=0$, (b) the triangular lattice at $\mu = -7.0t$ and (c) the continuum circular dispersion at $\mu=0$.}\label{isogeometric curves}
\end{figure}

\noindent 
The scheme adopted for the RG involves decoupling initially states whose energy is highest with respect to Fermi energy $E_{F}=\mu$, followed by ones closer to the Fermi surface. This is ensured by defining parallel curves \textit{isogeometric} to the Fermi surface (see Fig.\ref{isogeometric curves}). The wave-vectors $\mathbf{k}_{\Lambda\hat{s}}=\mathbf{k}_{F}(\hat{s})+\Lambda\hat{s}$ are relabelled by the distance $\Lambda$ from the Fermi surface and the unit normal vector to it, $\hat{s}=\nabla\epsilon_{\mathbf{k}}/|\nabla\epsilon_{\mathbf{k}}||_{\epsilon_{\mathbf{k}}=E_{F}}$. 
The renormalization group flow of the Hamiltonian, $H_{(j-1)}=U_{(j)}H_{(j)}U^{\dagger}_{(j)}$, takes place via disentangling an entire isogeometric curve by a unitary operation $U_{(j)}=\prod_{l}U_{(j,l)}$,
where the collective coordinates $l=(\hat{s},\sigma)$, $(j,l)\equiv \mathbf{k}_{\Lambda_{j}\hat{s}}\sigma$ represent the state labelled by the quantum numbers $\mathbf{k}_{\Lambda_{j}\hat{s}}\sigma$. $U_{(j,l)}$ represents an individual unitary operation that decouples a single electronic state $(j,l)\equiv \mathbf{k}_{\Lambda_{j}\hat{s}}\sigma$ and has the form
\begin{eqnarray}
U_{(j,l)} &=&\frac{1}{\sqrt{2}}[1+\eta^{\dagger}_{(j,l)}-\eta_{(j,l)}]
\label{unitary_op_jth_curve}~,
\end{eqnarray}
where the form of the e-h transition operators $\eta_{(j,l)}$ are shown in eq.\eqref{e-h transition operator}. Following the RG flow equations eq.\eqref{vertexRGflows} obtained from the Hamiltonian renormalization, we will investigate various parameter regimes for the microscopic $H_{SFIM}$ theory, culminating in the RG phase diagram for the model. In keeping with our discussion of the influence of 3-particle vertices on the 1-particle self-energy and 2-particle correlation energy, we truncate the RG flow equation heirarchy in eq.\eqref{vertexRGflows} to six-point vertices. 
\subsection{Derivation of the RG equation for 1 particle self energy $\Sigma_{\mathbf{k}\sigma}(\omega)$} \label{1pSelfEnergy}
\par\noindent
Translational invariance of $H_{SFIM}$ ensures that the two-point vertices are purely number diagonal in momentum space: $\Gamma^{2,(j)}_{\alpha\alpha'}(\omega_{(j)})=\epsilon_{\Lambda,\hat{s}}+\Sigma^{(j)}_{\Lambda,\hat{s}}(\omega_{(j)})$. Further, from eq.\eqref{vertexRGflows}, the RG flow of $\Gamma^{2}_{\alpha\alpha'}$ can be attributed to the contribution from 4- and 6-point vertices to the RG flow of the single-particle self-energy
\begin{eqnarray}
\hspace*{-1cm}
\Delta\Sigma^{(j)}_{\Lambda,\hat{s},\sigma}(\omega_{(j)}) &=& \sum_{\gamma}\Gamma^{4,(j)}_{\alpha\gamma}(\omega_{(j)})G_{\gamma\gamma'}^{6,(j)}(\omega_{(j)})\Gamma^{4,(j)}_{\gamma'\alpha'}(\omega_{(j)})+\sum_{\gamma_{1}}\Gamma^{6,(j)}_{\alpha_{1}\gamma_{1}}(\omega_{(j)})G_{\gamma_{1}\gamma'_{1}}^{10,(j)}(\omega_{(j)})\Gamma^{6,(j)}_{\gamma_{1}'\alpha_{1}'}(\omega_{(j)})~.~~~~~~\label{1-p-self-energy}
\end{eqnarray}
Here, $G_{\gamma\gamma'}^{6,(j)}(\omega_{(j)})$ and $G_{\gamma\gamma'}^{10,(j)}(\omega_{(j)})$ are Green's function operators that contain the kinetic and interaction energies of three particles and five particles respectively, with explicit forms given by
\begin{eqnarray}
G_{\gamma\gamma'}^{6,(j)}(\omega_{(j)}) &=& \frac{8\tau_{\mathbf{k}_{\Lambda_{j}\hat{s}_{1}}\sigma'}\sigma_{\mathbf{p}-\mathbf{k}_{\Lambda_{j}\hat{s}_{1}}\sigma'}\sigma_{\mathbf{p}-\mathbf{k}_{\Lambda\hat{s}}\sigma}}{\omega_{(j)}-h^{1}_{(j)}-h^{2}_{(j)}-h^{3}_{(j)}}~,\nonumber\\
h^{1}_{(j)}&=&(\epsilon_{\mathbf{k}_{\Lambda_{j}\hat{s}_{1}}}+\Sigma^{(j)}_{\mathbf{k}_{\Lambda_{j}\hat{s}_{1}}}(\omega_{(j)}))\tau_{\mathbf{k}_{\Lambda_{j}\hat{s}_{1}}\sigma'}-(\epsilon_{\mathbf{p}-\mathbf{k}_{\Lambda_{j}\hat{s}_{1}}}+\Sigma^{(j)}_{\mathbf{p}-\mathbf{k}_{\Lambda_{j}\hat{s}_{1}}}(\omega_{(j)}))\sigma_{\mathbf{p}-\mathbf{k}_{\Lambda_{j}\hat{s}_{1}}\sigma'}\nonumber\\
&-&(\epsilon_{\mathbf{p}-\mathbf{k}_{\Lambda\hat{s}}}+\Sigma^{(j)}_{\mathbf{p}-\mathbf{k}_{\Lambda\hat{s}}}(\omega_{(j)}))\sigma_{\mathbf{p}-\mathbf{k}_{\Lambda\hat{s}}\sigma}\nonumber\\
h^{2}_{(j)}&=&\Gamma^{4,(j)}_{(\mathbf{k}_{\Lambda\hat{s}},\sigma'),(\mathbf{p}-\mathbf{k}_{\Lambda\hat{s}},\sigma)}
\sigma_{\mathbf{p}-\mathbf{k}_{\Lambda_{j}\hat{s}_{1}},\sigma'}
\sigma_{\mathbf{p}-\mathbf{k}_{\Lambda\hat{s}},\sigma}+\Gamma^{4,(j)}_{(\mathbf{k}_{\Lambda_{j}\hat{s}_{1}},\sigma'),(\mathbf{p}-\mathbf{k}_{\Lambda\hat{s}},\sigma)}
\tau_{\mathbf{k}_{\Lambda_{j}\hat{s}_{1}},\sigma'}
\sigma_{\mathbf{p}-\mathbf{k}_{\Lambda\hat{s}},\sigma}
\nonumber\\
&+&\Gamma^{4,(j)}_{(\mathbf{k}_{\Lambda_{j}\hat{s}_{1}},\sigma'),(\mathbf{p}-\mathbf{k}_{\Lambda_{j}\hat{s}_{1}},\sigma)}
\tau_{\mathbf{k}_{\Lambda_{j}\hat{s}_{1}},\sigma'}
\sigma_{\mathbf{p}-\mathbf{k}_{\Lambda_{j}\hat{s}},\sigma}\nonumber\\
h^{3}_{(j)}&=&\Gamma^{6,(j)}_{(\mathbf{k}_{\Lambda_{j}\hat{s}_{1}}\sigma'),(\mathbf{p}-\mathbf{k}_{\Lambda\hat{s}},\sigma),(\mathbf{p}-\mathbf{k}_{\Lambda_{j}\hat{s}_{1}},\sigma')}\sigma_{\mathbf{p}-\mathbf{k}_{\Lambda_{j}\hat{s}_{1}},\sigma'}
\sigma_{\mathbf{p}-\mathbf{k}_{\Lambda\hat{s}},\sigma}\tau_{\mathbf{k}_{\Lambda_{j}\hat{s}_{1}},\sigma'}~.\label{six_point_green}
\end{eqnarray}
and
\begin{eqnarray}
G^{10,(j)}_{\gamma\gamma'}(\omega_{(j)})&=&\frac{32\prod_{\substack{i=1\\ \mathbf{k}_{i}\sigma_{i}\in\gamma}}^{4}\sigma_{\mathbf{k}_{i},\sigma_{i}}\tau_{\mathbf{k}_{\Lambda_{j}\hat{s}_{1}},\sigma'}}{\omega - h_{(j)}}~,\nonumber\\
h_{(j)} &=& \sum_{(\mathbf{k}_{i},\sigma)\in\rho\subset \gamma}\Gamma^{2,(j)}_{\rho\rho'}\sigma_{\mathbf{k}_{i},\sigma}+\sum_{\substack{((\mathbf{k}_{i},\sigma),\\(\mathbf{k}_{j},\sigma'))\in\rho\subset\gamma}}\Gamma^{4,(j)}_{\rho\rho'}\sigma_{\mathbf{k}_{i},\sigma}\sigma_{\mathbf{k}_{j},\sigma'}\nonumber\\
&+&\sum_{\substack{((\mathbf{k}_{i},\sigma),\\(\mathbf{k}_{j},\sigma'),\\
(\mathbf{k}_{l},\sigma''))\in\rho\subset\gamma}}\Gamma^{6,(j)}_{\rho\rho'}\sigma_{\mathbf{k}_{i},\sigma}\sigma_{\mathbf{k}_{j},\sigma'}\tau_{\mathbf{k}_{\Lambda_{j}\hat{s}_{1}},\sigma'}~,~~~~
\end{eqnarray}
respectively. The labels $\alpha$, $\gamma$ and $\alpha'$ are defined similar to that discussed below eq.\eqref{cluster decomposition}, as the collection of pairwise labels $\lbrace(\mathbf{k},\sigma,\mu)\rbrace$, such that $l$ represents the electronic state labels and $\mu=0/1$ represent the electron annihilation/creation operator.
The labels $\alpha=\lbrace ((\mathbf{k}_{\Lambda\hat{s}},\sigma),\mu_{1}),((\mathbf{p}-\mathbf{k}_{\Lambda\hat{s}},\sigma'),\mu_{2})\rbrace$, $\gamma=\lbrace ((\mathbf{k}_{\Lambda_{j}\hat{s}},\sigma),\mu_{1}),((\mathbf{p}-\mathbf{k}_{\Lambda_{j}\hat{s}},\sigma'),\mu_{2})\rbrace$. $\alpha'$ and $\gamma'$ involve the same electronic states with $\mu_{1},\mu_{2}$ replaced by the complement $\bar{\mu}_{1,2}=1-\mu_{1,2}$. Similar definitions exist for $\alpha_{1}$, $\gamma_{1}$ and $\alpha_{1}'$, $\gamma_{1}'$. In the above equation, the operators $\tau_{\mathbf{k}\sigma}=n_{\mathbf{k}\sigma}-\frac{1}{2}$ represents decoupled degrees of freedom that commute with the Hamiltonian, while the operators $\sigma_{\mathbf{k}\sigma}=n_{\mathbf{k}\sigma}-\frac{1}{2}$ belong to the coupled space and do not commute with the Hamiltonian. We now derive the four-point and six-point vertex RG flow equations.
\subsection{Derivation of RG flow equations for 
$\Gamma^{4}_{\alpha\beta}$ and $\Gamma^{6}_{\alpha\beta}$}
\label{2pAnd3pScattVert}
\pin 
Using the RG flow equation heirarchy of eq.\eqref{vertexRGflows}, the RG flow equation for the four-point vertex at the $l$-th step and along a given isogeometric curve $j$ is given by
\begin{eqnarray}
\Delta\Gamma^{4,(j,l)}_{\alpha\beta}(\omega_{(j,l)})&=&\Gamma^{4,(j,l)}_{\alpha\beta}(\omega_{(j,l)})-\Gamma^{4,(j,l-1)}_{\alpha\beta}(\omega_{(j,l-1)})\nonumber\\
&=&\sum_{\gamma}\Gamma^{4,(j,l)}_{\alpha\gamma}G^{4,(j,l)}_{\gamma\gamma'}(\omega_{(j,l)})\Gamma^{4,(j,l)}_{\gamma'\beta}+\sum_{\gamma}\Gamma^{6,(j,l)}_{\alpha\gamma}G^{8,(j,l)}_{\gamma\gamma'}(\omega_{(j,l)})\Gamma^{6,(j,l)}_{\gamma'\beta}~.~~~~~~\label{4-pointvertexRGflow}
\end{eqnarray} 
Here, $G^{4,(j,l)}_{\gamma\gamma'}(\omega_{(j,l)})$, $G^{6,(j,l)}_{\gamma\gamma'}(\omega_{(j,l)})$, $G^{8,(j,l)}_{\gamma\gamma'}(\omega_{(j,l)})$ are four-, six- and eight-point Green's functions respectively. We also note that the four-point off-diagonal vertex scattering between normals $\hat{s}$ and $\hat{s}'$ takes places either directly ($\hat{s}\to\hat{s}'$) or via correlated (i.e., higher order) tangential scattering processes (e.g., $\hat{s}\to\hat{s}_{1}\to\hat{s}'$, $\hat{s}\to\hat{s}_{1}\to\hat{s}_{2}\to\hat{s}'$). Taken together, the first term in eq.\eqref{4-pointvertexRGflow}) involves the contribution to the four-point vertex RG flow $\Delta\Gamma^{4,(j)}$ in eq\eqref{4-pointvertexRGflow} due to other four-point vertices (including the effects of correlated scattering between states residing on the same isogeometric curve ($j$))
\begin{eqnarray}
\Delta\Gamma^{4,(j)}_{\alpha\beta}(\omega_{(j)})
&\to & \sum_{\gamma}\Gamma^{4,(j)}_{\alpha\gamma}G^{4,(j)}_{\gamma\gamma'}\Gamma^{4,(j)}_{\gamma'\beta}+\sum_{\gamma_{1}}\Gamma^{4,(j)}_{\alpha\gamma}G^{4,(j)}_{\gamma\gamma'}\Gamma^{4,(j)}_{\gamma'\gamma_{1}}G^{4,(j)}_{\gamma_{1}\gamma_{1}'}\Gamma^{4,(j)}_{\gamma_{1}'\beta}
+ ...\nonumber\\
&=&\hspace*{-0.3cm}\sum_{\gamma_{i}}\hspace*{-0.1cm}\prod_{i=1}^{n-1}\Gamma^{4,(j)}_{\alpha\gamma_{i}}G^{4,(j)}_{\gamma_{i}\gamma'_{i}}\Gamma^{4,(j)}_{\gamma'_{i}\gamma'_{i+1}}\hspace*{-0.1cm}G^{4,(j)}_{\gamma_{n}\gamma'_{n}}\Gamma^{4,(j)}_{\gamma'_{n}\beta}~.~
\end{eqnarray}
In the above expression, the renormalization contribution from the $k$-correlated 4-point off-diagonal scattering vertex (i.e., $\Gamma^{4}G^{4}\Gamma^{4}\ldots_{k-\text{times}}\ldots\Gamma^{4}G^{4}\Gamma^{4}$) for a $D$-dimensional system scales with volume as $L^{(k-1)(D-1)}L^{-kD}=L^{-D-k+1}$. Thus, in the thermodynamic limit (TL, $L\to \infty$), the leading contribution comes from the $k=2$ scattering (in comparison to all $k>2$ processes). The same holds true for the $\Gamma^{6}G^{8}\Gamma^{6}$ and $\Gamma^{6}G^{6}\Gamma^{4}$ renormalization contributions to $\Gamma^{4,(j)}$ in eq.\eqref{4-pointvertexRGflow}, with volume dependence $L^{-(k+1)D-k+1}$ and $L^{-3kD}$ respectively. 
\par\noindent
Accounting for the leading contributions in the thermodynamic limit, a compact form of the flow equations for 4-point vertices and 6 point off-diagonal vertices is given by
\begin{eqnarray}
\Delta\Gamma^{4,(j)}_{\alpha\beta}(\omega_{(j)})&=&
\sum_{\gamma}\Gamma^{4,(j)}_{\alpha\gamma}G^{4,(j)}_{\gamma\gamma'}(\omega_{(j)})\Gamma^{4,(j)}_{\gamma'\beta}+\sum_{\gamma}\Gamma^{6,(j)}_{\alpha\gamma}G^{8,(j)}_{\gamma\gamma'}(\omega_{(j)})\Gamma^{6,(j)}_{\gamma'\beta},~\nonumber\\
\Delta\Gamma^{6,(j)}_{\alpha\beta}(\omega_{(j)})&=&\sum_{\gamma}\Gamma^{4,(j)}_{\alpha\gamma}G^{2,(j)}_{\gamma\gamma'}(\omega_{(j)})\Gamma^{4,(j)}_{\gamma'\beta}+\sum_{\gamma}\Gamma^{6,(j)}_{\alpha\gamma}G^{6,(j)}_{\gamma\gamma'}(\omega_{(j)})\Gamma^{6,(j)}_{\gamma'\beta}~\nonumber\\
&+&\sum_{\gamma}\Gamma^{6,(j)}_{\alpha\gamma}G^{4,(j)}_{\gamma\gamma'}(\omega_{(j)})\Gamma^{4,(j)}_{\gamma'\beta}~.\label{4and6 particle vertex}
\end{eqnarray} 
Finally, the RG flow equations for the 6-point diagonal vertices are given by
\begin{eqnarray}
\Delta\Gamma^{6,(j)}_{\alpha\alpha'}(\omega_{(j)})&=&\sum_{\gamma}\Gamma^{4,(j)}_{\alpha\gamma}G^{2,(j)}_{\gamma\gamma'}(\omega_{(j)})\Gamma^{4,(j)}_{\gamma'\beta}+\sum_{\gamma}\Gamma^{6,(j)}_{\alpha\gamma}G^{6,(j)}_{\gamma\gamma'}(\omega_{(j)})\Gamma^{6,(j)}_{\gamma'\beta}~.\label{six-point-D-vertices}
\end{eqnarray}
In the above flow equations, the Green's functions $G^{2,(j)}_{\gamma\gamma'}(\omega_{(j)})$, $G^{4,(j)}_{\gamma\gamma'}(\omega_{(j)})$, $G^{6,(j)}_{\gamma\gamma'}(\omega_{(j)})$ and $G^{8,(j)}_{\gamma\gamma'}(\omega_{(j)})$ contain the energy contributions from one-particle self energies, as well as two- and three-particle correlation energies. The form of the two- particle Green's function $G^{4,(j)}_{\gamma\gamma'}(\omega_{(j)})$ is given by
\begin{eqnarray}
&&G^{4,(j)}_{\gamma\gamma'}(\omega_{(j)}) = \frac{4\tau_{\mathbf{k}_{\Lambda_{j}\hat{s}_{1}},\sigma'}\sigma_{\mathbf{p}-\mathbf{k}_{\Lambda_{j}\hat{s}_{1}},\sigma}}{\omega_{(j)}-h^{1}_{(j)}-h^{2}_{(j)}}\nonumber\\
&&h^{1}_{(j)}=\epsilon_{\mathbf{k}_{\Lambda_{j}\hat{s}_{1}}}\tau_{\mathbf{k}_{\Lambda_{j}\hat{s}_{1}},\sigma'}+\epsilon_{\mathbf{p}-\mathbf{k}_{\Lambda_{j}\hat{s}_{1}}}\sigma_{\mathbf{p}-\mathbf{k}_{\Lambda_{j}\hat{s}_{1}},\sigma}, h^{2}_{(j)}=\Gamma^{4,(j)}_{(\mathbf{k}_{\Lambda_{j}\hat{s}_{1}},\sigma'),(\mathbf{p}-\mathbf{k}_{\Lambda_{j}\hat{s}_{1}},\sigma)}\tau_{\mathbf{k}_{\Lambda_{j}\hat{s}_{1}},\sigma'}\tau_{\mathbf{p}-\mathbf{k}_{\Lambda_{j}\hat{s}_{1}},\sigma}~.~~~~~~~~\label{4-point-green}
\end{eqnarray}
The four-particle Green's function $G^{8,(j)}_{\gamma\gamma'}(\omega_{(j)})$ can be obtained similarly, while the three-particle Green's function $G^{6,(j)}_{\gamma\gamma'}(\omega_{(j)})$ has already been given earlier.
\pin
An important point can now be made. When relevant under RG flow, the contribution of six-point scattering vertices $\Gamma^{6,(j)}_{\alpha\beta}$ is responsible for the dynamical mixing of opposite spin electron-electron and electron-hole configurations. This feature results from the non-commutativity between the composite electron creation operator $(1-\hat{n}_{\mathbf{k}\sigma})c^{\dagger}_{\mathbf{k}'\sigma'}$ and the ee/eh pseudospin pair operators~\cite{anderson1958random}, $c^{\dagger}_{\mathbf{k}\sigma}c^{\dagger}_{\mathbf{k}'\sigma'}$ and $c^{\dagger}_{\mathbf{k}\sigma}c_{\mathbf{k}'\sigma'}$. In order to incorporate this effect within the four-point vertex RG equations, we follow Refs.\cite{anirbanmotti,anirbanurg1} and perform an $\omega$-dependent rotation, $\tan^{-1}(\sqrt{\frac{1-p}{p}})$, in the space of the electron/hole configurations of the pair of electronic states: $|1_{\mathbf{k}_{\Lambda\hat{s}}}1_{\mathbf{p}-\mathbf{k}_{\Lambda\hat{s}}}\rangle$ and $|1_{\mathbf{k}_{\Lambda\hat{s}}}0_{\mathbf{p}-\mathbf{k}_{\Lambda\hat{s}}}\rangle$. This is manifested in the RG equations of $\Gamma^{4,(j)}$, obtained in a rotated basis of occupied and unoccupied electronic states
\begin{eqnarray}
\Delta\Gamma^{4,(j)}_{\alpha\beta}(\omega_{(j)})&=&\sum_{\gamma}\bigg[\frac{p\Gamma^{4,(j)}_{\alpha\gamma}\Gamma^{4,(j)}_{\gamma\beta}}{\omega_{(j)} -\epsilon^{(j)}_{p,\mathbf{k}_{\Lambda_{j}\hat{s}'},\mathbf{p}}-\frac{2p-1}{4}\Gamma^{4,(j)}_{\gamma\gamma'}}-\frac{(1-p)\Gamma^{4,(j)}_{\alpha_{1}\gamma_{1}}\Gamma^{4,(j)}_{\gamma_{1}\beta_{1}}}{\omega_{(j)} -\epsilon^{(j)}_{p,\mathbf{k}_{\Lambda_{j}\hat{s}'},\mathbf{p}'}-\frac{2p-1}{4}\Gamma^{4,(j)}_{\gamma_{1}\gamma'_{1}}}\bigg]~,\nonumber\\
&+&\sum_{\gamma}\Gamma^{6,(j)}_{\alpha\gamma}G^{8,(j)}_{\gamma\gamma'}(\omega_{(j)})\Gamma^{6,(j)}_{\gamma'\beta}~,\label{4-particle vertex}
\end{eqnarray} 
where the ee/eh hybridized pair-dispersion is given by
\begin{eqnarray}
\epsilon^{(j)}_{p,\mathbf{k}_{\Lambda_{j}\hat{s}'},\mathbf{p}} &=& \frac{p}{2}(\epsilon^{(j)}_{\mathbf{k}_{\Lambda_{j}\hat{s}'}}+\epsilon^{(j)}_{\mathbf{p}-\mathbf{k}_{\Lambda_{j}\hat{s}'}})+\frac{1-p}{2}(\epsilon^{(j)}_{\mathbf{k}_{\Lambda_{j}\hat{s}'}}-\epsilon^{(j)}_{\mathbf{p}-\mathbf{k}_{\Lambda_{j}\hat{s}'}})~.\label{hybridizedKE}
\end{eqnarray}
In the above RG equation, the indices $\alpha,\beta,\gamma$, $\alpha',\beta',\gamma_{1}$ are related as follows: for $\alpha=\lbrace(\mathbf{k},\sigma,1),(\mathbf{p}-\mathbf{k},\sigma',1)\rbrace$, $\beta=\lbrace(\mathbf{k}',\sigma,0),(\mathbf{p}-\mathbf{k}',\sigma',0)\rbrace$ and $\gamma=\lbrace(\mathbf{k}_{\Lambda_{j}\hat{s}_{1}},\sigma,0),(\mathbf{p}-\mathbf{k}_{\Lambda_{j}\hat{s}_{1}},\sigma',0)\rbrace$, the indices $\alpha_{1}=\lbrace(\mathbf{k},\sigma,1),(\mathbf{p}-\mathbf{k},\sigma',0)\rbrace$, $\beta_{1}=\lbrace(\mathbf{k}',\sigma,0),(\mathbf{p}-\mathbf{k},\sigma',1)\rbrace$ and $\gamma_{1}=\lbrace(\mathbf{k}_{\Lambda_{j}\hat{s}},\sigma,0),(\mathbf{p}+\mathbf{k}_{\Lambda_{j}\hat{s}}-\mathbf{k}'-\mathbf{k},\sigma',1)\rbrace$. In order to manifest the dominant effect of the six-point off-diagonal vertices, the hybridisation parameter $p(\omega)$ is chosen so as to maximize the spin-charge hybridized Green's function $G^{p,j}(\omega_{(j)})=(\omega_{(j)} -\epsilon^{(j)}_{p,\mathbf{k}_{\Lambda_{j}\hat{s}'},\mathbf{p}})^{-1}$,
\begin{eqnarray}
p:=p'~s.t.~G^{p,j}(\omega_{(j)})=\max_{0<p'<1}G^{p',j}(\omega_{(j)})~. \label{p}
\end{eqnarray}
With these RG equations in place, we have laid the platform for investigating the low energy fixed point Hamiltonians of various quantum fluctuation energy scales and parameter regimes. 
\subsection{RG flows towards Fermi liquid and BCS fixed points}\label{conditionRG}
\par
\begin{figure*}
\resizebox{!}{0.5\textwidth}
{\begin{tikzpicture}
[sibling distance=4.5cm]
  \node {$H_{(j)}$}
    child {node{$\Sigma^{(j)}$}}
    child {[sibling distance=6cm] 
        node {$\Gamma^{4,(j)}$}
      child {[sibling distance=3cm]
         node {$\Gamma_{\alpha\alpha'}^{4,(j)}$}
        child{[sibling distance=1.4cm] 
           node {$V^{\sigma\sigma',(j)}_{\mathbf{k},\mathbf{p}}$ $\ldots$}
          child{node {$V^{\sigma\sigma,(j)}_{\mathbf{k},\mathbf{p}}$}}
          child{node {$V^{\sigma~-\sigma,(j)}_{\mathbf{k},\mathbf{p}}\ldots$}}}
        child{[sibling distance=1.4cm]
           node {$\ldots$ $V^{\sigma\sigma',(j)}_{\mathbf{k},\mathbf{p}'}$}
           child{node {$V^{\sigma\sigma,(j)}_{\mathbf{k},\mathbf{p}'}$}}
          child{node {$V^{\sigma~-\sigma,(j)}_{\mathbf{k},\mathbf{p}'}$}}
        }}
        child{[sibling distance=3cm]
           node {$\Gamma_{\alpha\beta}^{4,(j)}$}
        child{[sibling distance=1.4cm]
           node {$V^{\sigma\sigma',(j)}_{\mathbf{k},\mathbf{k}',\mathbf{p}}$ $\ldots$}
          child{node {$V^{\sigma\sigma,(j)}_{\mathbf{k},\mathbf{k}',\mathbf{p}}$}}
          child{node {$V^{\sigma~-\sigma,(j)}_{\mathbf{k},\mathbf{k}',\mathbf{p}}\ldots$}}}
        child{[sibling distance=1.4cm]
           node {$\ldots$ $V^{\sigma\sigma',(j)}_{\mathbf{k},\mathbf{k}',\mathbf{p}'}$}
           child{node {$V^{\sigma\sigma,(j)}_{\mathbf{k},\mathbf{k}',\mathbf{p}'}$}}
          child{node {$V^{\sigma~-\sigma,(j)}_{\mathbf{k},\mathbf{k}',\mathbf{p}'}$}}
        }}}
    child {[sibling distance=1cm] 
       node {$\Gamma^{6,(j)}$}
      child {node {$\Gamma^{6,(j)}_{\alpha\alpha'}$}}
      child {node {$\Gamma^{6,(j)}_{\alpha\beta}$}}};~.
\end{tikzpicture}}
\caption{Tree tensor network representation of the $2$-point, $4$-point and $6$-point vertices, where the $4$-point and $6$-point vertex tensor are comprised of diagonal and off-diagonal parts. The $4$-point vertex tensor is further decomposed in various total pair-momentum $\mathbf{p}$ and spin-pair $\sigma\sigma'$ channels. Note that $\alpha:=\lbrace(l,\mu)\rbrace$ and $\alpha':=\lbrace(l,\bar{\mu})\rbrace$ and $\beta:=\lbrace(l',\bar{\mu})\rbrace$, where the index $l$ represents a collection of labels marking states in the coupled space and $\mu=1/0$ refers to a particular state being either occupied or unoccupied (converse for $\bar{\mu}$).}\label{246vertices-tree}
\end{figure*}
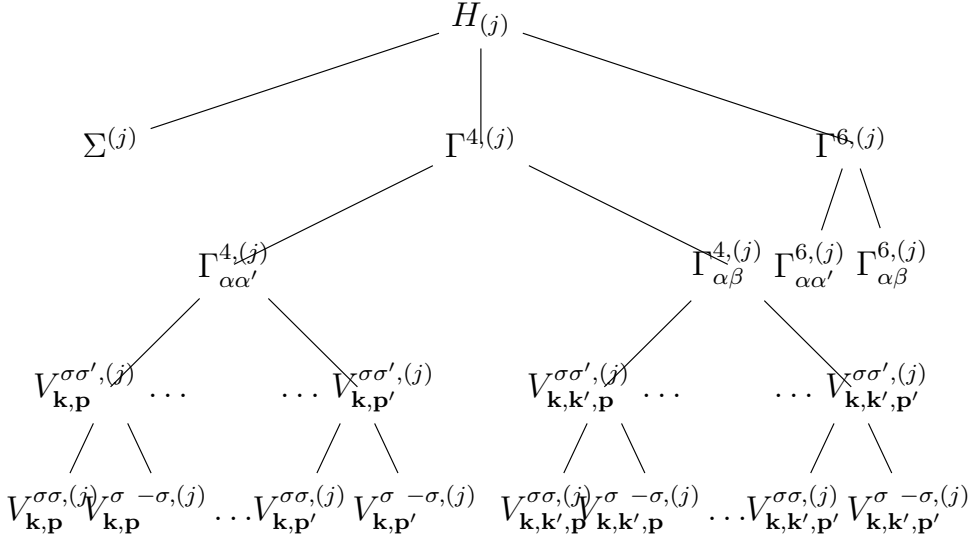

\noindent
We begin with an illustration in Fig.\ref{246vertices-tree} of various 2-, 4- and 6-point vertices obtained from a tree-like decomposition of the heirarchical Hamiltonian RG flow. This will be seen to assist us in describing the RG flows towards various stable phases of fermionic quantum matter obtained across several parameter regimes.
The first level of branching involves the separation into the 2-point (self-energy $\Sigma$), 4-point ($\Gamma^{4}$) and 6-point ($\Gamma^{6}$) vertices. A second level of branching involves that of the 4- and 6-point vertices $(\Gamma^{4,(j)}, \Gamma^{6,(j)})$ into diagonal $(\Gamma^{4,(j)}_{\alpha\alpha'},\Gamma^{6,(j)}_{\alpha\alpha'})$ and off-diagonal $(\Gamma^{4,(j)}_{\alpha\beta},\Gamma^{6,(j)}_{\alpha\beta})$ elements. In order to make the tensorial dependence on momenta and spin indices explicit, the diagonal and off-diagonal 4-point vertex elements $(\Gamma^{4,(j)}_{\alpha\alpha'}$, $\Gamma^{4,(j)}_{\alpha\beta})$ are further decomposed into different pair-momenta $\mathbf{p}$ and spin-pair channels $\sigma,\sigma'=\sigma,\sigma$ and $\sigma,\sigma'=\sigma, -\sigma$ as $\Gamma^{4,(j)}_{\alpha\alpha'}\equiv V^{\sigma\sigma',(j)}_{\mathbf{k}\mathbf{p}}$, $\Gamma^{4,(j)}_{\alpha\beta}\equiv V^{\sigma\sigma',(j)}_{\mathbf{k}\mathbf{k}'\mathbf{p}}$ respectively. The notation $\alpha\beta$ and $\alpha\alpha'$ employed for the 4-point vertices are 4-element sets chosen as follows: 
\begin{eqnarray}
\alpha\beta &=& \lbrace (\mathbf{k},\sigma,1),(\mathbf{p}-\mathbf{k},\sigma',1),(\mathbf{p}-\mathbf{k}',\sigma',0),(\mathbf{k}',\sigma',0)\rbrace~,\nonumber\\
\alpha\alpha' &=& \lbrace (\mathbf{k},\sigma,\mu_{1}),(\mathbf{p}-\mathbf{k},\sigma',\mu_{2}),(\mathbf{p}-\mathbf{k},\sigma',\bar{\mu}_{2}),(\mathbf{k},\sigma,\bar{\mu}_{1})\rbrace~,
\end{eqnarray}
where the first two indices label the state and $\mu=1,0$ represents creation or annhilation operation on that state and $\bar{\mu}$ is its compliment.
\pin 
Similarly, the 6-point vertices are decomposed into diagonal $\Gamma^{6,(j)}_{\alpha\alpha'}\equiv R^{\sigma\sigma'\sigma,(j)}_{\mathbf{k}\mathbf{k}'\mathbf{k}''}$ and off-diagonal vertices $\Gamma^{6,(j)}_{\alpha\beta}\equiv R^{\sigma\sigma'\sigma,(j)}_{\mathbf{k}\mathbf{k}_{1}\mathbf{k}_{2}\mathbf{k}_{3}\mathbf{k}_{4}}$. Again, the notation $\alpha\beta$ and $\alpha\alpha'$ for the six point vertices are 6-element sets chosen as follows: 
\begin{eqnarray}
\alpha\beta &=& \bigg\lbrace(\mathbf{k},\sigma,1),(\mathbf{p}-\mathbf{k},\sigma',1),(\mathbf{p}'-\mathbf{k}',\sigma',1),(\mathbf{p}-\mathbf{k}',\sigma',0),(\mathbf{k}'',\sigma',0),(\mathbf{p}'-\mathbf{k}'',\sigma,0) \bigg\rbrace\nonumber\\
\alpha\alpha'&=&\bigg\lbrace(\mathbf{k},\sigma,\mu_{1}),(\mathbf{p}-\mathbf{k},\sigma',\mu_{2}),(\mathbf{p}-\mathbf{k}',\sigma',\mu_{3}),(\mathbf{p}-\mathbf{k}',\sigma',\bar{\mu}_{3}),(\mathbf{k},\sigma,\bar{\mu}_{1}),(\mathbf{p}-\mathbf{k},\sigma',\bar{\mu}_{2}) \bigg\rbrace~.~~~
\end{eqnarray}
\par
\begin{figure}
\centering
\begin{tikzpicture}
[sibling distance=5.5cm]
  \node {$H_{(j)}$}
    child {[blue] node[blue]{$\Sigma^{(j)}$}}
    child {[blue][sibling distance=7cm] 
        node[blue] {$\Gamma^{4,(j)}$}
      child {[blue,sibling distance=2.9cm]
         node[blue] {$\Gamma_{\alpha\alpha'}^{4,(j)}$}
        child{[sibling distance=1.8cm] 
           node {$V^{\sigma\sigma',(j)}_{\mathbf{k},\mathbf{p}}$ $\ldots$}
          child{node {$V^{\sigma\sigma,(j)}_{\mathbf{k},\mathbf{p}}$}}
          child{node {$V^{\sigma~-\sigma,(j)}_{\mathbf{k},\mathbf{p}}\ldots$}}}
        child{[sibling distance=1.7cm]
           node {$\ldots$ $V^{\sigma\sigma',(j)}_{\mathbf{k},\mathbf{p}'}$}
           child{node {$V^{\sigma\sigma,(j)}_{\mathbf{k},\mathbf{p}'}$}}
          child{node {$V^{\sigma~-\sigma,(j)}_{\mathbf{k},\mathbf{p}'}$}}
        }}
        child{[red][sibling distance=3.5cm]
           node[red] {$\Gamma_{\alpha\beta}^{4,(j)}$}
        child{[sibling distance=1.8cm]
           node {$V^{\sigma\sigma',(j)}_{\mathbf{k},\mathbf{k}',\mathbf{p}}$ $\ldots$ }
          child{node {$V^{\sigma\sigma,(j)}_{\mathbf{k},\mathbf{k}',\mathbf{p}}$}}
          child{node {$V^{\sigma~-\sigma,(j)}_{\mathbf{k},\mathbf{k}',\mathbf{p}}\ldots$}}}
        child{[sibling distance=1.8cm]
           node {$\ldots$ $V^{\sigma\sigma',(j)}_{\mathbf{k},\mathbf{k}',\mathbf{p}'}$}
           child{node {$V^{\sigma\sigma,(j)}_{\mathbf{k},\mathbf{k}',\mathbf{p}'}$}}
          child{node {$V^{\sigma~-\sigma,(j)}_{\mathbf{k},\mathbf{k}',\mathbf{p}'}$}}
        }}}
    child[red] {[sibling distance=1.5cm] 
       node {$\Gamma^{6,(j)}$}
      child {node {$\Gamma^{6,(j)}_{\alpha\alpha'}$}}
      child {node {$\Gamma^{6,(j)}_{\alpha\beta}$}}};~.
\end{tikzpicture}
\caption{Tree tensor diagram representing the Fermi liquid (regime-I in Table \ref{Table-conditions-I}). The off-diagonal $4$-point vertices are RG irrelevant (i.e., flow towards zero) and are represented in red. Further, diagonal and off-diagonal $6$-point vertices are irrelevant and are represented in red colour. The $1$-particle self-energy and the 2-particle Hartree contribution is RG relevant, approaches fixed point values, and are reprented in blue.}\label{246vertices-tree-FL}
\end{figure}
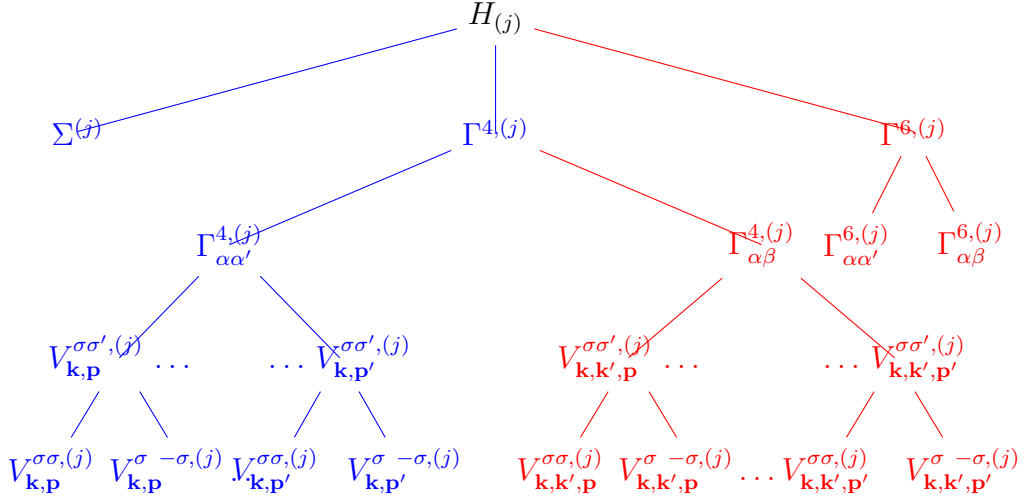

\begin{figure}
\centering
\includegraphics[width=0.6\textwidth]{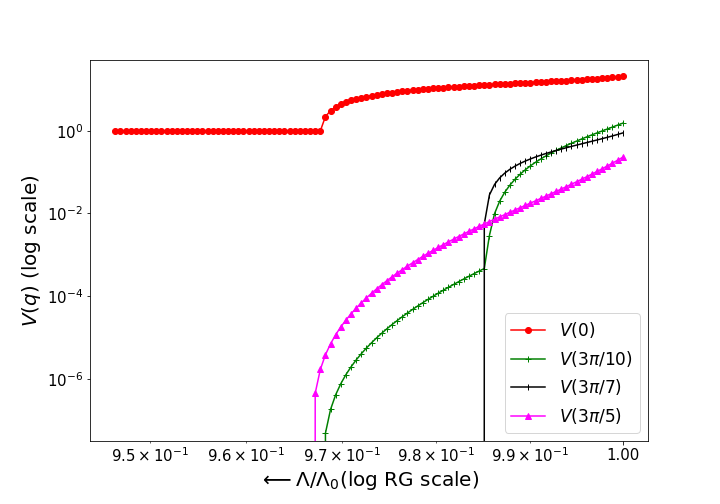}
\caption{Log-log plot for renormalization of two-particle vertices $V(q)$ with momentum transfer $|\mathbf{q}|=0,3\pi/10,3\pi/7$ and $3\pi/5$. The $x$-axis is the RG scale $\Lambda/\Lambda_{0}$ and the $y$-axis is the magnitude of the scattering vertex $V(q)$. For the numerical evaluation, we have taken a system volume of $1024\times 1024$ lattice sites, $\omega=\epsilon_{\Lambda_{0}}+0.1$ ($\Lambda_{0}=\pi/20$) and $V(q)=\eta^{2}V(0)/(q^{2}+\eta^{2})$ with $\eta=0.2$, $V(0)=20.5$.}\label{2particleVertices}
\end{figure}

\begin{figure}
\hspace*{-0.5cm}
\includegraphics[width=0.5\textwidth]{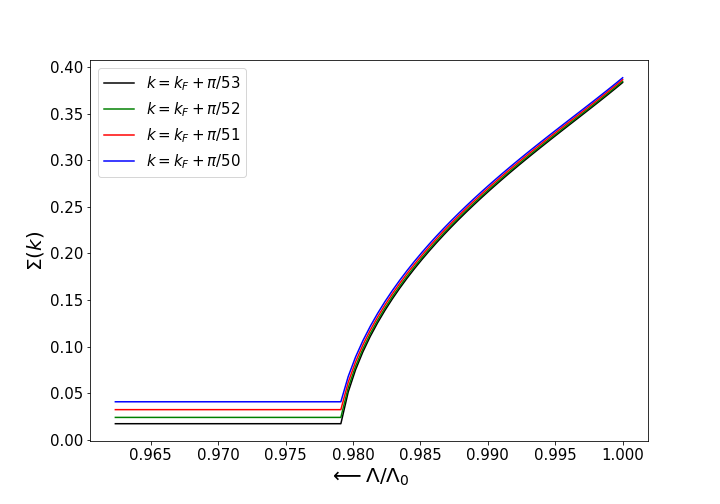}
\includegraphics[width=0.5\textwidth]{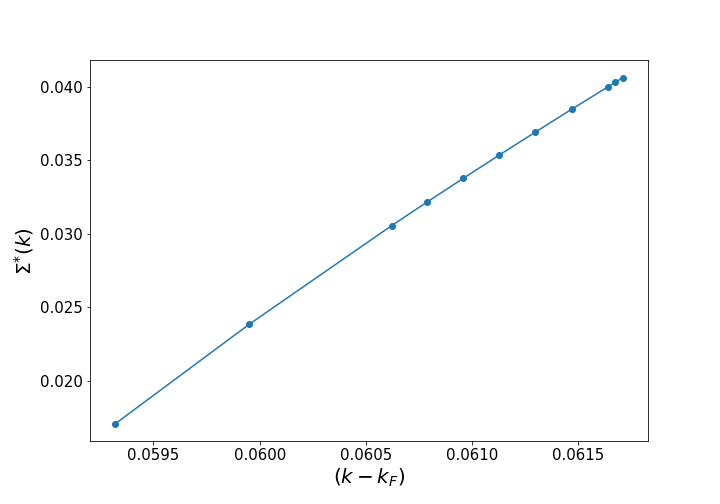}
\caption{Left Panel: RG flow of the single-particle self-energy for four different wave-vectors $\mathbf{k}$ close to $\mathbf{k}_{F}$. Right Panel: Variation of renormalized single-particle self-energy at the IR fixed point upon approacing the Fermi surface with lowering $|\mathbf{k}-\mathbf{k}_{F}|$.}\label{SelfEnergy}
\end{figure}

\begin{figure}
\hspace*{-1.5cm}
\includegraphics[width=0.6\textwidth]{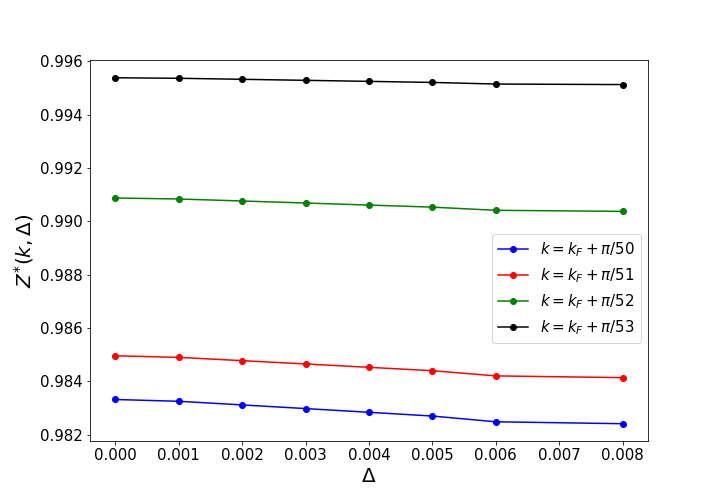}
\includegraphics[width=0.6\textwidth]{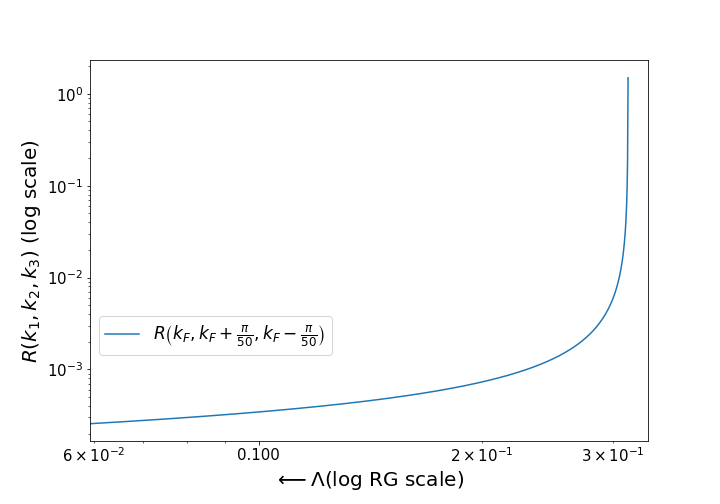}
\caption{Left Panel: Variation of the quasiparticle residue for eletronic states $\mathbf{k}$ with lowering excitation energy scale $\Delta$ about $E_{F}$. Right Panel: RG flow of the number diagonal three-particle vertex with wave-vectors $k_{F}$, $k_{F}+\frac{\pi}{50}$, $k_{F}-\frac{\pi}{50}$.}\label{3pEnergyAndResidue}
\end{figure}

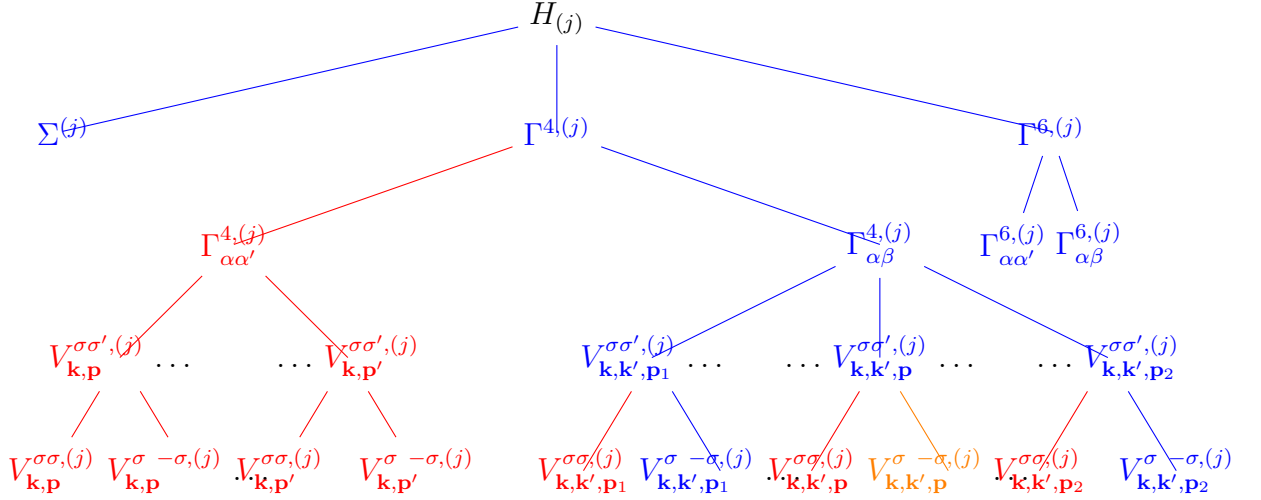
\begin{figure}
\centering
\begin{tikzpicture}
[sibling distance=6.5cm]
  \node {$H_{(j)}$}
    child {[blue] node[blue]{$\Sigma^{(j)}$}}
    child {[blue][sibling distance=8.5cm] 
        node[blue] {$\Gamma^{4,(j)}$}
      child {[red,sibling distance=3cm]
         node[red] {$\Gamma_{\alpha\alpha'}^{4,(j)}$}
        child{[sibling distance=1.8cm] 
           node {$V^{\sigma\sigma',(j)}_{\mathbf{k},\mathbf{p}}$ $\color{black} \ldots$}
          child{node {$V^{\sigma\sigma,(j)}_{\mathbf{k},\mathbf{p}}$}}
          child{node {$V^{\sigma~-\sigma,(j)}_{\mathbf{k},\mathbf{p}}$ $\color{black} \ldots$}}}
        child{[sibling distance=1.8cm]
           node { $\color{black} \ldots$ $V^{\sigma\sigma',(j)}_{\mathbf{k},\mathbf{p}'}$}
           child{node {$V^{\sigma\sigma,(j)}_{\mathbf{k},\mathbf{p}'}$}}
          child{node {$V^{\sigma~-\sigma,(j)}_{\mathbf{k},\mathbf{p}'}$}}
        }}
        child{[blue][sibling distance=3cm]
           node[blue] {$\Gamma_{\alpha\beta}^{4,(j)}$}
        child{[sibling distance=1.8cm]
           node {$V^{\sigma\sigma',(j)}_{\mathbf{k},\mathbf{k}',\mathbf{p}_{1}}$ $\color{black} \ldots$}
          child{[red]node {$V^{\sigma\sigma,(j)}_{\mathbf{k},\mathbf{k}',\mathbf{p}_{1}}$}}
          child{[blue]node[blue] {$V^{\sigma~-\sigma,(j)}_{\mathbf{k},\mathbf{k}',\mathbf{p}_{1}}$ $\color{black} \ldots$}}}
          child{[sibling distance=1.8cm]
           node { $\color{black} \ldots$ $V^{\sigma\sigma',(j)}_{\mathbf{k},\mathbf{k}',\mathbf{p}}$ $\color{black} \ldots$}
          child{[red]node {$V^{\sigma\sigma,(j)}_{\mathbf{k},\mathbf{k}',\mathbf{p}}$}}
          child{[orange]node[orange] {$V^{\sigma~-\sigma,(j)}_{\mathbf{k},\mathbf{k}',\mathbf{p}}$ $\color{black} \ldots$}}}
        child{[sibling distance=1.8cm]
           node { $\color{black} \ldots$ $V^{\sigma\sigma',(j)}_{\mathbf{k},\mathbf{k}',\mathbf{p}_{2}}$}
           child{[red]node {$V^{\sigma\sigma,(j)}_{\mathbf{k},\mathbf{k}',\mathbf{p}_{2}}$}}
          child{[blue]node {$V^{\sigma~-\sigma,(j)}_{\mathbf{k},\mathbf{k}',\mathbf{p}_{2}}$}}
        }}}
    child[blue] {[sibling distance=1cm] 
       node {$\Gamma^{6,(j)}$}
      child {node {$\Gamma^{6,(j)}_{\alpha\alpha'}$}}
      child {node {$\Gamma^{6,(j)}_{\alpha\beta}$}}};~.
\end{tikzpicture}
\caption{Tree tensor diagram representing the zero-momentum ($\mathbf{p}=0$) pairing reduced BCS model (regime-II in Table \ref{Table-conditions-I}), and the SPDW state composed of finite-momentum ($\mathbf{p}$) pseudospin-pairs interacting via XY interaction (regime-IV in Table \ref{Table-conditions-I}). The blue branches and the nodes of the vertex tensors represent the relevant scattering vertices, while orange branches and nodes represent the dominant RG relevant scattering vertices. The red branches and nodes represent RG irrelevant scattering vertices.}\label{246vertices-tree-XY-BCS}
\end{figure}

\noindent
Given the complex tensorial structure of the vertices, a simplified representation is needed by which families of RG flow equations can be characterized into different phases. Thus, we define the quantities
 \begin{eqnarray}
 r^{\sigma\sigma',(j)}_{\mathbf{p}}&=& \frac{sgn(V^{\sigma\sigma',(j)}_{\mathbf{k}\mathbf{k}'\mathbf{p}})\sqrt{\sum\limits_{\mathbf{k}\mathbf{k}'}|V^{\sigma\sigma',(j)}_{\mathbf{k}\mathbf{k}'\mathbf{p}}|^{2}}}{\sqrt{\sum\limits_{\mathbf{k}\neq \mathbf{k}'}|V^{\sigma\sigma',(j)}_{\mathbf{k}\mathbf{k}'\mathbf{p}}|^{2}+\sum\limits_{\mathbf{k}}|V^{\sigma\sigma',(j)}_{\mathbf{k}\mathbf{p}}|^{2}}}~,\nonumber\\
 s^{\sigma\sigma'\sigma,(j)}&=& \frac{\sqrt{\sum\limits_{\mathbf{k}\mathbf{k}'\mathbf{k}''}|R^{\sigma\sigma'\sigma,(j)}_{\mathbf{k},\mathbf{k}'\mathbf{k''}}|^{2}}}{\sqrt{\sum\limits_{\mathbf{k}\mathbf{k}'\mathbf{k}''}|R^{\sigma\sigma'\sigma,(j)}_{\mathbf{k}\mathbf{k}'\mathbf{k''}}|^{2}+\sum\limits_{\substack{\mathbf{k}\mathbf{k}_{1}\mathbf{k}_{2}\\ \mathbf{k}_{3}\mathbf{k}_{4}}}|R^{\sigma\sigma'\sigma,(j)}_{\mathbf{k}\mathbf{k}_{1}\mathbf{k}_{2}\mathbf{k}_{3}\mathbf{k}_{4}}|^{2}}}~.~~~\label{2and3particleVertexcond}
\end{eqnarray}
Importantly, we add that our analysis is confined to models with off-diagonal 4-point vertices that are either $V^{\sigma\sigma'}_{\mathbf{k}\mathbf{k}'\mathbf{p}}<0$ or $V^{\sigma\sigma'}_{\mathbf{k}\mathbf{k}'\mathbf{p}}>0$ for all $\mathbf{k},\mathbf{k}',\mathbf{p}$. It is clear from this assumption that $-1<r^{\sigma\sigma',(j)}_{\mathbf{p}}<1$, $0<s^{\sigma\sigma'\sigma,(j)}<1$. 
As shown in Tables \ref{Table-conditions-I} and \ref{Table-conditions-II}, we can now tabulate the stable fixed point theories corresponding to various low-energy regimes by using the relevance, irrelevance and dominance criteria of various $\Gamma^{2,(j)}$, $\Gamma^{4,(j)}$ and $\Gamma^{6,(j)}$ vertex RG flows. Further, the RG flows for regimes (I-III) in Table \ref{Table-conditions-I} and regimes (IV and V) in Table \ref{Table-conditions-II} are represented via tree diagrams in Figs.\ref{246vertices-tree-FL}-\ref{246vertices-tree-XXZ-BCS}. Below, we discuss various low energy fixed point theories arising from these RG flow equations.\\
\begin{table}
\hspace*{-1cm}
\begin{tabular}{|c|c|c|}
\hline
&\textbf{Cases}&$r^{\sigma\sigma',(j)}_{\mathbf{p}}$, $s^{\sigma\sigma'\sigma,(j)}$ \\
\hline
I & \parbox{.5\textwidth}{\begin{enumerate}
            \item $\frac{V^{\sigma\sigma',(j)}_{\mathbf{k}_{\Lambda_{j}\hat{s}},\mathbf{p}}}{4}>\omega_{(j)}-\frac{1}{2}(\epsilon^{(j)}_{\mathbf{k}_{\Lambda_{j}\hat{s}}}+\epsilon^{(j)}_{\mathbf{p}-\mathbf{k}_{\Lambda_{j}\hat{s}}})>0$\\
            \item $V^{\sigma\sigma',(j)}_{\mathbf{k}_{\Lambda_{j}\hat{s}},\mathbf{p}}>V^{\sigma\sigma',(j)}_{\mathbf{k}_{\Lambda_{j}\hat{s}},\mathbf{k},\mathbf{p}}$
        \end{enumerate}} & \parbox{.5\textwidth}{\begin{enumerate}\item $r^{\sigma\sigma',(j)}_{\mathbf{p}}\to 0+$ for all $\mathbf{p}$\\
        \item $s^{\sigma\sigma'\sigma,(j)}\to 0+$ 
        \end{enumerate}}\\\hline
II & \parbox{.5\textwidth}{\begin{enumerate}\item $\omega_{(j)}-\frac{1}{2}(\epsilon^{(j)}_{\mathbf{k}_{\Lambda_{j}\hat{s}}}+\epsilon^{(j)}_{-\mathbf{k}_{\Lambda_{j}\hat{s}}})<0$ \\ \item $\epsilon^{(j)}_{\mathbf{k}_{\Lambda_{j}\hat{s}}}>E_{F}$\\ \item $V^{\sigma,\sigma',(j)}_{\mathbf{k},\mathbf{p}}>0$\\ \item $V^{\sigma,-\sigma,(j)}_{\mathbf{k}_{\Lambda_{j}\hat{s}},\mathbf{k},\mathbf{p}}<0$\\ \item $V^{\sigma,\sigma,(j)}_{\mathbf{k}_{\Lambda_{j}\hat{s}},\mathbf{k},\mathbf{p}}>0$\\ \item $|V^{\sigma,-\sigma,(j)}_{\mathbf{k}_{\Lambda_{j}\hat{s}},\mathbf{p}}|<|V^{\sigma,-\sigma,(j)}_{\mathbf{k}_{\Lambda_{j}\hat{s}},\mathbf{k},\mathbf{p}}|$
\end{enumerate}} & \parbox{.5\textwidth}{\begin{enumerate}\item $r^{\sigma,-\sigma,(j)}_{\mathbf{p}=0}\to -1$\\  \item $\frac{V^{\sigma,-\sigma}_{\mathbf{k}\mathbf{k}'\mathbf{p}'}}{V^{\sigma,-\sigma}_{\mathbf{k}\mathbf{k}'\mathbf{p}=0}}\bigg\vert_{L\to \infty}\to 0+$~\text{for}~ $\epsilon_{\mathbf{p}'-\mathbf{k}}>E_{F}$\\ \item $r^{\sigma,-\sigma,(j)}_{\mathbf{p}'}\to 0+$ for $\epsilon_{\mathbf{p}'-\mathbf{k}}<E_{F}$\\ \item  $r^{\sigma,\sigma,(j)}_{\mathbf{p}}\to 0$ for all $\mathbf{p}$\\ \item  $\frac{s^{\sigma\sigma'\sigma,(j)}}{r_{\mathbf{p}=0}^{\sigma,-\sigma,(j)}}\bigg\vert_{L\to \infty}\to 0+$ \end{enumerate}}  \\
\hline
III & \parbox{.5\textwidth}{\begin{enumerate}\item $\omega_{(j)}-\frac{1}{2}(\epsilon^{(j)}_{\mathbf{k}_{\Lambda_{j}\hat{s}}}+\epsilon^{(j)}_{-\mathbf{k}_{\Lambda_{j}\hat{s}}})<0$\\  \item $\epsilon^{(j)}_{\mathbf{k}_{\Lambda_{j}\hat{s}}}>E_{F}$\\ \item $V^{\sigma,\sigma',(j)}_{\mathbf{k},\mathbf{p}}>0$\\ \item $V^{\sigma,-\sigma,(j)}_{\mathbf{k}_{\Lambda_{j}\hat{s}},\mathbf{k},\mathbf{p}}<0$\\ \item $V^{\sigma,\sigma,(j)}_{\mathbf{k}_{\Lambda_{j}\hat{s}},\mathbf{k},\mathbf{p}}>0$\\ \item $|V^{\sigma,-\sigma,(j)}_{\mathbf{k}_{\Lambda_{j}\hat{s}},\mathbf{p}}|>|V^{\sigma,-\sigma,(j)}_{\mathbf{k}_{\Lambda_{j}\hat{s}},\mathbf{k},\mathbf{p}}|$\end{enumerate}} & \parbox{.5\textwidth}{\begin{enumerate}\item $r^{\sigma,-\sigma,(j)}_{\mathbf{p}=0}\to -r$, $r<1$\\  \item $\frac{V^{\sigma,-\sigma}_{\mathbf{k}\mathbf{k}'\mathbf{p}'}}{V^{\sigma,-\sigma}_{\mathbf{k}\mathbf{k}'\mathbf{p}=0}}\bigg\vert_{L\to \infty}\to 0+$~\text{for}~ $\epsilon_{\mathbf{p}'-\mathbf{k}}>E_{F}$\\ \item $r^{\sigma,-\sigma,(j)}_{\mathbf{p}'}\to 0+$ for $\epsilon_{\mathbf{p}'-\mathbf{k}}<E_{F}$\\ \item  $r^{\sigma,\sigma,(j)}_{\mathbf{p}}\to 0$ for all $\mathbf{p}$\\ \item $\frac{s^{\sigma\sigma'\sigma,(j)}}{r_{\mathbf{p}=0}^{\sigma,-\sigma,(j)}}\bigg\vert_{L\to \infty}\to 0+$\end{enumerate}}\\
\hline   
\end{tabular}
\caption{Table representing various parameter space regimes (Column II) and the associated flow of quantities describing nature of fixed point theory (Column III). Regimes I, II, III leads to fixed point Hamiltonians $H^{*}_{FL}$, $H^{*,XY}_{RBCS}$, $H^{*,XXZ}_{RBCS}$ respectively in the main text.}\label{Table-conditions-I}
\end{table}
\begin{table}
\hspace*{-1cm}
\begin{tabular}{|c|c|c|}
\hline
&\textbf{Cases}&$r^{\sigma\sigma',(j)}_{\mathbf{p}}$, $s^{\sigma\sigma'\sigma,(j)}$\\
\hline
IV & \parbox{.5\textwidth}{\begin{enumerate}\item $\omega_{(j)}-\frac{1}{2}(\epsilon^{(j)}_{\mathbf{k}_{\Lambda_{j}\hat{s}}}+\epsilon^{(j)}_{\mathbf{p}-\mathbf{k}_{\Lambda_{j}\hat{s}}})<0$ \\ \item $\epsilon^{(j)}_{\mathbf{k}_{\Lambda_{j}\hat{s}}}>E_{F}$\\ \item $V^{\sigma,\sigma',(j)}_{\mathbf{k},\mathbf{p}}>0$\\ \item $V^{\sigma,-\sigma,(j)}_{\mathbf{k}_{\Lambda_{j}\hat{s}},\mathbf{k},\mathbf{p}}<0$ \\ \item $V^{\sigma,\sigma,(j)}_{\mathbf{k}_{\Lambda_{j}\hat{s}},\mathbf{k},\mathbf{p}}>0$ \\ \item $|V^{\sigma,-\sigma,(j)}_{\mathbf{k}_{\Lambda_{j}\hat{s}},\mathbf{p}}|<|V^{\sigma,-\sigma,(j)}_{\mathbf{k}_{\Lambda_{j}\hat{s}},\mathbf{k},\mathbf{p}}|$\end{enumerate}} & \parbox{.5\textwidth}{\begin{enumerate}\item $r^{\sigma,-\sigma,(j)}_{\mathbf{p}}\to -1$ \\ \item $\frac{V^{\sigma,-\sigma}_{\mathbf{k}\mathbf{k}'\mathbf{p}'}}{V^{\sigma,-\sigma}_{\mathbf{k}\mathbf{k}'\mathbf{p}}}\bigg\vert_{L\to \infty}\to 0+$ ~\text{for}~ $\epsilon_{\mathbf{p}'-\mathbf{k}}>E_{F}$\\ \item $r^{\sigma,-\sigma,(j)}_{\mathbf{p}'}\to 0+$ for $\epsilon_{\mathbf{p}'-\mathbf{k}}<E_{F}$\\ \item  $r^{\sigma,\sigma,(j)}_{\mathbf{p}}\to 0$ for all $\mathbf{p}$\\ \item $\frac{s^{\sigma\sigma'\sigma,(j)}}{r_{\mathbf{p}}^{\sigma,-\sigma,(j)}}\bigg\vert_{L\to \infty}\to 0+$\end{enumerate}}  \\
\hline
V & \parbox{.5\textwidth}{\begin{enumerate}\item $\omega_{(j)}-\frac{1}{2}(\epsilon^{(j)}_{\mathbf{k}_{\Lambda_{j}\hat{s}}}+\epsilon^{(j)}_{
\mathbf{p}-\mathbf{k}_{\Lambda_{j}\hat{s}}})<0$ \\ \item $\epsilon^{(j)}_{\mathbf{k}_{\Lambda_{j}\hat{s}}}>E_{F}$\\ \item $V^{\sigma,\sigma',(j)}_{\mathbf{k},\mathbf{p}}>0$\\ \item $V^{\sigma,-\sigma,(j)}_{\mathbf{k}_{\Lambda_{j}\hat{s}},\mathbf{k},\mathbf{p}}<0$\\ \item $V^{\sigma,\sigma,(j)}_{\mathbf{k}_{\Lambda_{j}\hat{s}},\mathbf{k},\mathbf{p}}>0$\\ \item $|V^{\sigma,-\sigma,(j)}_{\mathbf{k}_{\Lambda_{j}\hat{s}},\mathbf{p}}|>|V^{\sigma,-\sigma,(j)}_{\mathbf{k}_{\Lambda_{j}\hat{s}},\mathbf{k},\mathbf{p}}|$\end{enumerate}} & \parbox{.5\textwidth}{\begin{enumerate}\item $r^{\sigma,-\sigma,(j)}_{\mathbf{p}}\to -r$, $r<1$\\  \item $\frac{V^{\sigma,-\sigma}_{\mathbf{k}\mathbf{k}'\mathbf{p}'}}{V^{\sigma,-\sigma}_{\mathbf{k}\mathbf{k}'\mathbf{p}}}\bigg\vert_{L\to \infty}\to 0+$~\text{for}~ $\epsilon_{\mathbf{p}'-\mathbf{k}}>E_{F}$\\ \item $r^{\sigma,-\sigma,(j)}_{\mathbf{p}'}\to 0+$ for $\epsilon_{\mathbf{p}'-\mathbf{k}}<E_{F}$\\ \item  $r^{\sigma,\sigma,(j)}_{\mathbf{p}}\to 0$ for all $\mathbf{p}$\\ \item  $\frac{s^{\sigma\sigma'\sigma,(j)}}{r_{\mathbf{p}}^{\sigma,-\sigma,(j)}}\bigg\vert_{L\to \infty}\to 0+$\end{enumerate}}\\
\hline
VI & \parbox{.5\textwidth}{\begin{enumerate}\item $\frac{1}{2}\epsilon^{(j)}_{\mathbf{k}_{\Lambda_{j}\hat{s}}}<\omega_{(j)}<\frac{1}{2}(\epsilon^{(j)}_{\mathbf{k}_{\Lambda_{j}\hat{s}}}+\epsilon^{(j)}_{
-\mathbf{k}_{\Lambda_{j}\hat{s}}})$\\  \item $\epsilon^{(j)}_{\mathbf{k}_{\Lambda_{j}\hat{s}}}>E_{F}$\\ \item $V^{\sigma,\sigma',(j)}_{\mathbf{k}_{\Lambda_{j}\hat{s}},\mathbf{k},\mathbf{p}}>0$\end{enumerate}}
& \parbox{.5\textwidth}{\begin{enumerate}\item $s^{\sigma,-\sigma,\sigma,(j)}\to 1$\\  \item $s^{\sigma,\sigma,\sigma,(j)} = 0$\\
\item $r_{\mathbf{p}}^{\sigma\sigma',(j)}\to 0$
\end{enumerate}}
\\
\hline
\end{tabular}
\caption{Table representing various parameter space regimes (Column II) and the associated flow of quantities describing nature of fixed point theory (Column III). Regimes IV, V, VI leads to fixed point Hamiltonians $H^{*,XY}_{SPDW}$, $H^{*,XXZ}_{SPDW}$, $H^{*}_{MFL}$ respectively in main text.}\label{Table-conditions-II}
\end{table}
\pin
\textbf{I. \textit{The Fermi Liquid}}\\
Fermi liquid theory~\cite{landau1959theory} arises in the low-energy regime I in Table \ref{Table-conditions-I} due to vanishing of all the non-zero momentum $\mathbf{k}-\mathbf{k}'$ scattering vertices $V^{\sigma\sigma'}_{\mathbf{k},\mathbf{k}',\mathbf{p}}$. We discuss the details of this RG flow here. The condition 1 in regime I provides the ranges for fluctuation scale $\omega_{(j)}$ and number diagonal vertex $V^{\sigma\sigma',(j)}_{\mathbf{k},\mathbf{p}}$ for which the Green's function $G^{4,(j)}_{\gamma\gamma'}$ appearing in eq.\eqref{4and6 particle vertex} picks up a negative signature 
\begin{eqnarray}
\hspace*{-0.8cm}G^{4,(j)}_{\gamma\gamma'}&=&\left(\omega_{(j)}-\frac{1}{2}(\epsilon^{(j)}_{\mathbf{k}_{\Lambda_{j}\hat{s}}}+\epsilon^{(j)}_{\mathbf{p}-\mathbf{k}_{\Lambda_{j}\hat{s}}})-\frac{V^{\sigma\sigma',(j)}_{\mathbf{k},\mathbf{p}}}{4}\right)^{-1}\hspace*{-0.5cm}<0~,\label{condFL}
\end{eqnarray}
leading to the RG irrelevance for both the off-diagonal and diagonal vertices, i.e., $(\Delta V^{\sigma\sigma',(j)}_{\mathbf{k},\mathbf{k}',\mathbf{p}}, \Delta V^{\sigma\sigma',(j)}_{\mathbf{k},\mathbf{p}})\to 0$. Further, condition 2 in regime I ensures that in the limit of large system sizes ($L\to \infty$), the off-diagonal vertices vanishes $V^{\sigma\sigma',(j)}_{\mathbf{k},\mathbf{k}',\mathbf{p}}\to 0$, whereas the diagonal vertices reach a intermediate value given by the fixed point condition (corresponding to a vanishing of the denominator in eq.\eqref{4and6 particle vertex})
\begin{eqnarray}
\omega_{(j^{*})} -\frac{1}{2}\left(\epsilon^{(j^{*})}_{\mathbf{k}_{\Lambda_{j^{*}}\hat{s}}}+\epsilon^{(j^{*})}_{\mathbf{p}-\mathbf{k}_{\Lambda_{j^{*}}\hat{s}}}\right) &=& \frac{1}{4}V^{\sigma\sigma',(j^{*})}_{\mathbf{k}_{\Lambda_{j^{*}}\hat{s}}\mathbf{p}}~.~\label{FL_fixed point}
\end{eqnarray}
The tree diagram Fig. \ref{246vertices-tree-FL} represents the vertex tensor RG flow, where the blue branches and nodes represent vertices whose magnitudes flow towards a finite value at the stable fixed point. On the other hand, the magnitudes of the red branches flow towards towards zero. As $V^{\sigma\sigma'}_{\mathbf{k},\mathbf{k}',\mathbf{p}}$ is RG irrelevant, the 6-point vertices $R^{\sigma\sigma'\sigma''}$ also do not contribute in the limit of $L\to \infty$. This results in the quantities $r^{\sigma\sigma',(j)}_{\mathbf{p}}\to 0$ and $s^{\sigma\sigma'\sigma,(j)}\to 0$. Thus, the theory at the Fermi liquid fixed point is free of all 2-particle as well as higher order off-diagonal vertices, leading to Landau's Fermi liquid Hamiltonian~\cite{landau1959theory}
\begin{eqnarray}
H^{*}_{FL}(\omega) &=& \sum_{\Lambda<\Lambda_{j^{*}},\hat{s}}\epsilon_{\mathbf{k}_{\Lambda\hat{s}}}\hat{n}_{\mathbf{k}_{\Lambda\hat{s}}\sigma}+\sum_{\Lambda <\Lambda_{j^{*}},\hat{s},\mathbf{p}} V^{\sigma\sigma',(j^{*})}_{\mathbf{k}_{\Lambda\hat{s}}\mathbf{p}}(\omega_{(j*)})\hat{n}_{\mathbf{k}_{\Lambda\hat{s}}\sigma}\hat{n}_{\mathbf{p}-\mathbf{k}_{\Lambda\hat{s}}\sigma'}~.~~~~~\label{HFL}
\end{eqnarray}
\pin
For a quantitative demonstration of the Fermi liquid fixed point theory from our URG analysis, we numerically analyse below the URG equations for various 2- and 3-particle vertices, the 1-particle self-energy ($\Sigma$) and the quasiparticle residue ($Z$). For this, we consider a screened interaction potential $V(|\mathbf{q}|)=\eta^{2}V(0)/(q^{2}+\eta^{2})$, with $V(0)=20.5$, $\eta=0.2$, a two-dimensional circular Fermi surface with $|k_{F}|=\pi/20$ and $\omega=\epsilon_{\Lambda_{0}}+0.1$ (Regime-I of Table \ref{Table-conditions-I}) and a system volume represented by a $k$-space grid of $1024$ lattice sites $\times 1024$ lattice sites. In Fig.\ref{2particleVertices}, the 2-particle scattering vertices $V(|\mathbf{q}|)$ with non-zero ($\mathbf{q}\neq 0$) momentum transfer (red, green and orange curves) are found to be irrelevant under RG flow. On the other hand, the $|\mathbf{q}|=0$ vertices $V(0)$ (red curve, corresponding to the couplings associated with terms like $\hat{n}_{\mathbf{k}\sigma}\hat{n}_{\mathbf{k}'\sigma'}$) attain a finite value at the IR fixed point. In this way, we numerically confirm the RG flow towards the effective Fermi liquid Hamiltonian  $H^{*}_{FL}(\omega)$ given in eq.\eqref{HFL}. In Fig. \ref{SelfEnergy} (left panel), we see that the 1-particle self-energy $\Sigma^{(j)}_{k}$ renormalizes to a finite value $\Sigma^{*}_{k}$ at the RG fixed point,  and that the $|k\rangle$ states closer to the Fermi surface ($k_{F}$) have a lower $\Sigma^{*}_{k}$ (Fig.\ref{SelfEnergy} (right panel)). Fig.\ref{3pEnergyAndResidue} (left panel) shows that the quasiparticle residue $Z(k,\Delta)\to 1$ upon approaching the Fermi energy $\Delta\to 0$, demonstrating the existence of well-defined Landau quasiparticles in the neighbourhood of the Fermi surface. Fig. \ref{3pEnergyAndResidue} (right panel) shows the RG irrelevance of 2-electron 1-hole scattering vertices (which constitute the primary decay channel for the Landau quasiparticles). Taken together, these results  verify numerically the phenomenology of the Landau Fermi liquid theory (eq.\eqref{HFL}) as derived from the URG analysis of the $H_{SFIM}$ model (eq.\eqref{HSFIM}).
\pin
In a companion manuscript~\cite{anirbanurg1}, we derive the form of the renormalized Friedel's scattering phase shift starting from the exponential representation of the unitary operator $\Delta N=Tr(\log(U_{(j)}))=i\frac{\pi}{4}Tr(\eta_{(j)}-\eta^{\dagger}_{(j)})$. Further, we note that this is similar to Langer and Ambegaokar's definition of scattering phase shift~\cite{langer1961friedel}. As all off-diagonal terms are RG irrelevant in the Fermi liquid, $\eta^{\dagger}$ and $\eta$ both vanish at the RG fixed point. As a consequence, the Friedel's phase shift  for the Fermi liquid is given by $\Delta N=0$, i.e., upon placing a test charge near the Fermi surface, no electrons are permanently displaced from within it and the Luttinger volume is preserved~\cite{seki2017topological,martin1982fermi}. The incompressibility displays the topological protection for the Fermi surface ($FS$) associated with the Volovik invariant~\cite{volovik2003universe}, as shown in Ref.\cite{anirbanurg1}. Further, we obtain a
vanishing thermal scale in eq.\eqref{Thermal scale}, $T=0K$, corresponding to $\omega_{(j^{*})} =0$ and $\Lambda_{(j^{*})}=0$ for the Fermi liquid.
\\ 
\pin
\textbf{II. \textit{Reduced BCS theory-XY interaction}}\\
\begin{figure}
\hspace*{-2cm}
\begin{tikzpicture}
[sibling distance=6.5cm]
  \node {$H_{(j)}$}
    child {[blue] node[blue]{$\Sigma^{(j)}$}}
    child {[blue][sibling distance=10cm] 
        node[blue] {$\Gamma^{4,(j)}$}
      child {[blue,sibling distance=3cm]
         node[blue] {$\Gamma_{\alpha\alpha'}^{4,(j)}$}
        child{[sibling distance=1.8cm] 
           node {$V^{\sigma\sigma',(j)}_{\mathbf{k},\mathbf{p}_{1}}$ $\color{black}$ $\color{black} \ldots$}
          child{[red]node[red] {$V^{\sigma\sigma,(j)}_{\mathbf{k},\mathbf{p}_{1}}$}}
          child{[blue]node[blue] {$V^{\sigma~-\sigma,(j)}_{\mathbf{k},\mathbf{p}_{1}}$ $\color{black} \ldots$}}}
        child{[sibling distance=1.8cm] 
           node {$\color{black} \ldots$ $V^{\sigma\sigma',(j)}_{\mathbf{k},\mathbf{p}}$ $\color{black} \ldots$}
          child{[red]node[red]{$V^{\sigma\sigma,(j)}_{\mathbf{k},\mathbf{p}}$}}
          child{[orange]node[orange]{$V^{\sigma~-\sigma,(j)}_{\mathbf{k},\mathbf{p}}\ldots$}}}  
        child{[sibling distance=1.8cm]
           node {$\color{black} \ldots$ $V^{\sigma\sigma',(j)}_{\mathbf{k},\mathbf{p}_{2}}$}
           child{[red] node[red] {$V^{\sigma\sigma,(j)}_{\mathbf{k},\mathbf{p}_{2}}$}}
          child{node {$V^{\sigma~-\sigma,(j)}_{\mathbf{k},\mathbf{p}_{2}}$}}
        }}
        child{[blue][sibling distance=3.2cm]
           node[blue] {$\Gamma_{\alpha\beta}^{4,(j)}$}
        child{[sibling distance=1.8cm]
           node {$V^{\sigma\sigma',(j)}_{\mathbf{k},\mathbf{k}',\mathbf{p}_{1}} \ldots$}
          child{[red]node {$V^{\sigma\sigma,(j)}_{\mathbf{k},\mathbf{k}',\mathbf{p}_{1}}$}}
          child{[blue]node[blue] {$V^{\sigma~-\sigma,(j)}_{\mathbf{k},\mathbf{k}',\mathbf{p}_{1}}$ $\ldots$}}}
          child{[sibling distance=1.8cm]
           node {$\ldots V^{\sigma\sigma',(j)}_{\mathbf{k},\mathbf{k}',\mathbf{p}} \ldots$}
          child{[red]node[red]{$V^{\sigma\sigma,(j)}_{\mathbf{k},\mathbf{k}',\mathbf{p}}$}}
          child{[orange]node[orange] {$V^{\sigma~-\sigma,(j)}_{\mathbf{k},\mathbf{k}',\mathbf{p}_{1}}\ldots$}}}
        child{[sibling distance=1.8cm]
           node {$V^{\sigma\sigma',(j)}_{\mathbf{k},\mathbf{k}',\mathbf{p}_{2}}$}
           child{[red]node[red] {$V^{\sigma\sigma,(j)}_{\mathbf{k},\mathbf{k}',\mathbf{p}_{2}}$}}
          child{[blue]node[blue] {$V^{\sigma~-\sigma,(j)}_{\mathbf{k},\mathbf{k}',\mathbf{p}_{2}}$}}
        }}}
    child[blue] {[sibling distance=1cm] 
       node {$\Gamma^{6,(j)}$}
      child {node {$\Gamma^{6,(j)}_{\alpha\alpha'}$}}
      child {node {$\Gamma^{6,(j)}_{\alpha\beta}$}}};~.
\end{tikzpicture}
\caption{Tree tensor diagram representing the zero-momentum ($\mathbf{p}=0$) pairing reduced BCS model (regime-III in Table \ref{Table-conditions-I}) and the SPDW state made of finite-momentum ($\mathbf{p}$) pseudospin-pairs interacting via XXZ interaction (regime-V in Table \ref{Table-conditions-I}). The blue branches and the nodes of the vertex tensors represent the relevant scattering vertices, while orange branches and nodes represent the dominant RG relevant scattering vertices. The red branches and nodes represent RG irrelevant scattering vertices.}\label{246vertices-tree-XXZ-BCS}
\end{figure}
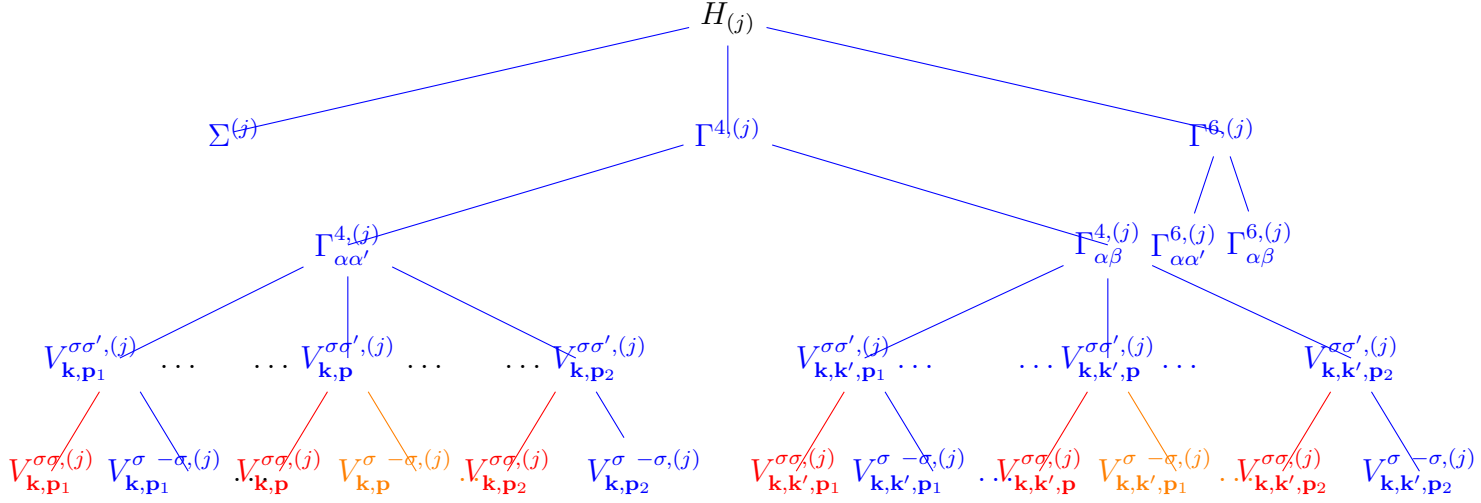  
\begin{figure}
\centering
\includegraphics[width=\textwidth]{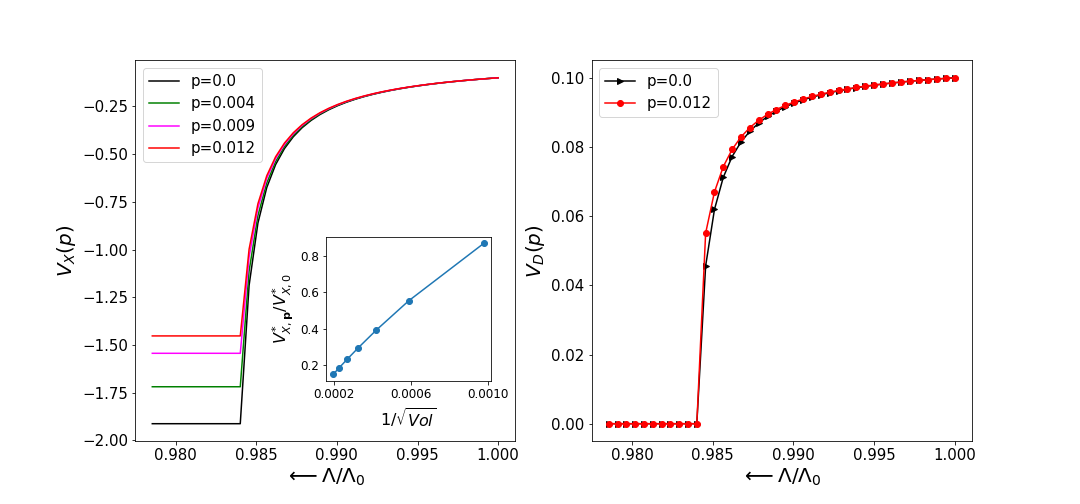}
\caption{Left/right panel represents the RG flow of two-particle number off-diagonal/diagonal scattering vertices for $p=0$ ($V^{(j)}_{X,p=0}$) and $p\neq 0$ finite-momentum pairs ($(V^{(j)}_{X,p'})$). Numerical evaluations are for system size $=1024\times 1024$ and bare $V_{D,\mathbf{p}}=-0.1$, $V_{X,\mathbf{p}}=-0.25$ (in units of $t$), $\omega=\epsilon_{\Lambda_{0}}-0.5$. Left panel inset: Finite size scaling plot of the ratio $V^{*}_{X,p=0.004}/V^{*}_{X,p=0}$ with $1/\sqrt{Vol}$.}\label{XY_BCS_1}
\end{figure}

\begin{figure}
\hspace*{-0.7cm}
\includegraphics[width=0.6\textwidth]{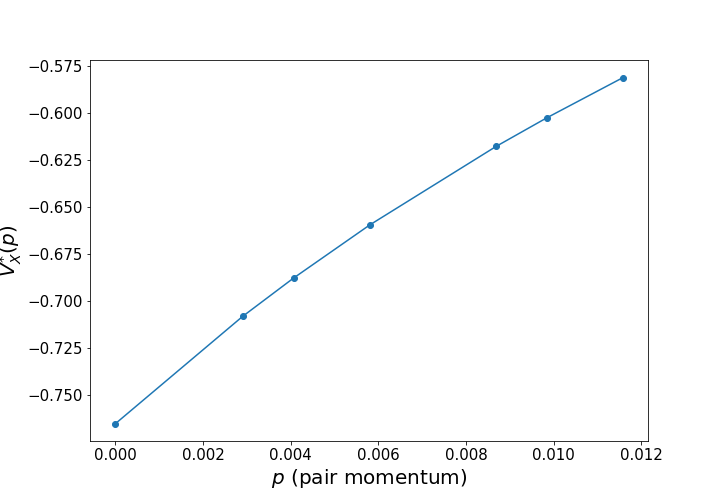}
\includegraphics[width=0.6\textwidth]{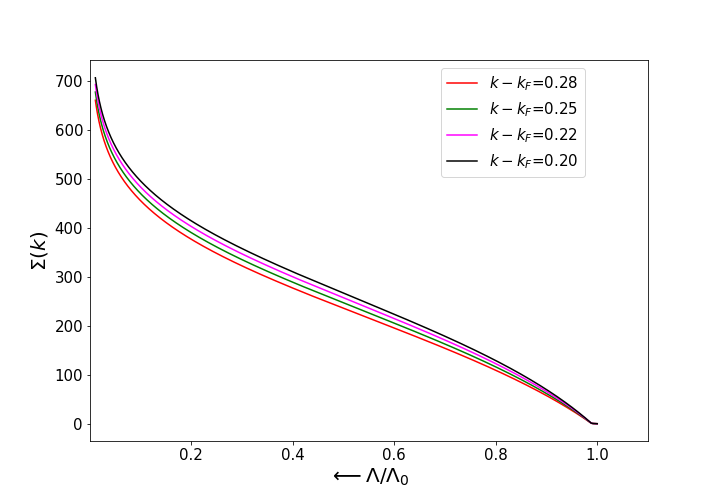}
\caption{Left panel shows the variation of electronic pair-scattering vertex magnitude attained at RG fixed point $V_{X}^{*}(\mathbf{p})$ as a function of pair-momentum $p$. Right panel shows the RG flow of the single-particle self-energy for four different $|k\rangle$ states close to Fermi momentum $k_{F}$.}\label{XY_BCS_2}
\end{figure}

The reduced BCS theory~\cite{bardeen1957theory} constitutes attractive interactions between opposite-spin pairs of electrons with zero net-momentum $\mathbf{p}=0$. This theory is an outcome of (i) RG irrelevance for the 2-particle number-diagonal vertices, (ii) dominant RG flow for zero-momentum $\mathbf{p}=0$ electronic pairs and (iii) the sub-dominance (or RG irrelevance) of the $\mathbf{p}\neq 0$ pair-scattering vertices. The above features of the RG flows is an outcome of the conditions listed as 1-6 in regime II of Table \ref{Table-conditions-I}. Conditions 1-3 imply that the Green's function 
\begin{eqnarray}
G^{4,(j)}_{\gamma\gamma'}\equiv G^{4,(j)}_{\mathbf{k}_{\Lambda_{j}\hat{s}},\mathbf{p}}&=&\left(\omega_{(j)}-\frac{1}{2}(\epsilon_{\mathbf{k}_{\Lambda_{j}\hat{s}}}+\epsilon_{\mathbf{p}-\mathbf{k}_{\Lambda_{j}\hat{s}}})\right)^{-1},
\end{eqnarray}
has a negative signature: $G^{4,(j)}_{\mathbf{k}_{\Lambda_{j}\hat{s}},\mathbf{p}}=-|G^{4,(j)}_{\mathbf{k}_{\Lambda_{j}\hat{s}},\mathbf{p}}|$.
As $G^{4,(j)}_{\mathbf{k}_{\Lambda_{j}\hat{s}},\mathbf{p}}$ appears in the RG flow eq.\eqref{4-particle vertex} (for $p=0$),  scattering between opposite-spin pairs due to attractive couplings ($V^{\sigma,-\sigma,(j)}_{\mathbf{k},\mathbf{k}',\mathbf{p}}$) become RG relevant, while repulsive Hartree interactions ($V^{\sigma,\sigma',(j)}_{\mathbf{k},\mathbf{p}}$) are RG irrelevant. Further, condition 4 ensures that all scattering vertices with identical spin ($V^{\sigma,\sigma,(j)}_{\mathbf{k},\mathbf{k}',\mathbf{p}}$) are RG irrelevant. Given the range of the fluctuation scale $\omega_{(j)}$ in condition 1, and for $\epsilon^{(j)}_{\mathbf{p}'-\mathbf{k}}>E_{F}$, the 2-particle Green's function for $\mathbf{p}>0$ momentum pairs $G^{4,(j)}_{\gamma\gamma'}=G^{4,(j)}_{\mathbf{k}_{\Lambda_{j}\hat{s}},\mathbf{p}}$ is smaller in magnitude compared to the zero-momentum pair ($\mathbf{p}=0$)
\begin{eqnarray}
|G^{4,(j)}_{\mathbf{k}_{\Lambda_{j}\hat{s}},\mathbf{p}}|&<&|G^{4,(j)}_{\mathbf{k}_{\Lambda_{j}\hat{s}},\mathbf{p}=0}|~.
\end{eqnarray}   
This Green's function appears in the RG flow equation eq.\eqref{4-particle vertex} for the vertices $V^{\sigma,-\sigma,(j)}_{\mathbf{k},\mathbf{k}',\mathbf{p}}$, rendering it sub-dominant compared to the vertex $V^{\sigma,-\sigma,(j)}_{\mathbf{k},\mathbf{k}',\mathbf{p}=0}$. As a result, $r^{\sigma,-\sigma,(j)}_{\mathbf{p}}/r^{\sigma,-\sigma,(j)}_{\mathbf{p}=0}\to 0$, as shown in regime-II  Table \ref{Table-conditions-I}. Condition 6 ensures that the number-diagonal interaction $V^{\sigma,-\sigma,(j)}_{\mathbf{k},\mathbf{p}}\to 0$. Finally, as the RG flow leads to dominance of only a given pair-momentum vertices compared to all others, the renormalization of the  6- point vertices $\Gamma^{6,(j)}$ that arise out of the interplay between different pair-momentum vertices in eq.\eqref{4and6 particle vertex} is sub-dominant compared to the $\mathbf{p}=0$ vertices. Thus, they are represented by the ratio $s^{\sigma\sigma'\sigma,(j)}/r^{\sigma,-\sigma,(j)}_{\mathbf{p}=0}|_{L\to \infty}\to 0+$ in regime II. 
All of these features finally lead to the fixed point condition
\begin{eqnarray}
\omega_{(j*)}=\frac{1}{2}(\epsilon^{(j*)}_{\mathbf{k}_{\Lambda_{j*}\hat{s}}}+\epsilon^{(j*)}_{-\mathbf{k}_{\Lambda_{j*}\hat{s}}})=\epsilon^{(j*)}_{\mathbf{k}_{\Lambda_{j*}\hat{s}}}~,
\end{eqnarray}
where we have used the band symmetry $\epsilon_{\mathbf{k}}=\epsilon_{-\mathbf{k}}$. 
\pin
Further, at this RG fixed point, the 1-particle self-energy $\Sigma^{(j)}_{\mathbf{k}}=\epsilon^{(j)}_{\mathbf{k}}-\epsilon_{\mathbf{k}}$ diverges. This can be seen as follows. The RG flow equation for $\Sigma^{(j)}_{\mathbf{k}}$ (eq.\eqref{1-p-self-energy}) now has a dominant contribution from the zero pair-momentum scattering vertices
\begin{equation}
\Delta\Sigma^{(j)}_{\mathbf{k}} = \left(V^{\sigma,-\sigma,(j)}_{\mathbf{k}_{\Lambda_{j}\hat{s}},\mathbf{k},\mathbf{p}=0}\right)^{2}(\omega_{(j)}-\epsilon^{(j)}_{\mathbf{k}_{\Lambda_{j}\hat{s}}}+\frac{1}{2}\epsilon_{\mathbf{k}}+\frac{1}{2}\Sigma^{(j)}_{\mathbf{k}})^{-1}~.\label{divSelfEnergy}
\end{equation}
From this relation, we see that the self-energy is RG relevant.  For the electronic states labelled $\mathbf{k}$ (i.e., residing within the emergent window), the self-energy RG flow equation has a fixed point at $\Sigma^{(j^{*})}_{\mathbf{k}}\to \infty$ as $\omega_{(j^{*})}=\frac{1}{2}\epsilon^{(j^{*})}_{\mathbf{k}_{\Lambda_{j*}\hat{s}}}$, indicating the breakdown of the Landau quasiparticles of the Fermi liquid. 
As discussed in an earlier section, the diverging self-energy corresponds to zeros in the single-particle Green's function
\begin{eqnarray}
G(\mathbf{k},\omega)=\frac{1}{\omega-\epsilon_{\mathbf{k}}-\Sigma_{\mathbf{k}}}\to 0~.\label{breakdown1pGreen}
\end{eqnarray}
This indicates the breakdown of the Luttinger volume sum-rule, i.e., $N_{e}\neq \sum_{\mathbf{k}\sigma}G(\mathbf{k},\omega)$~. 
Instead, we find that the total number of Cooper pairs $N_{CP}$ within the low-energy window  equals the net Friedel's phase shift, $\Delta N=Tr(\log(U_{(j)}))=N_{CP}\in 2\mathbb{Z}$, i.e., two electronic states are lost for each bound pair. This provides a way for taking accounts for the Luttinger surface of zeros~\cite{gros2006determining}: the Friedel phase shift compensates precisely the mismatch observed via the accumulation of topological phases arising from the non-commutativity of the twist and translation operators~\cite{oshikawa2000topological,hastings2004lieb} (as shown in a companion work~\cite{anirbanurg1}).
\par\noindent
The effective Hamiltonian, $H^{*,XY}_{RBCS}(\omega)$, at the stable fixed point of the flow has the form
\begin{eqnarray}
H^{*,XY}_{RBCS}(\omega) &=&\sum_{\mathbf{k}}\epsilon^{(j^{*})}_{\mathbf{k}_{\Lambda\hat{s}}}A^{z}_{\mathbf{k}_{\Lambda\hat{s}}}-\sum_{\mathbf{k}_{\Lambda\hat{s}}} V^{\sigma,-\sigma,(j^{*})}_{\mathbf{k}_{\Lambda\hat{s}},\mathbf{k}_{\Lambda'\hat{s}'},0} A^{+}_{\mathbf{k}_{\Lambda\hat{s}}}A^{-}_{\mathbf{k}_{\Lambda'\hat{s}'}}~,\label{HBCS}\hspace*{1cm}
\end{eqnarray}
where the set $\alpha' = \lbrace(\mathbf{k}_{\Lambda'\hat{s}'},\sigma);(-\mathbf{k}_{\Lambda'\hat{s}'},-\sigma)\rbrace$. This is the generalized reduced BCS Hamiltonian/pairing-force model~ \cite{bardeen1957theory,dukelsky2004colloquium}, where the pseudospin $\mathbf{A}_{\mathbf{k}_{\Lambda\hat{s}}}$ components are defined as 
~\cite{anderson1958random} 
\begin{eqnarray}
A^{z}_{\mathbf{k}_{\Lambda\hat{s}}}&=&\frac{1}{2}(\hat{n}_{\mathbf{k}_{\Lambda\hat{s}}\sigma}+\hat{n}_{-\mathbf{k}_{\Lambda\hat{s}}-\sigma}-1), A^{+}_{\mathbf{k}_{\Lambda\hat{s}}} = c^{\dagger}_{\mathbf{k}\sigma}c^{\dagger}_{-\mathbf{k}-\sigma}, A^{-}_{\mathbf{k}_{\Lambda\hat{s}}} = c_{-\mathbf{k}-\sigma}c_{\mathbf{k}\sigma}~.\label{pseudospin_charge}
\end{eqnarray}
In order to verify quantitatively the effective theory given in eq.\ref{HBCS}, we numerically simulated the RG equations for the bare couplings $V_{X,p}=-0.25$, $V_{D,p}=0.1$ and the fluctuation energy scale $\omega=\epsilon_{\Lambda_{0}}-0.5$ (Regime-II in Table~\ref{Table-conditions-I}) and an identical $k$-space grid as mentioned earlier. Fig.\ref{XY_BCS_1} (left panel) represents the RG flow for the two-particle off-diagonal scattering vertices involving electronic pairs with net-momentum $p=0.0$, $0.004$, $0.009$ and $0.012$ respectively. The inset in the left panel of Fig.\ref{XY_BCS_1} shows that the ratio $V^{*}_{X,p'}/V^{*}_{X,p=0}$ diminishes with increasing system volume (which we have taken to range from $1024\times 1024$ lattice to a $5000\times 5000$ lattice), indicating the dominance of $p=0$ momentum scattering vertices at low-energies and describing the condensation of Cooper pair degrees of freedom. Fig.\ref{XY_BCS_1} (right panel) shows that all the number-diagonal scattering vertices are RG irrelevant, and vanish along the RG flow. As seen in Fig.\ref{XY_BCS_2} (left panel), we find that the $p=0$ momentum electronic pair scattering vertices have the highest magnitude $|V^{*}_{X,p=0}|>|V^{*}_{X,p'}|$ ($V^{*}_{X,0}=V^{\sigma,-\sigma}_{k,k',0}$, $V^{*}_{X,p'}=V^{\sigma,-\sigma}_{k,k',p'}$) at low-energies, and the magnitude of $V^{*}_{X,p'}$ monotonically decreases with increasing pair-momentum ($p$). The relevance of off-diagonal $p=0$ momentum scattering vertices, together with the RG irrelevance of number-diagonal scattering vertices, describes the effective Hamiltonian $H^{*,XY}_{RBCS}$ (eq.\eqref{HBCS}) at the RG fixed point. 
Finally, Fig.\ref{XY_BCS_2} (right panel) shows a divergent renormalized self-energy $\Sigma_{k}$ (eq.\eqref{divSelfEnergy}) for the $|k\rangle$ states, where $|k-k_{F}|<\Lambda^{*}$ and $\Lambda^{*}$ is width of the momentum-space shell around the erstwhile Fermi surface.
\pin
The condensation of the pseudospins (i.e., Cooper pairs~\cite{cooper1956bound} with the electronic spins locked into singlets) in this subspace is described by the fixation of the pseudospin angular momentum value to $\frac{3}{4}$
 \begin{eqnarray}
 \mathbf{A}^{2}_{\mathbf{k}_{\Lambda\hat{s}}}=\frac{3}{4}(\hat{n}_{\mathbf{k}_{\Lambda\hat{s}}\sigma}+\hat{n}_{-\mathbf{k}_{\Lambda\hat{s}}-\sigma}-1)^{2}=\frac{3}{4}~,~ \Lambda<\Lambda_{j^{*}}~.~~~~\label{constraint_condense_1}
 \end{eqnarray} 
Thus, the emergence of the constraint $\hat{n}_{\mathbf{k}_{\Lambda\hat{s}}\sigma}=\hat{n}_{-\mathbf{k}_{\Lambda\hat{s}}-\sigma}$ describes the phenomenon of condensation of Cooper pairs within the low-energy window of the BCS fixed point theory. For the case of a spherical Fermi surface, i.e., $\epsilon_{\Lambda\hat{s}}=\epsilon_{\Lambda}$, the phase described by the BCS reduced model (eq.\eqref{HBCS}) will persist upto the thermal scale (from eq\eqref{Thermal scale})
\begin{eqnarray}
T^{*} = \frac{1}{\pi k_{B}}\left[2\epsilon^{0}_{\Lambda_{j^{*}}}+(\epsilon^{0}_{\Lambda_{j^{*}}}-\omega)\log |\frac{\omega-\epsilon^{0}_{\Lambda_{j^{*}}}}{\omega +\epsilon^{0}_{\Lambda_{j^{*}}}}|\right]~.\label{critical_scale}
\end{eqnarray}
The above equation is obtained from the self-energy of the electronic state at the momentum-space boundary of the emergent phase in eq.\eqref{Thermal scale}: $\Sigma^{(j^{*})}_{\Lambda_{j^{*}}} = \omega - \epsilon^{0}_{\Lambda_{j^{*}}}$~, where $\epsilon^{0}_{\Lambda_{j^{*}}}$ is the bare dispersion magnitude. The temperature scale $T^{*}$ is greater than the critical temperature $T_{c}$ obtained from the BCS mean-field solution~\cite{bardeen1957theory}, and indicates the presence of pairing in the ground state of the reduced BCS Hamiltonian but without the off-diagonal long-ranged order (ODLRO) that characterises the phase-stiff BCS ground state. We will present further insights on the ground state properties of this quantum liquid in Sec.\ref{gauge_theory}. Finally, the tree decomposition of the vertices representing this phase is shown in Fig. \ref{246vertices-tree-XY-BCS}.\\
\par\noindent
\textbf{III. \textit{Reduced BCS theory-XXZ interaction}}\\
\begin{figure}
\includegraphics[width=\textwidth]{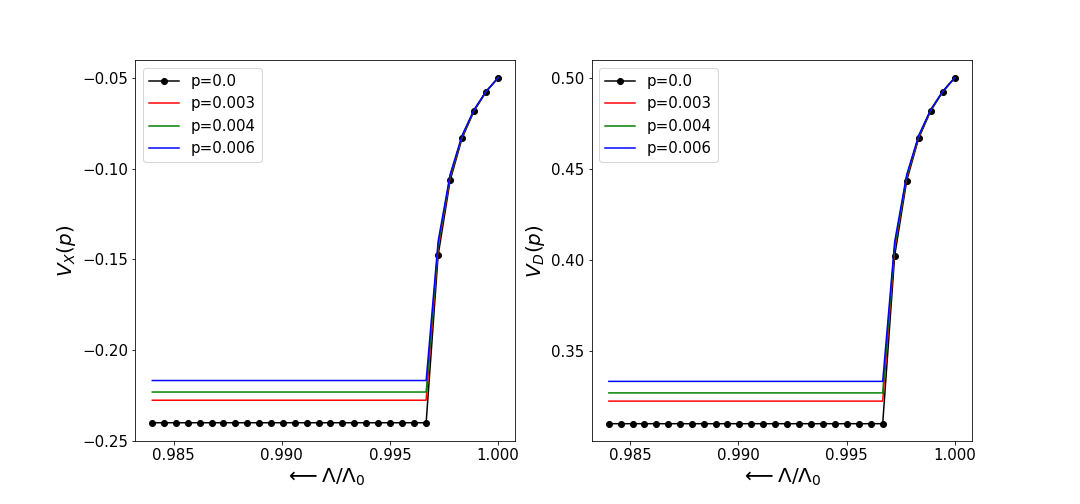}
\caption{Left/right panel represents the RG flow of two-particle number off-diagonal/diagonal scattering vertices for $p=0$ ($V^{(j)}_{X,p=0}$) and $p\neq 0$ finite-momentum pairs ($(V^{(j)}_{X,p'})$). Numerical evaluations are for system volume $1024\times 1024$ and with bare $V_{D,\mathbf{p}}=0.5$, $V_{X,\mathbf{p}}=-0.05$ (in units of $t$), $\omega=\epsilon_{\Lambda_{0}}-0.5$.}\label{XXZ_BCS}
\end{figure}

For the case when $V^{\sigma\sigma'}_{\mathbf{k},\mathbf{p}}>V^{\sigma\sigma'}_{\mathbf{k},\mathbf{k}',\mathbf{p}}$ (regime-IV in Table \ref{Table-conditions-II}), the diagonal vertices do not vanish under RG. Here, the fixed point condition becomes
\begin{eqnarray}
\frac{1}{2}\left(\epsilon^{(j^{*})}_{\mathbf{k}_{\Lambda
^{*}\hat{s}}}+\epsilon^{(j^{*})}_{-\mathbf{k}_{\Lambda^{*}\hat{s}}}\right)-\omega=\frac{1}{4}V^{\sigma,-\sigma,(j^{*})}_{\mathbf{k}_{\Lambda_{*}\hat{s}},0}.\label{energy balance cond_XXZ}
\end{eqnarray}
As an outcome, the fixed point is described by a modified XXZ pseudospin Hamiltonian
\begin{eqnarray}
H^{*,XXZ}_{RBCS}(\omega) &=&\sum_{\mathbf{k}}\epsilon^{(j^{*})}_{\mathbf{k}_{\Lambda\hat{s}}}A^{z}_{\mathbf{k}_{\Lambda\hat{s}}}-\sum_{\Lambda,\Lambda'<\Lambda^{*}}V^{\sigma,-\sigma,(j^{*})}_{\mathbf{k}_{\Lambda\hat{s}},\mathbf{k}_{\Lambda'\hat{s}'},0}A^{+}_{\mathbf{k}_{\Lambda\hat{s}}}A^{-}_{\mathbf{k}_{\Lambda'\hat{s}'}}+\sum_{\mathbf{k}_{\Lambda\hat{s}},\mathbf{p}}V^{\sigma,-\sigma,(j^{*})}_{\mathbf{k}_{\Lambda\hat{s}},\mathbf{p}} A^{z}_{\mathbf{k}_{\Lambda\hat{s}}}A^{z}_{\mathbf{p}-\mathbf{k}_{\Lambda\hat{s}}}~,~~~\label{XXZ_BCS_reduced_theory}
\end{eqnarray}
where $V^{\sigma,-\sigma,(j^{*})}_{\mathbf{k}_{\Lambda\hat{s}},\mathbf{p}}$ is the value of the Ising coupling at the fixed point. The RG flow features for this phase is represented via the tree diagram Fig.\ref{246vertices-tree-XXZ-BCS}. In this phase, finite magnitudes for both the number-diagonal as well as off-diagonal interactions lead to the quantities $r^{\sigma,-\sigma,(j*)}_{\mathbf{p}=0}= -r$ (where $r<1$). The left and right panels of Fig.\ref{XXZ_BCS} represent the RG flows for the 2 particle off-diagonal and diagonal scattering vertices respectively, involving electronic pairs with net-momentum $p=0.0$, $0.003$, $0.004$ and $0.006$. The bare couplings $V_{X,p}=-0.05$, $V_{D,p}=0.5$ ($|V_{X,p}|<|V_{D,p}|$) and fluctuation scale $\omega =\epsilon_{\Lambda_{0}}-0.5$ (Regime-III in Table~\ref{Table-conditions-I}), and a system volume as mentioned earlier. As the low-energy fixed point in this regime is dominated by $p=0$ momentum electronic pair scattering vertices: $|V^{*}_{X,p=0}|>|V^{*}_{X,p'}|$, $|V^{*}_{D,p=0}|>|V^{*}_{D,p'}|$, the resulting theory is described by the presence of both Ising and XY interactions between pseudospins (eq.\eqref{XXZ_BCS_reduced_theory}).
\\
\par\noindent
\textbf{IV. \textit{Reduced BCS theory for finite momentum pairs-XY regime}}\\
\begin{figure}
\includegraphics[width=\textwidth]{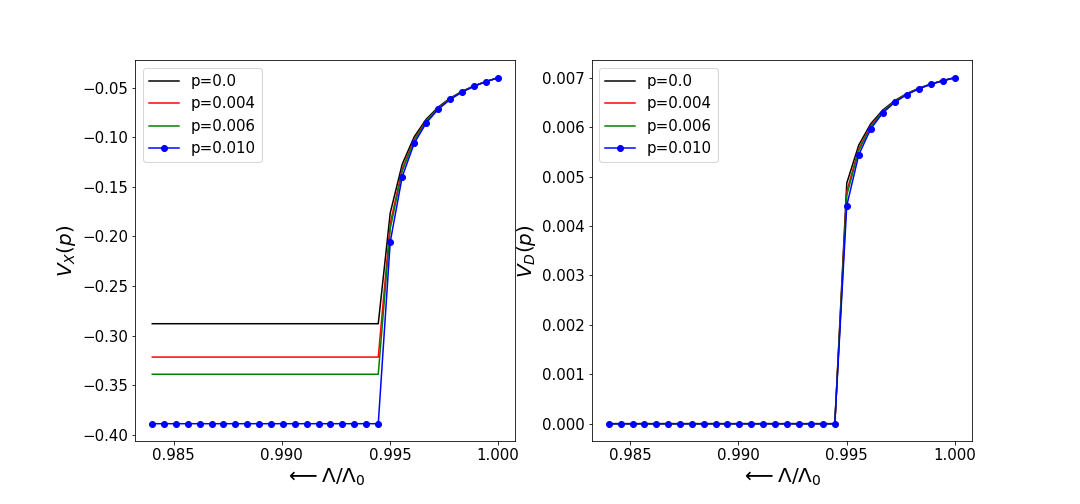}
\caption{Left/right panel represents the RG flow of two-particle number off-diagonal/diagonal scattering vertices respectively for $p=0$ ($V^{(j)}_{X,p=0}$) and $p\neq 0$ finite-momentum pairs ($(V^{(j)}_{X,p'})$). Numerical evaluations are for system volume $1024\times 1024$ and with bare $V_{D,\mathbf{p}}=0.007$, $V_{X,\mathbf{p}}=-0.05$ (in units of $t$), $\omega=\frac{1}{2}(\epsilon_{k_{\Lambda_{0}}}+\epsilon_{0.01-\mathbf{k}_{\Lambda_{0}}})-0.5$.}\label{XY-SPDW}
\end{figure}
In regime IV of Table \ref{Table-conditions-II}, for the fluctuation energy scale lying in the range
\begin{eqnarray}
\frac{1}{2}(\epsilon^{(j)}_{\mathbf{k}_{\Lambda_{j}\hat{s}}}+\epsilon^{(j)}_{\mathbf{p}-\mathbf{k}_{\Lambda_{j}\hat{s}}})>\omega >\frac{1}{2}(\epsilon^{(j)}_{\mathbf{k}_{\Lambda_{j}\hat{s}}}+\epsilon^{(j)}_{-\mathbf{k}_{\Lambda_{j}\hat{s}}})~,
\end{eqnarray}
finite $\mathbf{p}$ pair-momentum pseudospins attain a reduced BCS theory like fixed-point Hamiltonian. The fixed point is given by the condition
\begin{eqnarray}
\frac{1}{2}\left(\epsilon^{(j^{*})}_{\mathbf{k}_{\Lambda
^{*}\hat{s}}}+\epsilon^{(j^{*})}_{\mathbf{p}-\mathbf{k}_{\Lambda^{*}\hat{s}}}\right)-\omega =0~, \label{energy balance cond_PDW}
\end{eqnarray}
and with an effective fixed point Hamiltonian described by 
\begin{eqnarray}
H^{*,XY}_{SPDW}(\omega) &=&\sum_{\mathbf{k}}\epsilon^{(j^{*})}_{\mathbf{k}_{\Lambda\hat{s}},\mathbf{p}}A^{z}_{\mathbf{k}_{\Lambda\hat{s}},\mathbf{p}}\nonumber\\
&-& \sum_{\mathbf{k}_{\Lambda\hat{s}},\Lambda<\Lambda^{*}} V^{\sigma,-\sigma}_{\mathbf{k}_{\Lambda\hat{s}},\mathbf{k}_{\Lambda'\hat{s}'},\mathbf{p}} A^{+}_{\mathbf{k}_{\Lambda\hat{s}},\mathbf{p}}A^{-}_{\mathbf{k}_{\Lambda'\hat{s}'},\mathbf{p}},~~~~\label{HPDW}
\end{eqnarray}
where the set $\nu =\lbrace (\mathbf{k}_{\Lambda\hat{s}},\sigma);(\mathbf{p}-\mathbf{k}_{\Lambda\hat{s}},-\sigma)\rbrace$ corresponds to a pair of electronic states with net momentum $\mathbf{p}$. The ground state of $H^{*,XY}_{SPDW}$ is composed of symmetry-unbroken pair-density waves (SPDWs)~\cite{fulde1964superconductivity,larkin1965sov}. The pseudospin vector components for such finite-momentum pair of electrons are defined as
\begin{eqnarray}
A^{+}_{\mathbf{k}_{\Lambda\hat{s}},\mathbf{p}}&=&c^{\dagger}_{\mathbf{k}_{\Lambda\hat{s}}\sigma}c^{\dagger}_{\mathbf{p}-\mathbf{k}_{\Lambda\hat{s}} -\sigma}, A^{-}_{\mathbf{k}_{\Lambda\hat{s}},\mathbf{p}}=A^{+\dagger}_{\mathbf{k}_{\Lambda\hat{s}},\mathbf{p}}, A^{z}_{\mathbf{k}_{\Lambda\hat{s}},\mathbf{p}}=\frac{1}{2}[A^{+}_{\mathbf{k}_{\Lambda\hat{s}},\mathbf{p}},A^{-}_{\mathbf{k}_{\Lambda\hat{s}},\mathbf{p}}]~.\label{finite_momentum_pseudo_spin_pairs}
\end{eqnarray}\par\noindent
Given that Ising terms are absent from the effective Hamiltonian, 
we obtain the quantity $r^{\sigma,-\sigma,(j*)}_{\mathbf{p}}= -1$ for the fixed point theory. The RG flow features for this phase is represented via the tree diagram in Fig.\ref{246vertices-tree-XY-BCS}. A numerical evaluation of the RG flow is shown in Fig.\ref{XY-SPDW} left and right panels for 2-particle off-diagonal and number-diagonal scattering vertices respectively, and involving electronic pairs with net momentum $p=0.0$, $0.004$, $0.006$, $0.01$. The bare couplings $V_{X,p}=-0.05$, $V_{D,p}=0.5$ ($|V_{X,p}|<|V_{D,p}|$) and fluctuation scale $\omega =\frac{1}{2}(\epsilon_{k_{\Lambda_{0}}}+\epsilon_{p-k_{\Lambda_{0}}})-0.5$ (Regime-IV in Table~\ref{Table-conditions-II}), and system volume as mentioned earlier. In this regime, we find that the off-diagonal scattering vertices $V^{*}_{X,p}=V^{\sigma,-\sigma}_{k,k',p}$ with the largest non-zero pair-momentum (here, the curve for $p=0.01$ in Fig.\ref{XY-SPDW} (left panel)) dominate the low energy physics. However, we find the Ising interactions to be RG irrelevant for all pair momenta (Fig.\ref{XY-SPDW} (right panel)), and the phase is described in terms of $p=0.01$ momentum pseudospin pairs interacting via XY interactions (eq.\eqref{XXZ_BCS_reduced_theory}).
\\
\pin
\textbf{V. \textit{Reduced BCS theory for finite momentum pairs-XXZ regime}}\\
\begin{figure}
\includegraphics[width=\textwidth]{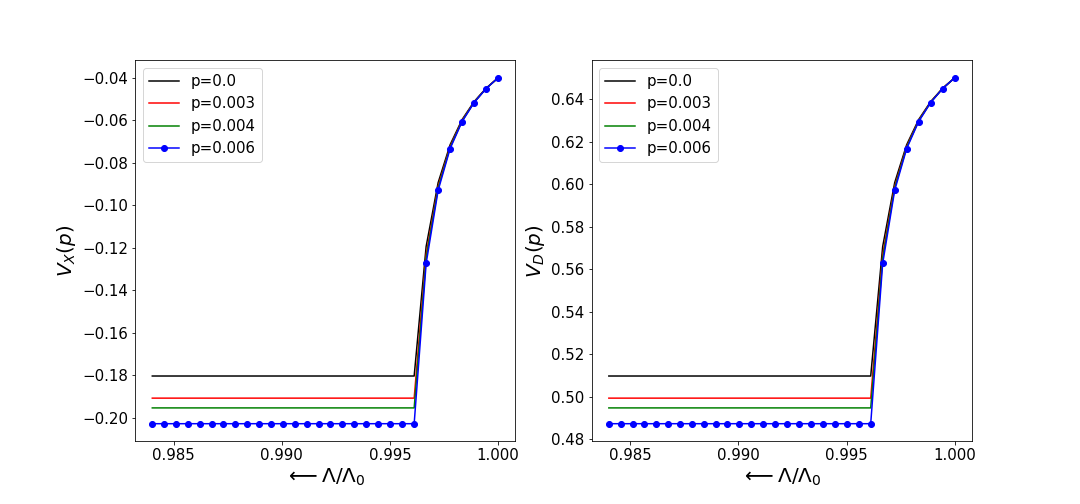}
\caption{Left/right panel represents the RG flow of two-particle number off-diagonal/diagonal scattering vertices respectively for $p=0$ ($V^{(j)}_{X,p=0}$) and $p\neq 0$ finite-momentum pairs ($(V^{(j)}_{X,p'})$). Numerical evaluations are for system volume $1024\times 1024$ and with bare $V_{D,\mathbf{p}}=0.64$, $V_{X,\mathbf{p}}=-0.04$ (in units of $t$), $\omega=\frac{1}{2}(\epsilon_{k_{\Lambda_{0}}}+\epsilon_{0.01-\mathbf{k}_{\Lambda_{0}}})-0.5$.}\label{XXZ-SPDW}
\end{figure}
Similarly, in regime V in Table \ref{Table-conditions-II}, we obtain a phase composed of finite-momentum pseudospins interacting via XXZ interaction. The effective Hamiltonian describing this phase is
\begin{eqnarray}
H^{*,XXZ}_{SPDW}(\omega)&=&\sum_{\mathbf{k}}\epsilon^{(j^{*})}_{\mathbf{k}_{\Lambda\hat{s}}}A^{z}_{\mathbf{k}_{\Lambda\hat{s}}}\nonumber\\
&-&\sum_{\Lambda,\Lambda'<\Lambda^{*}}V^{\sigma,-\sigma,(j^{*})}_{\mathbf{k}_{\Lambda\hat{s}},\mathbf{k}_{\Lambda'\hat{s}'},0}A^{+}_{\mathbf{k}_{\Lambda\hat{s}},\mathbf{p}}A^{-}_{\mathbf{k}_{\Lambda'\hat{s}'},\mathbf{p}}\nonumber\\
&+&\sum_{\mathbf{k}_{\Lambda\hat{s}},\mathbf{p}}V^{\sigma,-\sigma,(j^{*})}_{\mathbf{k}_{\Lambda\hat{s}},\mathbf{p}'} A^{z}_{\mathbf{k}_{\Lambda\hat{s}},\mathbf{p}}A^{z}_{\mathbf{p}'-\mathbf{k}_{\Lambda\hat{s}},\mathbf{p}}~.\label{XXZ-PDW}
\end{eqnarray}
Finite magnitudes for both the number-diagonal and off-diagonal interactions leads to the quantity $r^{\sigma,-\sigma,(j*)}_{\mathbf{p}=0}= -r$, where $r<1$. The tree diagram Fig.\ref{246vertices-tree-XXZ-BCS} represents the corresponding vertex tensor RG flow. As shown in Fig.\ref{XXZ-SPDW} (left and right panels), a numerical evaluation of the RG flow for 
bare coupling $V_{X,p}=-0.04$, $V_{D,p}=0.64$ and $\omega=\frac{1}{2}(\epsilon_{\Lambda_{0}}+\epsilon_{0.01-k_{\Lambda_{0}}})-0.5$ (Regime-V in Table\ref{Table-conditions-II}) reveals that at the IR fixed point, both off-diagonal and number-diagonal renormalized couplings attain a finite magnitude. The low-energy fixed point theory is, therefore, dominated by finite-momentum pseudospin pairs interacting by a XXZ interaction (eq.\eqref{XXZ-PDW}).
\\
\par\noindent
\textbf{VI. \textit{Tensor network representation of the reduced BCS model and Fermi liquid theory}}\\
The above fixed point Hamiltonians for regimes I-V Table\ref{Table-conditions-I} and \ref{Table-conditions-II} can be broadly classified into gapless and gapped phases. The Fermi liquid corresponds to the gapless phase containing purely number-diagonal interactions, such that $H^{*}_{FL}$ eq\eqref{HFL} is purely number-diagonal in Fock space and various terms in it commute. Therefore, all the number operators $\hat{n}_{\mathbf{k}\sigma}$ corresponding to states $|\mathbf{k}\sigma\rangle$ (lying within the window whose boundaries are given by the states $\mathbf{k}_{\Lambda^{*}\hat{s}}$, eq.\eqref{FL_fixed point}) commute with $H_{FL}$, such that their  eigenvalues correspond to integrals of motion. Following our demonstration of a tensor network representation for the unitary RG flow in \cite{anirbanurg1}, the RG flow towards Fermi liquid fixed point is displayed as a tensor network in Fig.\ref{TN-FL} below; the features of the emergent Fermi liquid theory discussed above are clearly visible at the final layer of the network.
\begin{figure}[h!]
\centering
\includegraphics[scale=4]{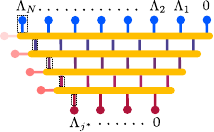}
\caption{Vertex tensor network representation of the RG flow towards the Fermi liquid. The blue legs labelled $0,\ldots,\Lambda_{n}$ represent the holographic boundary composed of electronic states, the yellow blocks represent nonlocal unitary disentanglers that map the boundary states to the bulk with lowering energies (UV to IR, varying from light to deeper shade of red). The variation in the colour of the input legs into the subsequent unitary operators (yellow blocks) depicts the change of the entanglement content  within the remnant coupled electronic states as the RG flows from UV to IR. The final unitary transformation layer leads to a theory comprised of decoupled legs labelled $0,\ldots,\Lambda_{j^{*}}$. These are represented in brown, and each leg has an integral of motion associated with it.}\label{TN-FL}
\end{figure}
\par
\begin{figure}[h!]
\centering
\includegraphics[scale=4]{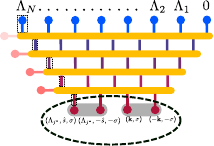}
\caption{Entanglment holographic mapping network representation of the RG flow towards the reduced BCS theory. Pairs of brown legs (in grey boxes) represents pairs of electronic states with zero net-momentum and zero net-spin. The dashed oval represents the interaction between the pseudospins.}\label{TN-BCS}
\end{figure}

\noindent
The reduced BCS theory $H^{RBCS}_{(j^{*}),XY}$ (eq.\eqref{HBCS}) and its variants $H^{RBCS}_{(j^{*}),XXZ}$, $H^{SPDW}_{(j^{*}),XY}$, $H^{SPDW}_{(j^{*}),XY}$ all correspond to gapped condensates. In contrast to that shown for the Fermi liquid, the tensor network representation of the RG flow towards such gapped ground states displays an emergent pairing of the legs in the final layer. The pairing of legs $\mathbf{k}\sigma$ and $-\mathbf{k}-\sigma$ can be seen in the grey boxes in Fig.\ref{TN-BCS}, while the emergent condensate as a whole is encircled in the black dashed line. The dashed oval in Fig. \ref{TN-BCS} represents the XY and Ising interaction between this pseudospins.
Having achieved the Fermi liquid and BCS regimes, we will turn below
towards looking for more exotic states of matter, such as the marginal Fermi liquid and gapped condensate ground states that involve hybridised spin- and charge-pseudospin pairing.
\subsection{The Marginal Fermi liquid}
In this subsection, we explore the possibility of a metallic phase different from the Fermi liquid being found within the parameter space of the $H_{SFIM}$ model. For this, one possible distinguishing feature could be the nature of long-lived excitations in the proximity of the Fermi surface that replace the Landau quasiparticles of the Fermi liquid. Thus, we investigate the physics of the lowest-order decay channel of 1-particle (Landau quasiparticle) excitations, i.e., 2-electron 1-hole composites with a net charge $e$ and net spin $1/2$. Although 6-point (or 3-particle) scattering vertices are absent in the bare Hamiltonian $H_{SFIM}$ (eq.\eqref{SFIM}), they are generated under RG~\cite{anirbanurg1}.
Such 6-point diagonal/off-diagonal scattering terms describe the interaction between the 2-electron 1-hole composites. These interactions bring about a log-divergence in the 1-particle self-energy~\cite{anirbanurg1}, and require therefore a controlled RG treatment to study the nature of the resulting metallic phase.
\begin{figure}
\includegraphics[width=\textwidth]{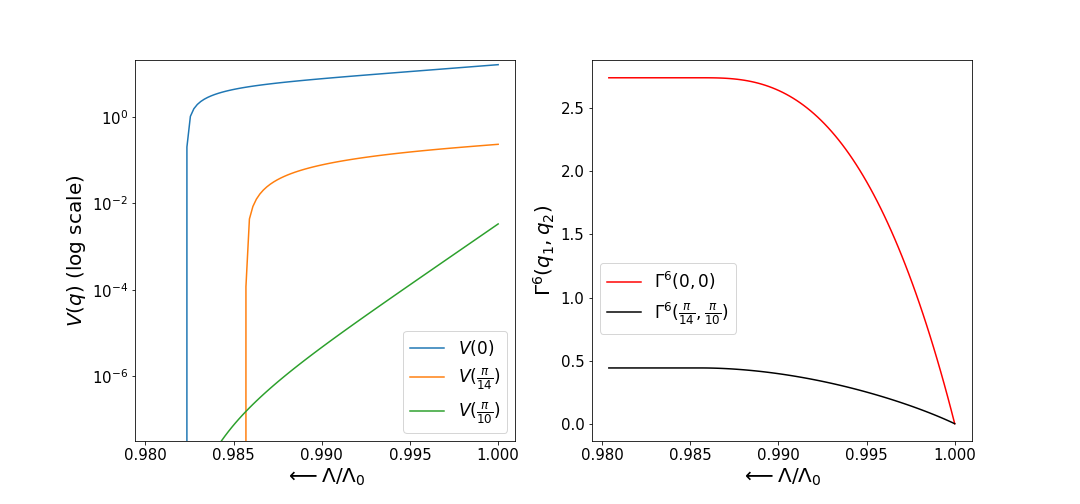}
\caption{Left panel: RG flows for finite-momentum transfer $q\neq0$ off-diagonal two-particle forward scattering vertices ($V(q)$, orange and green curves) and number-diagonal ($q=0$, blue curve) scattering vertices, both represented in log-scale on the $y$-axis. Right panel: RG flows for the number off-diagonal $\Gamma^{6}(q_{1},q_{2})$ three-particle scattering vertex (black curve), and two-particle one-hole number-diagonal ($\Gamma^{6}(0)$) scattering vertex (red curve).}\label{MFL-I}
\end{figure}
\par\noindent
We will now see that the conditions 1-3 listed under regime VI in Table \ref{Table-conditions-II} correspond to the formation of the non-Fermi liquid gapless phase best described as a marginal Fermi liquid. We begin by exploring the implications of these conditions using 4- and 6-point vertex flow equations in eq.\eqref{4and6 particle vertex}. To proceed, we first note the form of the six point diagonal and off diagonal vertices
\begin{eqnarray}
H^{6}_{(j)}&=&\sum_{\substack{\mathbf{k},\mathbf{k}'\mathbf{k}'',\\
\mathbf{p},\mathbf{p}'}}R^{\sigma\sigma'\sigma}_{\mathbf{k}\mathbf{k}'\mathbf{k}''\mathbf{p}\mathbf{p}'}\bigg(c^{\dagger}_{\mathbf{k},\sigma}c^{\dagger}_{\mathbf{p}-\mathbf{k},\sigma'}c^{\dagger}_{\mathbf{p}'-\mathbf{k}',\sigma'}c_{\mathbf{p}-\mathbf{k}',\sigma'}c_{\mathbf{k}'',\sigma'}c_{\mathbf{p}'-\mathbf{k}'',\sigma}\bigg)\nonumber\\
&+&\sum_{\mathbf{k},\mathbf{k}',\mathbf{p}}R^{\sigma\sigma'\sigma}_{\mathbf{k}\mathbf{k}'\mathbf{p}}\tau_{\mathbf{k}\sigma}\tau_{\mathbf{p}-\mathbf{k}\sigma'}\tau_{\mathbf{p}-\mathbf{k}'\sigma'}~.\label{terms-six-point}
\end{eqnarray}
Then, the RG flow equations for the 6-point diagonal (eq.\eqref{six-point-D-vertices}) and off-diagonal vertices (eq.\eqref{2and3particleVertexcond}) are given by
\begin{eqnarray}
\Delta\Gamma^{6,(j)}_{\alpha\beta}&=& \Gamma^{4,(j)}_{\alpha\gamma}G^{2,(j)}_{\gamma\gamma'}\Gamma^{4,(j)}_{\gamma\beta}-\Gamma^{6,(j)}_{\alpha\gamma}G^{6,(j)}_{\gamma\gamma'}\Gamma^{6,(j)}_{\gamma\beta}\nonumber\\
&+&\Gamma^{6,(j)}_{\alpha\gamma}G^{4,(j)}_{\gamma\gamma'}\Gamma^{4,(j)}_{\alpha\gamma},\label{six-point-OD}\\
\Delta \Gamma^{6,(j)}_{\alpha\alpha'}&=&\Gamma^{4,(j)}_{\alpha\gamma}G^{2,(j)}_{\gamma\gamma'}\Gamma^{4,(j)}_{\gamma'\alpha'}-\Gamma^{6,(j)}_{\alpha\gamma}G^{6,(j)}_{\gamma\gamma'}\Gamma^{6,(j)}_{\gamma'\alpha'}~.~~~~\label{six-point-D}
\end{eqnarray}
In the above expressions, $G^{6,(j)}_{\gamma\gamma'}$ is obtained from eq.\eqref{six_point_green} in the 2-electron 1-hole eigenconfiguration of the three-fermion string
$\tau_{\mathbf{k}_{\Lambda_{j}\hat{s}_{1}}\sigma'}\sigma_{\mathbf{p}-\mathbf{k}_{\Lambda_{j}\hat{s}_{1}}\sigma'}\sigma_{\mathbf{p}-\mathbf{k}_{\Lambda\hat{s}}\sigma}=-\frac{1}{8}$, leading to a negative sign in the RG equations given above in eqs.\eqref{six-point-OD} and \eqref{six-point-D}. Now, for the fluctuation energy in the range (regime VI condition 1)
\begin{eqnarray}
\frac{1}{2}\epsilon_{\mathbf{k}_{\Lambda\hat{s}}}<\omega <\frac{1}{2}(\epsilon_{\mathbf{k}_{\Lambda\hat{s}}} +\epsilon_{\mathbf{p}-\mathbf{k}_{\Lambda\hat{s}}}), \text{ for }\epsilon^{(j)}_{\mathbf{p}-\mathbf{k}_{\Lambda\hat{s}}}>0~,~~~\label{three_particle_regime}
\end{eqnarray}
we have $G^{4,(j)}_{\gamma\gamma'}<0$ (eq\eqref{4-point-green}). Following eq.\eqref{4-particle vertex}, this results in the 4-point vertex RG flow being irrelevant: $\Delta\Gamma^{4,(j)}_{\alpha\beta}<0$.
\pin
Importantly, note that $G^{4,(j)}_{\gamma\gamma'}<0$ leads to an additional negative contribution in the RG equation (eq.\eqref{six-point-OD}) for the off-diagonal 6-point vertices $\Gamma^{6,(j)}_{\alpha\beta}$, while such a term is absent in the RG equation for the diagonal 6-point vertices $\Gamma^{6,(j)}_{\alpha\alpha'}$ (eq.\eqref{six-point-D}). This extra negative contribution leads to $\Delta \Gamma^{6,(j)}_{\alpha\alpha'}<\Delta \Gamma^{6,(j)}_{\alpha\beta}$. We now argue that the above inequality implies $\Gamma^{6,(j)}_{\alpha\alpha'}<\Gamma^{6,(j)}_{\alpha\beta}$. For this, we first note that the 6-point vertices are generated only at the first RG step from the 4-point vertices (eq.\eqref{2and3particleVertexcond}) as follows
\begin{eqnarray}
\Gamma^{6,(N-1)}_{\alpha\beta} &=& \Delta\Gamma^{6,(N)}_{\alpha\beta} = \Gamma^{4,(N)}_{\alpha\gamma}G^{4,(N)}_{\gamma\gamma'}\Gamma^{4,(N)}_{\gamma'\beta},\nonumber\\
\Gamma^{6,(N-1)}_{\alpha\alpha'} &=& \Delta\Gamma^{6,(N)}_{\alpha\alpha'} = \Gamma^{4,(N)}_{\alpha\gamma}G^{4,(N)}_{\gamma\gamma'}\Gamma^{4,(N)}_{\gamma'\alpha'}.\label{six-point-initialize}
\end{eqnarray}
As the 2-point Green's function $G^{2,\sigma,(j)}_{\gamma\gamma'}=(\omega_{(j)}-\frac{1}{2}\epsilon_{(j)})>0$ carries positive signature in the energy range of eq.\eqref{three_particle_regime}, the 3-particle interactions are repulsive in nature. It is then simple to observe from the above expression that the diagonal and the off-diagonal 6-point vertices have similar magnitude $\Gamma^{6,(N-1)}_{\alpha\beta} \sim \Gamma^{6,(N-1)}_{\alpha\alpha'}$. Then, from the discussion above, we conclude that under RG, the renormalised 6-point vertices satisfy
\begin{eqnarray}
\Gamma^{6,(j)}_{\alpha\beta}<\Gamma^{6,(j)}_{\alpha\alpha'}~.\label{ordering_of_vertices}
\end{eqnarray}
In order to numerically evaluate the renormalized six-point vertices generated under RG and their precise ordering (eq.\eqref{ordering_of_vertices}), we assume a simplified bare form of the vertices. $\Gamma^{6,(j)}(0,0)$ represents the strength of the number diagonal vertices (i.e., the second term in eq.\eqref{terms-six-point}), and $\Gamma^{6}(q_{1},q_{2})=\Gamma^{6,(j)}_{\alpha\beta}$($\mathbf{q}_{1}=\mathbf{k}-\mathbf{p}'+\mathbf{k}''$, $\mathbf{q}_{2}=\mathbf{p}-\mathbf{p}'$) the strength of the six-point off-diagonal vertex (i.e., the first term in eq.\eqref{terms-six-point}). As shown in the right panel of Fig.\ref{MFL-I}, both $\Gamma^{6}(0,0)$ and $\Gamma^{6}(q_{1},q_{2})$ grow under RG and saturate at fixed points $\Gamma^{6,*}(0,0)<\Gamma^{6,*}(q_{1},q_{2})$ with $\Lambda^{*}=0.98\Lambda_{0}$. Given that in the vicinity of the fixed point, both number-diagonal and off-diagonal four-point vertices vanish under RG (Fig.\ref{MFL-I} (left panel)), $\Gamma^{4,(j)}_{\alpha\beta}\to 0$, we find that the RG equations for the six-point diagonal and off-diagonal vertices attain a simplfied form
\begin{eqnarray}
\Delta \Gamma^{6,(j)}_{\alpha\beta} = -\Gamma^{6,(j)}_{\alpha\gamma}G^{6,(j)}_{\gamma\gamma'}\Gamma^{6,(j)}_{\gamma\beta},\nonumber\\
\Delta \Gamma^{6,(j)}_{\alpha\alpha'} = -\Gamma^{6,(j)}_{\alpha\gamma}G^{6,(j)}_{\gamma\gamma'}\Gamma^{6,(j)}_{\gamma\alpha'}~,\label{sixRGflows}
\end{eqnarray}
where, by using eq.\eqref{six_point_green}, $G^{6,\sigma\sigma'\sigma,(j)}_{\mathbf{k}_{\Lambda_{j}\hat{s}},\mathbf{p},\mathbf{p}'}$ ($\omega_{(j)}=\omega$) is given by
\begin{eqnarray}
G^{6,(j)}_{\gamma\gamma'}&=&\bigg(\omega -\epsilon^{(j)}_{\mathbf{k}_{\Lambda_{j}\hat{s}}}\tau_{\mathbf{k}_{\Lambda_{j}\hat{s}}\sigma} -\epsilon^{(j)}_{\mathbf{p}'-\mathbf{k}_{\Lambda_{j}\hat{s}}}\tau_{\mathbf{p}'-\mathbf{k}_{\Lambda_{j}\hat{s}},\sigma'}-\epsilon^{(j)}_{\mathbf{p}-\mathbf{k}_{\Lambda_{j}\hat{s}}}\tau_{\mathbf{p}'-\mathbf{k}_{\Lambda_{j}\hat{s}},\sigma}\nonumber\\
&-&\Gamma^{6,(j)}_{\gamma\gamma'}\tau_{\mathbf{k}_{\Lambda_{j}\hat{s}}\sigma}\tau_{\mathbf{p}-\mathbf{k}_{\Lambda_{j}\hat{s}},\sigma'}\tau_{\mathbf{p}'-\mathbf{k}_{\Lambda_{j}\hat{s}},\sigma} \bigg)^{-1}.\label{6-point green function}
\end{eqnarray}
\pin
In order to obtain the stable fixed point theory, we choose an intermediate configuration ($\hat{n}_{\mathbf{k}_{\Lambda_{j}\hat{s}}\sigma}=1$, $\hat{n}_{\mathbf{p}-\mathbf{k}_{\Lambda_{j}\hat{s}}\sigma}=1$, $\hat{n}_{\mathbf{p}'-\mathbf{k}_{\Lambda_{j}\hat{s}}\sigma}=0$) for the 6-point Green's function. The net configurational energy for such a composite 3-particle is given by a combination of their individual 1-particle energies and the net 3-particle energy
\begin{eqnarray}
E^{(j)} &=&E^{(j)}_{1}-\frac{1}{8}R^{\sigma\sigma'\sigma,(j)}_{\mathbf{k}'',\mathbf{p},\mathbf{p}'}~~,~~ E^{(j)}_{1} =\frac{1}{2}(\epsilon^{(j)}_{\mathbf{k}_{\Lambda_{j}\hat{s}}}+\epsilon^{(j)}_{\mathbf{p}-\mathbf{k}_{\Lambda_{j}\hat{s}}}-\epsilon^{(j)}_{\mathbf{p}'-\mathbf{k}_{\Lambda_{j}\hat{s}}})~.~~~~~~\label{tot energy 1,3}
\end{eqnarray} 
Thus, the 6-point Green's function eq.\eqref{six_point_green} is given by 
\begin{eqnarray}
G_{\gamma\gamma'}^{6,(j)}(\omega)=(\omega-E^{(j)})^{-1}~.\label{six_point_green_function}
\end{eqnarray}
\begin{figure}
\centering
\includegraphics[width=0.7\textwidth]{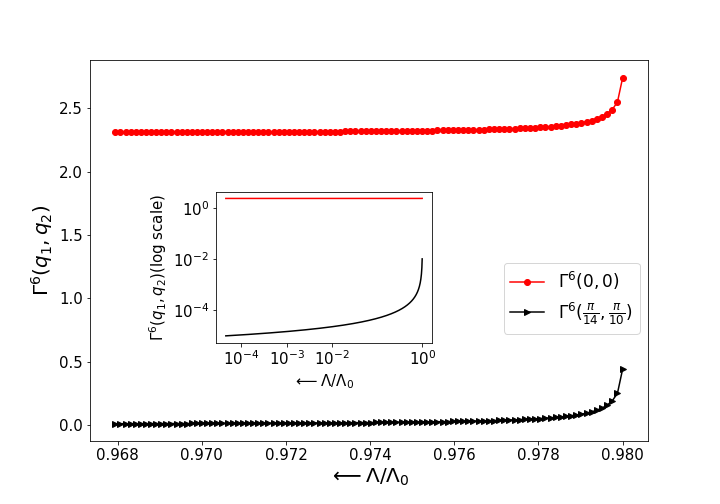}
\caption{RG flow of the six-point off-diagonal ($\Gamma^{6}(q_{1},q_{2})$, black curve)~/~diagonal ($\Gamma^{6}(0,0)$, red curve) scattering vertices for $\Lambda<0.98\Lambda_0$. The inset plot shows the same curves in log-scale for both $x$- and $y$-axes.}\label{MFL-II}
\end{figure}
\pin
The set $\gamma = \lbrace(\mathbf{k}_{\Lambda\hat{s}},\sigma,1),(\mathbf{k}^{'},\sigma',1),(\mathbf{k}^{''},-\sigma',0)\rbrace$, while the set $\gamma'$ involves the same momentum and spin labels, but with the 1s and 0s interchanged. In this basis, the RG flows for the off-diagonal 3-particle vertex and the diagonal 2-electron 1-hole dispersion are given by
\begin{eqnarray}
\Delta \Gamma^{6,(j)}_{\alpha\beta} = -\frac{\Gamma^{6,(j)}_{\alpha\gamma}\Gamma^{6,(j)}_{\gamma\beta}}{\omega - E_{1}^{(j)}+\frac{1}{8}\Gamma^{6,(j)}_{\gamma\gamma'}}, ~\Delta \Gamma^{6,(j)}_{\alpha\alpha'} = -\frac{\Gamma^{6,(j)}_{\rho\gamma}\Gamma^{6,(j)}_{\gamma\rho'}}{\omega - E_{1}^{(j)}+\frac{1}{8}\Gamma^{6,(j)}_{\gamma\gamma'}}~. \label{sixRGflows1}
\end{eqnarray}
\pin
Finally, from the RG flow of the couplings eq.\eqref{sixRGflows1}, the parameter range eq.\eqref{ordering_of_vertices} for the 6-point off-diagonal/diagonal vertices and for fluctuation energies $\omega<E^{(j)}_{1}$, we obtain the fixed point condition for a gapless phase with the 3-particle composite excitations proximate to the Fermi surface
\begin{eqnarray}
\frac{1}{8}\Gamma^{6,(j*)}_{\gamma\gamma'} &=& E^{(j{*})}_{1}-\omega~, \Gamma^{6,(j*)}_{\alpha\beta}=0.
\label{3-number_diag_fp}
\end{eqnarray}
Additionally, we note that given $\frac{1}{2}(\epsilon^{(j)}_{\mathbf{k}}+\epsilon^{(j)}_{\mathbf{p}-\mathbf{k}})>\omega_{(j)}>\frac{1}{2}\epsilon^{(j)}_{\mathbf{k}}$ (for $\epsilon^{(j)}_{\mathbf{p}-\mathbf{k}}>E_{F}$), the energy $E^{(j)}_{1}>\frac{1}{2}\epsilon^{(j)}_{\mathbf{k}_{\Lambda_{j}\hat{s}}}$ and the inequality $\omega<E^{(j)}_{1}$ is immediately satisfied for the energy of the electron-occupied states lying above $E_{F}$ and the energy of the hole configuration lying below $E_{F}$. 
\pin
At the fixed point theory, the dynamics of the states within the window $0<\Lambda<\Lambda_{j^{*}}$ is governed by the effective Hamiltonian 
\begin{eqnarray}
H_{MFL}^{*}(\omega) &=&\sum_{\Lambda<\Lambda_{j^{*}},\hat{s}}\epsilon^{*}_{\mathbf{k}_{\Lambda\hat{s}}}\hat{n}_{\mathbf{k}_{\Lambda\hat{s}}\sigma}+\sum_{\rho,\Lambda<\Lambda_{l^{*}},\mathbf{p}}\Gamma^{6,(j*)}_{\rho
\rho'}\hat{n}_{\mathbf{k}'',-\sigma}\hat{n}_{\mathbf{p}-\mathbf{k}_{\Lambda\hat{s}},\sigma}(1-\hat{n}_{\mathbf{k}_{\Lambda\hat{s}},\sigma}).~~~~~~\label{3-numberdiag}
\end{eqnarray}
For $\Lambda<0.98\Lambda_{0}$, the RG flows of six-point vertices have no contribution from two-particle vertices (as already observed in Fig.\ref{MFL-I}), and is generated purely by the six-point vertices (eq.\eqref{sixRGflows1}). The nature of the RG flow for the six-point vertices $\Gamma^{6}(q_{1},q_{2})$ and $\Gamma^{6}(0)$ below the RG scale $\Lambda<0.98\Lambda_{0}$ is thus obtained in Fig.\ref{MFL-II} from a numerical computation of eqs.\eqref{sixRGflows1}. The plots indicate vanishing of the six-point off-diagonal vertices $\Gamma^{6}(\pi/14,\pi/10)$ under RG, while the two-particle one-hole vertices $\Gamma^{6}(0,0)$ reach an RG fixed point with a finite (and large) value. The inset in Fig.\ref{MFL-II} shows that $\Gamma^{6}(\pi/14,\pi/10)$ reduces in magnitude from $O(1)$ to $O(10^{-4})$ (black curve), while $\Gamma^{6}(0,0)$ saturates at $O(1)$. In this way, we demonstrate numerically the MFL effective Hamiltonian (eq.\eqref{3-numberdiag}). We note that the MFL fixed point Hamiltonian is purely number-diagonal (similar to the Fermi liquid), translational invariant and has a gapless continuum spectrum that is a function of the wave-vector. All of this indicates the metallic nature of the ground state obtained at this new fixed point. 
\par\noindent
Next, we proceed to find the effect of such three-particle vertices on the 1-electron excitations in the neighborhood of the fixed point theory. For that, we note that the primary decay channel for the one-electron degrees of freedom due to three-particle off-diagonal scattering terms are three-electron two-hole excitations. Therefore, the electronic self-energy renormalizes via six-point vertices 
(eq.\eqref{1-p-self-energy}) 
\begin{eqnarray}
\Delta\Sigma^{(j)}_{\mathbf{k}_{\Lambda\hat{s}}}(\omega) = \sum_{\mathbf{k}'\mathbf{k}''\mathbf{p}}\frac{(\Gamma^{6,(j)}_{\alpha\beta})^{2}}{\omega  - E^{(j)}_{5,1}-E^{(j)}_{5,3}}~,\label{1-particle_self_energy_RG}
\end{eqnarray}
where
\begin{eqnarray}
E^{(j)}_{5,1}&=& E^{(j)}_{1}+\frac{1}{2}(\epsilon^{(j)}_{\mathbf{k}_{\Lambda_{1}\hat{s}}}+\epsilon^{(j)}_{\mathbf{p}-\mathbf{k}_{\Lambda_{1}\hat{s}}})~,\label{NetFiveCorrelationEnergy}
\end{eqnarray}
is the collective energy due to 4-p 1-h intermediate configuration of electronic states. Here, $E^{(j)}_{5,3}$ contains  the 2-electron 1-hole correlation energy term $-\frac{1}{8}\Gamma^{6,(j)}_{\gamma\gamma'}$, and $E^{(j)}_{1}$ is the net energy due to 3-electron 2-hole composite given by eq.\eqref{tot energy 1,3}. Now, using the 1-p self-energy RG flow equation eq.\eqref{1-particle_self_energy_RG}, and following Appendix \ref{Appendix-1pselfenergy}, we arrive at the form for the renormalized self-energy at fixed point ($j^{*}$)
\begin{eqnarray}
\Sigma^{(j^{*})}_{\mathbf{k}_{\Lambda_{l^{*}}\hat{s}}}(\omega) = N(0)\frac{(\Gamma^{6,(l^{*})}_{X,\mathbf{k}_{\Lambda_{l^{*}}\hat{s}}}(\omega))^{2}}{\Gamma^{6,(l^{*})}_{D,\mathbf{k}_{\Lambda_{l^{*}}\hat{s}}}(\omega)}\ln\bigg\vert\frac{\omega_{c}}{\omega}\bigg\vert,~~~~~~~\label{renormalized_self_energy}
\end{eqnarray}
where $\omega_{c}=\epsilon_{\mathbf{k}_{\Lambda_{(l^{*})}}}$ is the characteristic energy scale that is emergent from the RG fixed point eq.\eqref{3-number_diag_fp} and $N(0)$ is a dimensionless number equal to the total electronic state count at the FS.The ratio of the final fixed point 6-point off-diagonal/diagonal vertex strength for the states at $\Lambda_{l^{*}}$ distance from FS (eq.\eqref{renormalized_self_energy}) can be computed by investigating their RG equations in its neighborhood. Near the fixed point (and near the FS), the 6-point vertex flow equations (eq.\eqref{3-number_diag_fp} ) are simplified by using eq.\eqref{energy_space_constraint} and given by
\begin{eqnarray}
\frac{\Delta \Gamma^{6,(j)}_{X,\mathbf{k}_{\Lambda_{j}\hat{s}}}}{\Delta\log_{b}
\frac{\Lambda_{j}}{\Lambda_{0}}}=\frac{\Delta \Gamma^{6,(j)}_{D,\mathbf{k}_{\Lambda_{j}\hat{s}}}}{\Delta\log_{b}
\frac{\Lambda_{j}}{\Lambda_{0}}} = \frac{(\Gamma^{6,(j)}_{X,\mathbf{k}_{\Lambda_{j}\hat{s}}})^{2}}{\omega-\frac{1}{2}\epsilon^{(j)}_{\mathbf{k}_{\Lambda_{j\hat{s}}}}-\frac{1}{8}\Gamma^{6,(j)}_{D,\mathbf{k}_{\Lambda_{l}\hat{s}}}(\omega)}~,~~~~~~~\label{low_energy_flow_equations}
\end{eqnarray}
where $\Delta\log_{b}
\frac{\Lambda_{j}}{\Lambda_{0}}=1$ for $\Lambda_{j}=\Lambda_{0}b^{j}$. Here, $\Gamma^{6,(j)}_{X,\mathbf{k}_{\Lambda_{j}\hat{s}}}$/$\Gamma^{6,(j)}_{D,\mathbf{k}_{\Lambda_{j}\hat{s}}}$ represent the uniform pieces of the off-diagonal/diagonal parts of the three-particle vertex. From eq.\eqref{low_energy_flow_equations}, we obtain the RG invariant relation: $\Gamma^{6,(j)}_{X,\mathbf{k}_{\Lambda\hat{s}}} = \Gamma^{6,(j)}_{D,\mathbf{k}_{\Lambda\hat{s}}}+C$, where $C$ is the RG invariant. At the fixed point eq.\eqref{3-number_diag_fp}, $C=0$ and $\epsilon^{(j)}_{\Lambda\hat{s}}\to 0$ as $\Lambda\to 0$ (FS), leading to $\Gamma^{6,(l^{*})}_{D,\mathbf{k}_{\Lambda\hat{s}}} = -\omega$($\Lambda =\Lambda_{l^{*}}$). 
\par\noindent 
Thus, the self energy for states near the $FS$ has the universal $\mathbf{k}$-independent form
\begin{eqnarray}
\Sigma^{(l^{*})}(\omega) = N(0)\omega\ln\bigg\vert\frac{\omega}{\omega_{c}}\bigg\vert~.\label{frequency_dependent_self_energy}
\end{eqnarray}
\begin{figure}
\includegraphics[width=\textwidth]{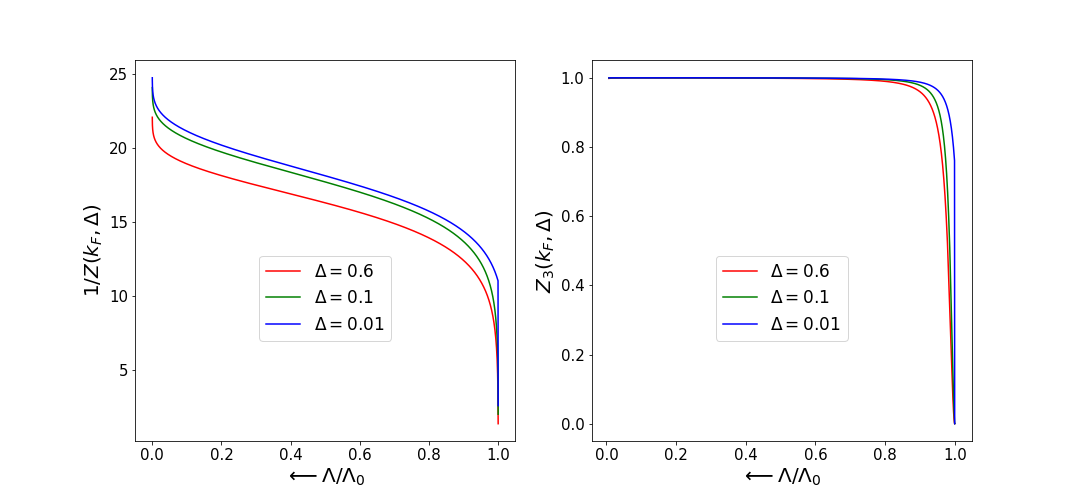}
\caption{Left panel: RG flow of the quasiparticle residue $Z_{1}(k_{F},\Delta)$. Right panel: RG flow of the 2-electron 1-hole residue $Z_{3}(k_{F},\Delta)$. $\Delta$ represents the probe energyscale about $E_{F}$.}\label{MFL-III}
\end{figure}
The real part of the self-energy in eq.\eqref{frequency_dependent_self_energy} has the same structure as the well known form of the self-energy for the \emph{marginal Fermi liquid metal}~\cite{varma2002singular}. Further, this result is a generic outcome for any single band dispersion, and emergent purely from repulsive 4-point and 6-point vertex RG flows. Using eq.\eqref{frequency_dependent_self_energy}, we obtain the imaginary part of the complex self energy, i.e., the scattering rate ($\tau$) as a function of the fluctuation scale $\omega$; using Kramers-Kronig relations eq.\eqref{Thermal scale}, we can connect $\tau$ to the temperature $T$
\begin{eqnarray}
k_{B}T = \frac{1}{N(0)}\Sigma^{im,(l^{*})}(\omega) = \hbar |\omega |=\hbar 2\pi\tau^{-1}.\label{Temparature_fluctuations_scale}
\end{eqnarray}
The finite $T$ resistivity per unit length $\rho(T)/L$ for layered 2d systems can be obtained from eq.\eqref{Temparature_fluctuations_scale} by replacing $N(0)=(2mE_{F})^{-1}\hbar^{2}(\Delta k)^{2}N_{e}(E_{F})$, particle density $n=N_{e}(E_{F})/L^{3}$ ($L^{3}$ is the volume in 3D) and the Fermi energy $E_{F}$ in terms of the Fermi Temperature ($T_{F}$) $E_{F}=k_{B}T_{F}$ 
\begin{eqnarray}
\frac{\rho}{L} =\frac{m}{ne^{2}L\tau} = \frac{h}{2e^{2}}\frac{T}{T_{F}}~.\label{T-linear}
\end{eqnarray}
Here, $\Delta k=2\pi L^{-1}$ is the momentum space lattice spacing, $L$ is the system length, $N_{e}(E_{F})$ number of electrons around FS that comprise the transport. This obtains a universal Planckian $T$-linear resistivity form starting from a very general microscopic single band model $H_{SFIM}$, and supports various experimental observations and theoretical proposals~\cite{homes2004universal,zaanen2004superconductivity,
legros2018universal}.
\par
\begin{figure}
\centering
\includegraphics[scale=4]{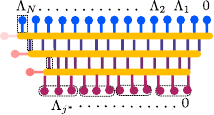}
\caption{EHM tensor network representation of the Marginal Fermi liquid. Each dotted circle (comprising three electronic legs) represents a long-lived composite 2 electron-1 hole excitation of the MFL proximate to the Fermi surface. }\label{TN-MFL}
\end{figure}

\noindent
Following eq.\eqref{frequency_dependent_self_energy} and eq.\eqref{Temparature_fluctuations_scale}, the quasiparticle residue has the following form at finite temperatures 
\begin{eqnarray}
Z_{1}(T)&=& \frac{1}{1-N(0)(1+ln|\frac{k_{B}T}{\hbar\omega_{c}}|)}~.\label{entanglement_spectral_weight_content}
\end{eqnarray}
\begin{figure}
\centering
\includegraphics[width=0.6\textwidth]{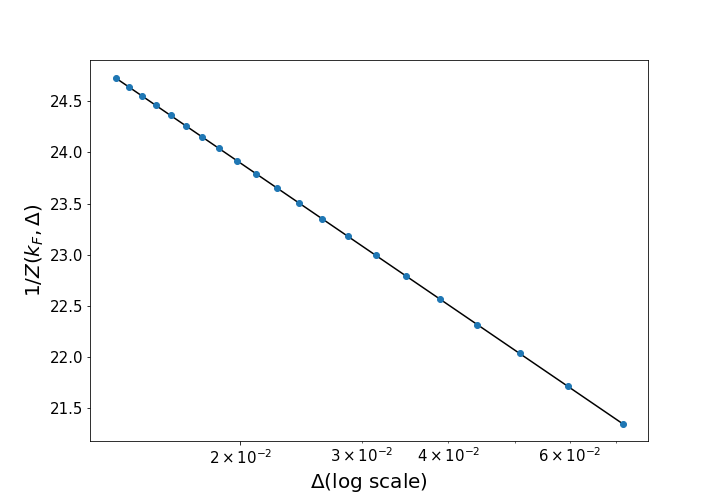}
\caption{Variation of the quasiparticle residue $Z_{1}(k_{F},\Delta)$ with energy scale $\Delta$ about the Fermi energy ($E_{F}$).}\label{MFL-IV}
\end{figure}
The left and right panels of Fig.\ref{MFL-III}  represent the renormalization of the quasi-particle residue $Z_{1}(k_{F},\Delta)$ and 2-particle 1-hole residue $Z_{3}(k_{F},\Delta)$. We find that, for $\Delta=0.6,0.1,0.001$, $Z_{1}(k_{F},\Delta)$ reduces under RG (left panel of Fig.\ref{MFL-III}), indicating the breakdown of the Landau quasiparticle picture. On the other hand, $Z_{3}(k_{F},\Delta)$ (right panel of Fig.\ref{MFL-III}) is seen to increase towards 1, indicating well-formed 2-electron 1-hole composites in the neighbourhood of the Fermi surface. Finally, Fig.\ref{MFL-IV} is a numerical verification of the logarithmic dependence of $Z_{1}(k_{F},\Delta)$ on the energy scale $\Delta(=k_{B}T)$ given in eq.\eqref{entanglement_spectral_weight_content}.
\pin
As the quasi-particle residue $0<Z<1$, the relation eq.\eqref{entanglement_spectral_weight_content} holds for $\omega<\omega_{c}/e$, corresponding to a temperature $T<\hbar\omega_{c}/(ek_{B})$. The vanishing of the quasiparticle residue, $Z\to 0$, leads to a integer Friedel's phase shift $\Delta N\in\mathbb{Z}$: a test electron binds together with a electron-hole pair, forming a three-particle composite. As the Hamiltonian eq.\eqref{3-numberdiag} is  diagonal, the residue of this 2-electron 1-hole composite approaches $1$ at the $FS$. We note that this was also shown for the parent MFL of the Mott insulating state in the 2D Hubbard model on the square lattice at $1/2$-filling in Ref.\cite{anirbanmotti}. We also present the tensor network representation of the RG flow towards the marginal Fermi liquid fixed point in Fig.\ref{TN-MFL}. 
\par\noindent
In this subsection, we found the parameter and fluctuation regime where three-particle off-diagonal vertices are RG irrelevant, while the 2-electron 1-hole dispersion achieves a finite value at the fixed point. This observation provides a perfect setting for the question: what are the primary instabilities of the marginal Fermi liquid metal? We present the answer to this question next.  
\subsection{RG flow into the spin/charge hybridized pseudospin-pairing force models}\label{MottLiquids}
\pin
We have seen earlier that the off-diagonal three particle vertices causes dynamical mixing between electron-electron and electron-hole pairs (eq.\eqref{p})~\cite{anirbanurg1}. The spin/charge backscattering processes in the mixed configuration leads to a two-particle self-energy containing log-divergences as leading corrections\cite{anirbanurg1}. In order to observe the condensation of various spin-charge mixed configurations, we now apply the Hamiltonian RG formalism on the $H_{SFIM}$. 
\par\noindent
If the Fermi surface is nested, there are Umklapp scattering vertices generically present in the Hamiltonian $H_{SFIM}$. They are denoted by $\Gamma^{4,(N)}_{c,\alpha\beta}$ where $\alpha$ and $\beta$ represent electronic pairs with pair momentum differing by $\mathbf{Q}$. Further, the index $\alpha = \lbrace\mathbf{k}_{\Lambda_{j}\hat{s}}\sigma,\mathbf{p}-\mathbf{k}_{\Lambda_{j}\hat{s}}-\sigma\rbrace$ and $\beta = \lbrace\mathbf{k}_{\Lambda_{j}\hat{s}}\sigma,\mathbf{Q}+\mathbf{p}-\mathbf{k}_{\Lambda_{j}\hat{s}}-\sigma\rbrace$. The spin backscattering vertices in $H_{SFIM}$ involve pairs of electronic states that exchange their spin orientations while scattering. These vertices are denoted by $\Gamma^{4,(N)}_{s,\alpha\beta}$ where $\alpha = \lbrace\mathbf{k}_{\Lambda_{j}\hat{s}}\sigma,\mathbf{p}-\mathbf{k}_{\Lambda_{j}\hat{s}}-\sigma\rbrace$, $\beta = \lbrace\mathbf{k}_{\Lambda_{j}\hat{s}} -\sigma,\mathbf{p}-\mathbf{k}_{\Lambda_{j}\hat{s}} \sigma\rbrace$. 
For the Umklapp and spin backscattering processes, the RG flow hierarchy yields
\begin{eqnarray}
\Delta H^{(j)}_{c} &=& \sum_{\mathbf{k},\mathbf{k'},\mathbf{p}}\frac{\Gamma^{4,(j)}_{c,\alpha\gamma}\Gamma^{4,(j)}_{c,\gamma\beta}\hat{n}_{\mathbf{k}_{\Lambda_{j}\hat{s}} \sigma}\hat{n}_{\mathbf{p}-\mathbf{k}_{\Lambda_{j}\hat{s}} -\sigma}}{\omega -\epsilon^{(j)}_{p,\mathbf{k}_{\Lambda_{j}\hat{s}'},\mathbf{p}}-\Gamma^{4,(j)}_{\gamma\gamma}\tau_{\mathbf{k}_{\Lambda_{j}\hat{s}}\sigma}\tau_{\mathbf{p}-\mathbf{k}_{\Lambda_{j}\hat{s}}\sigma'}}\nonumber\\
&\times & c^{\dagger}_{\mathbf{k}\sigma}c^{\dagger}_{\mathbf{p}-\mathbf{k} -\sigma}c_{\mathbf{Q}+\mathbf{p}-\mathbf{k}'-\sigma}c_{\mathbf{k}' \sigma},\nonumber\\
\Delta H^{(j)}_{s} &=&-\sum_{\mathbf{k},\mathbf{k'},\mathbf{p}}\frac{\Gamma^{4,(j)}_{s,\alpha\gamma}\Gamma^{4,(j)}_{s,\gamma\beta}\hat{n}_{\mathbf{k}_{\Lambda_{j}\hat{s}} \sigma}(1-\hat{n}_{\mathbf{p}-\mathbf{k}_{\Lambda_{j}\hat{s}} -\sigma})}{\omega -\epsilon^{(j)}_{p,\mathbf{k}_{\Lambda_{j}\hat{s}'},\mathbf{p}}-\Gamma^{4,(j)}_{\gamma\gamma}\tau_{\mathbf{k}_{\Lambda_{j}\hat{s}}\sigma}\tau_{\mathbf{p}-\mathbf{k}_{\Lambda_{j}\hat{s}}\sigma'}}\nonumber\\
&\times & c^{\dagger}_{\mathbf{k}_{1}\sigma}c_{\mathbf{k}_{1}-\mathbf{p}+2\mathbf{k}_{\Lambda_{j}\hat{s}}-\sigma}c^{\dagger}_{\mathbf{k}_{2}-\sigma}c_{\mathbf{k}_{2}-\mathbf{p}+2\mathbf{k}_{\Lambda_{j}\hat{s}}\sigma}~.
\end{eqnarray} 
The spin-type configuration 1e-1h/1h-1e  for the set $\alpha$ is constrained by the relation
\begin{eqnarray}
n_{\mathbf{k}_{\Lambda_{j}\hat{s}}\sigma}  + n_{\mathbf{p} - \mathbf{k}_{\Lambda_{j}\hat{s}} -\sigma} = 1~~\text{for}~~\epsilon_{\mathbf{k}_{\Lambda_{j}\hat{s}}}>E_{F}~, \label{intermediate_configuration_2}
\end{eqnarray}
and has an associated kinetic energy lying within the range $\epsilon_{\mathbf{k}_{\Lambda_{j}\hat{s}}}>\epsilon_{\mathbf{p}-\mathbf{k}_{\Lambda_{j}\hat{s}}}$,
$\omega >\frac{1}{2}(\epsilon_{\mathbf{k}_{\Lambda_{j}\hat{s}}}-\epsilon_{\mathbf{p}-\mathbf{k}_{\Lambda_{j}\hat{s}}})$.
The intermediate configuration of eq.\eqref{intermediate_configuration_2} leads to a spin-type pseudospin configuration $S^{-}_{\mathbf{k},\mathbf{p}} = c^{\dagger}_{\mathbf{k}\sigma}c_{\mathbf{p}-\mathbf{k}-\sigma},S^{+}_{\mathbf{k},\mathbf{p}} =(S^{-}_{\mathbf{k},\mathbf{p}})^{\dagger}$ and $S^{z}=\frac{1}{2}[S^{+},S^{-}]$, in which the 4-point vertexes are RG irrelevant. On the other hand, certain 6-point vertices are RG relevant in this regime, and can be represented in terms of charge- and spin-type pseudospins as follows
\begin{eqnarray}
 &&c^{\dagger}_{\mathbf{p}-\mathbf{p}''\sigma}c^{\dagger}_{\mathbf{p}''-\mathbf{k}_{\Lambda'\hat{s}'}-\sigma} c^{\dagger}_{\mathbf{k}_{\Lambda'\hat{s}'}\sigma}c_{\mathbf{k}_{\Lambda\hat{s}}\sigma}c_{\mathbf{p}'-\mathbf{k}_{\Lambda\hat{s}} -\sigma}c_{\mathbf{p}-\mathbf{p}'-\sigma} \nonumber\\
 &=&A^{+}_{\mathbf{k}_{\Lambda'\hat{s}'},\mathbf{p}''}A^{-}_{\mathbf{k}_{\Lambda\hat{s}},\mathbf{p}'}S^{-}_{\mathbf{p}-\mathbf{p}'',2\mathbf{p}-\mathbf{p}'-\mathbf{p}''}~.\label{3-particle pairing insabilties}
 \end{eqnarray}
The operators $A^{+/-}_{\mathbf{k},\mathbf{p}}$ corresponds to finite momentum pseudospin raising and lowering operators given in eq.\eqref{finite_momentum_pseudo_spin_pairs}. This spin-charge mixed representation of 6-point vertices is manifested in the dynamical mixing between the pseudospin state configuration (see discussion below eq.\eqref{4-particle vertex}, and leads to a hybridized pair-kinetic energy eq. \eqref{hybridizedKE} entering the 4-point Green's function. In turn, this 4-point Green's function enters the 4-point vertex flow equations given in eq.\eqref{4-particle vertex}.
\begin{figure}
\includegraphics[width=\textwidth]{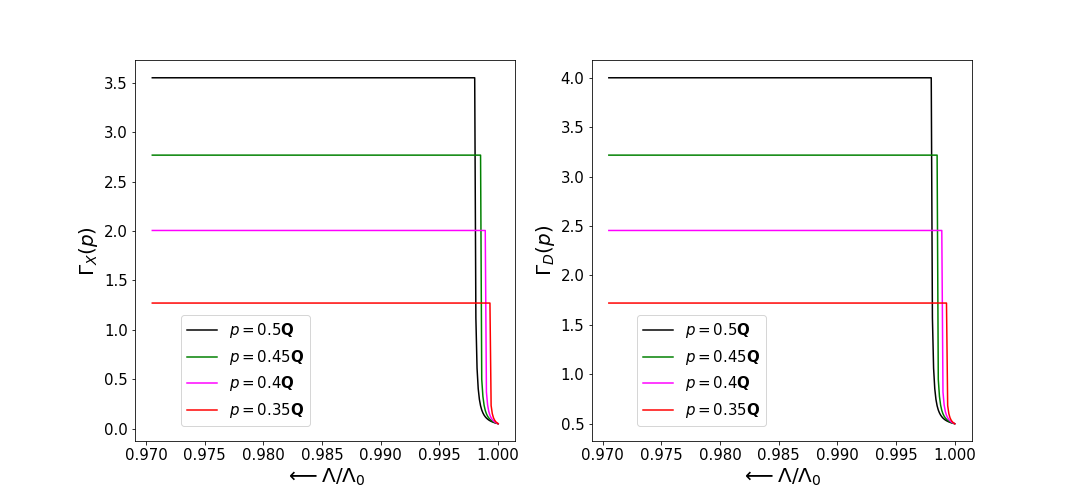}
\caption{Left/right panels show the RG flow for the off-diagonal/diagonal $\Gamma_{X}/\Gamma_{D}$ scattering vertices respectively for various finite momenta ($\mathbf{p}$) pairs. $\Gamma^{0}_{X}=0.1$ is the bare magnitude of the umklapp scattering vertex, $\Gamma^{0}_{D}=0.5$. The $k$-space grid (system volume) used is $1024\times 1024$. }\label{mottvertx}
\end{figure}
\par\noindent
For the $\mathbf{p}$-momentum opposite-spin pairs, the 2-particle backscattering processes (with $\Delta \mathbf{p} = 0$ (spin exchange) and $\Delta\mathbf{p}=2\pi$ (Umklapp)) given by
\begin{eqnarray}
\mathbf{k}_{\Lambda\hat{s}},\mathbf{p}-\mathbf{k}_{\Lambda\hat{s}}\leftrightarrow  -\mathbf{k}_{\Lambda_{j}\hat{s}},\mathbf{Q}_{1}+\mathbf{p}+\mathbf{k}_{\Lambda_{j}\hat{s}},~\mathbf{Q}_{1}=0,\mathbf{Q}~~~~~~
\end{eqnarray} 
produce a log-divergence in the composite self-energy (shown in a companion manuscript~\cite{anirbanurg1}), requiring a RG treatment once more. Using eq.\eqref{4-particle vertex}, the charge backscattering vertex flow equation is given by 
\begin{eqnarray}
&&\Delta\Gamma^{4,(j)}_{c,\alpha\beta}(\omega) = \frac{p\Gamma^{2,(j)}_{c,\alpha\gamma} \Gamma^{2,(j)}_{c,\gamma\beta}}{\omega -p\epsilon^{ee,(j)}_{\mathbf{k}_{\Lambda_{j}\hat{s}},\mathbf{p}}-p'\epsilon^{eh,(j)}_{\mathbf{k}_{\Lambda_{j}\hat{s}},\mathbf{p}}-\frac{p}{4}\Gamma^{2,(j)}_{c,\gamma\gamma}+\frac{p'}{4}\Gamma^{2,(j)}_{s,\gamma\gamma}},~~~~~~~~
\end{eqnarray}
where $p'=1-p$, $\epsilon^{ee,(j)}_{\mathbf{k}_{\Lambda_{j}\hat{s}},\mathbf{p}}=\epsilon^{ee,(j)}_{\mathbf{k}_{\Lambda_{j}\hat{s}},\mathbf{p}-\mathbf{k}_{\Lambda_{j}\hat{s}}}$ and 
$\epsilon^{eh,(j)}_{\mathbf{k}_{\Lambda_{j}\hat{s}},\mathbf{p}}=\epsilon^{eh,(j)}_{\mathbf{k}_{\Lambda_{j}\hat{s}},\mathbf{p}-\mathbf{k}_{\Lambda_{j}\hat{s}}}$. Similarly, the spin backscattering RG flow equation is given by $\Delta \Gamma^{4,(j)}_{s,\alpha\beta} = (\frac{p-1}{p})\Gamma^{4,(j)}_{c,\alpha\beta}$~.
\par\noindent
The stable fixed point is obtained from the condition
\begin{eqnarray}
\omega -p\epsilon^{ee,(j^{*})}_{\mathbf{k}_{\Lambda_{j}\hat{s}},\mathbf{p}}-p'\epsilon^{eh,(j^{*})}_{\mathbf{k}_{\Lambda_{j^{*}}\hat{s}},\mathbf{p}}=\frac{p}{4}\Gamma^{2,(j)}_{c,\gamma\gamma'}+\frac{p'}{4}\Gamma^{2,(j)}_{s,\gamma\gamma'}~,~~~~~~~
\end{eqnarray}
leading to the effective pseudospin XXZ Hamiltonian given by
\begin{eqnarray}
H^{*,XXZ}_{ML}(\omega) &=&\sum_{\mathbf{k}}\epsilon^{ee,(j^{*})}_{\mathbf{k}_{\Lambda\hat{s}},\mathbf{p}}A^{z}_{\mathbf{k}_{\Lambda\hat{s}},\mathbf{p}}+\epsilon^{eh,(j^{*})}_{\mathbf{k}_{\Lambda\hat{s}},\mathbf{p}}S^{z}_{\mathbf{k}_{\Lambda\hat{s}},\mathbf{p}}+\sum_{\mathbf{k}_{\Lambda\hat{s}},\mathbf{k}}\Gamma^{4,(j^{*})}_{s,\alpha\beta} [S^{+}_{\alpha}S^{-}_{\beta}+h.c.]+\Gamma^{4,(j^{*}),||}_{s,\alpha\beta}S^{z}_{\alpha}S^{z}_{\beta}\nonumber\\
&+& \sum_{\mathbf{k}_{\Lambda\hat{s}},\mathbf{k}} \Gamma^{4,(j^{*})}_{c,\alpha\beta} [A^{+}_{\alpha}A^{-}_{\beta}+h.c.]+\Gamma^{4,(j^{*}),||}_{\alpha\beta} A^{z}_{\alpha}A^{z}_{\beta}~,\label{pseudospin_ham_repulsive}
\end{eqnarray}
where $\Gamma^{4,(j^{*})}_{\mathbf{k}_{\Lambda\hat{s}},\mathbf{p},\mathbf{k},\mathbf{Q}}$ and $\Gamma^{4,(j^{*}),||}_{\mathbf{k}_{\Lambda\hat{s}},\mathbf{p},\mathbf{k},\mathbf{Q}}$ are the XY and Ising pseudospin couplings respectively. 
\begin{figure}[!ht]
\includegraphics[width=\textwidth]{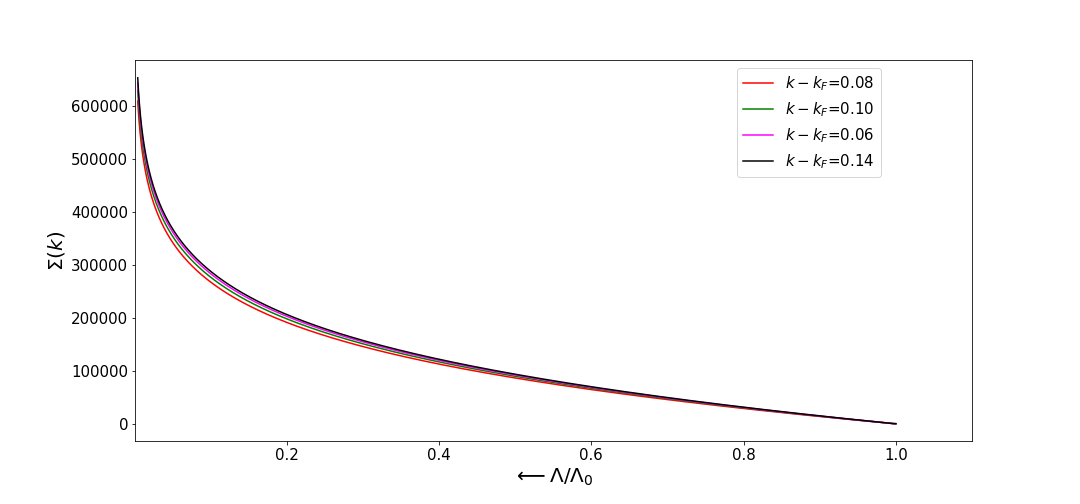}
\caption{RG flow of the self-energy $\Sigma(k)$ for $k$ wave-vectors close to Fermi surface.}\label{selfEnergy}
\end{figure}
\pin
Fig.\ref{mottvertx} shows the RG flow of the Umklapp scattering vertices connecting opposite sides of a nested Fermi surface for a tight-binding square lattice model at half-filling in two spatial dimensions for electronic pairs with net momentum $\mathbf{p}=\mathbf{Q}/2,0.45\mathbf{Q},0.4\mathbf{Q}$ and $0.35\mathbf{Q}$ (where $(\mathbf{Q}=\pi,\pi)$). The left and right panels show the renormalization of  the off-diagonal ($\Gamma_{c,X}=\Gamma_{c,\alpha\gamma}$) and diagonal ($\Gamma_{c,D}=\Gamma_{c,\gamma\gamma}$) vertices respectively. The bare couplings are taken to be $\Gamma_{c,X}^{0}=0.1$, $\Gamma_{c,D}^{0}=0.5$, and we consider a momentum-space grid of size $1024\times 1024$. As observed in Fig.\ref{mottvertx}, both off-diagonal and diagonal couplings renormalize to higher magnitudes at the low-energy RG fixed point, such that the resulting theory is described by eq.\eqref{pseudospin_ham_repulsive}. The associated hybridized pseudospin state space also involves fermion states pairing up in the mixed configuration regime (discussion above in eq.\eqref{hybridizedKE}), such that the fermionic states transmute into the mixed pseudospin states. The condensation of pseudopsins due to the nesting instability leads to a zero of the Greens function, $G(k,\Delta)=(\Delta-\epsilon_{k}-\Sigma_{k})^{-1}$ (note the similarity with the zero of $G(k,\Delta)$ for the BCS instability in eq.\eqref{breakdown1pGreen}), as can be seen from the divergent self-energy in Fig.\ref{selfEnergy} as the Fermi surface is approached. The net Friedel phase-shift accounts for the number of mixed pseudospins, as well as the number of bound states formed via the RG. This can be computed through the Luttinger surface of zeros (see discussion below eq.\eqref{breakdown1pGreen}). This pseudospin Hilbert space will persist upto a temparature scale $T^{*}$ (eq.\eqref{Thermal scale}) computed using the renormalized 1-particle self-energy at the fixed point. Finally, we note that the tensor network representation of the RG flows towards the Mott liquid fixed point is similar to that presented in Fig.(\ref{TN-BCS}) for the BCS reduced Hamiltonian displaying pair formation.
\subsection{RG Phase diagram for $H_{SFIM}$}
\pin
Having numerically verified the RG flows to various IR fixed point theories, we can now gather all our results into the form of a RG phase diagram. In order to characterize efficiently various phases obtained from the RG flows for $H_{SFIM}$, we define the following two quantities $r_{2}$ and $r_{3}$
\begin{eqnarray}
r_{2} &=& sgn(V^{\sigma\sigma'}_{\mathbf{k},\mathbf{k}',\mathbf{p}})\sqrt{\frac{\sum_{\alpha\neq\beta}\Gamma^{4,(N)}_{\alpha\beta}\Gamma^{4,(N)}_{\alpha\beta}}{\sum_{\alpha\neq\beta}\Gamma^{4,(N)}_{\alpha\beta}\Gamma^{4,(N)}_{\alpha\beta}+\sum_{\alpha}\Gamma^{4,(N)}_{\alpha\alpha}\Gamma^{4,(N)}_{\alpha\alpha}}}~,\nonumber\\
r_{3} &=& \sqrt{\frac{\sum_{\gamma\gamma'}\Gamma^{6,(N-1)}_{\gamma\gamma'}\Gamma^{6,(N-1)}_{\gamma\gamma'}}{\sum_{\gamma}\Gamma^{6,(N-1)}_{\gamma\gamma'}\Gamma^{6,(N-1)}_{\gamma\gamma'}+\sum_{\alpha\beta}\Gamma^{6,(N-1)}_{\alpha\beta}\Gamma^{6,(N-1)}_{\alpha\beta}}}~.\label{r2r3def}
\end{eqnarray}
The quantity $-1<r_{2}<1$ represent the ratio of (i) the root mean square magnitude (RMS) for bare 4-point off-diagonal (OD) vertices, and (ii) the sum of the mean squares of 4-point diagonal (D) and OD vertices. Thus, $r_{2}$ carries the $(+/-)$ sign for OD terms representing attractive/repulsive interactions respectively. Similarly, the quantity $0\leq r_{3}\leq 1$ is the ratio of (i) the RMS for 6-point D vertices, and (ii) the square root sum of the mean squares of 6-point D and OD vertices. Recall that the index $\alpha$ represents a set of two (momentum, spin) indices for 2-particle vertices ($\Gamma^{2}$) and a set of three (momentum, spin) indices for 3-particle vertices ($\Gamma^{3}$). We show below that the parameters $r_{2}$ and $r_{3}$ allow for an efficient encoding of the numerically evaluated RG flows shown earlier at various points.
\par
\begin{figure}[h!]
\centering
\includegraphics[scale=1.3]{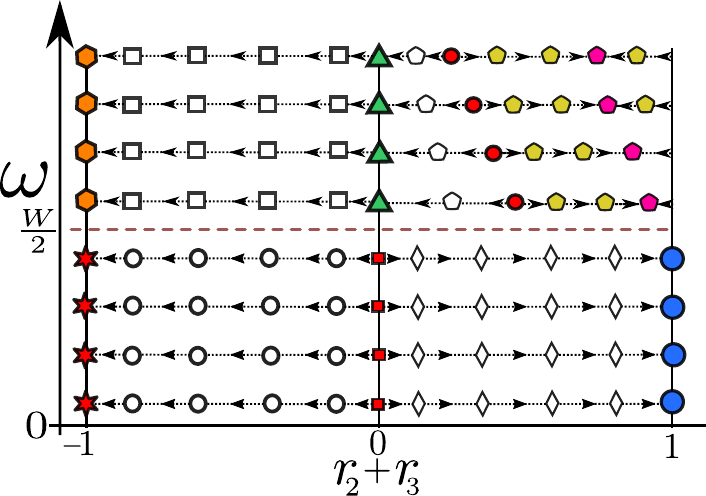} 
\caption{RG phase diagram for the single-band model of interacting electrons with translational invariance, $H_{SFIM}$. The y-axis represents the magnitude for quantum fluctuations ($\omega$), $W$ represents the bandwidth. The x-axis represents interaction due to two-particle ($r_{2}$) and three-particle ($r_{3}$) number diagonal/off-diagonal scattering processes. $-1<r_{2}<1$ with (+/-)sign representing nature of interaction, while $0<r_{3}<1$ (see text for detailed definitions). The magnitudes of $r_{2}$ and $r_{3}$ represent the relative magnitude of off-diagonal scattering in $2$-particle and $3$-particle scattering processes respectively. The red squares on line $r_{2}=r_{3}=0$ correspond to tight-binding metals. The unfilled squares and circles represent crossover RG flows involving symmetry unbroken $\mathbf{p}\neq 0$ (SPDW)/$\mathbf{p}=0$ (RBCS) gapped XXZ pseudospin theories respectively. These flows end at orange hexagon/red stars labelling corresponding stable fixed points with $XY$ pseudospin interaction. Unstable fixed points (red circles) lie between Ising- (unfilled pentagons) and XY- (yellow pentagons) dominated XXZ pseudospin Mott liquid theories arising out of charge and spin backscattering. The unfilled pentagons are crossover RG flows to the stable fixed point Fermi liquid (green triangles), while the yellow pentagons are crossovers to XXZ pseudospin gapped stable theories (magenta pentagons). Blue circles represent the marginal Fermi liquid theories residing at the end point of crossover RG flows involving dominant three-particle scattering (unfilled diamonds).}\label{Phase_Diagram_SFIM}
\end{figure}

\noindent
Fig.(\ref{Phase_Diagram_SFIM}) presents the RG phase diagram for $H_{SFIM}$ using $\omega$ (quantum fluctuations, y-axis) and $(r_{2},r_{3})$ (x-axis). For attractive couplings ($r_{2}<0$) and for $(\omega<\frac{W}{2})/(\omega>\frac{W}{2})$, the unfilled circles and squares represent crossover RG flows involving a XXZ symmetry-unbroken 
reduced BCS (eq.\eqref{XXZ_BCS_reduced_theory}) theory for $\mathbf{p}=0$ (Fig.\ref{XXZ_BCS}) and $\mathbf{p}\neq 0$ pairs (Fig.\ref{XXZ-SPDW}) respectively. The RG flows stop at stable fixed points (red stars/orange hexagons) given by symmetry unbroken XY BCS (eq.\eqref{HBCS}, Figs.\ref{XY_BCS_1} and \ref{XY_BCS_2}) and PDW (eq.\eqref{HPDW}, Fig.\ref{XY-SPDW}) theories respectively. The red squares ($r_{2}=r_{3}=0$) are unstable fixed points representing a tight-binding metal. In the repulsive regime $r_{2}\to 0, r_{3}>0$ and $\omega<\frac{W}{2}$, the unfilled diamond are crossover RG flows to three-particle theories with diagonal and off-diagonal terms. The blue circles are stable points representing the marginal Fermi liquid metal (eq.\eqref{3-numberdiag}, Figs.\ref{MFL-I} - \ref{MFL-IV}).
\pin
On the other hand, the Mott metal-insulator transition shown in the repulsive regime $r_{2}+r_{3}>0$ and $\omega>\frac{W}{2}$ is more complicated, due to the appearance of unstable fixed points (red circles) lying at intermediate coupling. These unstable fixed points separate RG flows (unfilled pentagons) to Fermi liquid theories (green triangles, eq.\eqref{HFL}, Figs.\ref{2particleVertices} - \ref{3pEnergyAndResidue}) with $r_{2}\to 0, r_{3}\to 0$ from those (yellow pentagons) towards XXZ pseudospin Mott liquids theories with finite values of $r_{2}$ and $r_{3}$ (magenta pentagons, eq.\eqref{pseudospin_ham_repulsive}, Figs.\ref{mottvertx} and \ref{selfEnergy}). As described in the previous subsection, the physics of nesting is responsible for the stabilisation of such Mott liquids. We recall that a recent RG analysis the half-filled 2D Hubbard model on the square lattice (whose underlying tight-binding Fermi surface is strongly nested) in Ref.\cite{anirbanmotti} identified the marginal Fermi liquid as being the parent metallic phase of the Mott liquid found therein. This leads us to conjecture that the unstable fixed point (red circles) gapless quantum critical theories lying at intermediate coupling correspond to a marginal Fermi liquid theory described by eq.\eqref{3-numberdiag}.
\subsection{Scaling features of holographic entanglement entropy bound for gapless and gapped quantum liquids}
\pin
In this section, we analyze the RG scaling relations for the holographic entanglement entropy bound for the various phases of the $H_{SFIM}$ model obtained by isolating a given region $R$. Among the IR fixed points, the gapless theories reached via URG comprise the number-diagonal Hamiltonians for the Fermi liquid (eq.\eqref{HFL}) and Marginal Fermi liquid phases (eq.\eqref{3-numberdiag}). For these cases, the ground state wavefunction obtained from the above low-energy Hamiltonians are separable in momentum-space. On the other hand, the gapped IR fixed point theories involve effective Hamiltonians that are number off-diagonal in momentum-space, e.g., the reduced BCS theory (eq.\eqref{HBCS}) and the Mott liquid (eq.\eqref{pseudospin_ham_repulsive}). Clearly, the ground states obtained from the latter low-energy effective Hamiltonians are highly entangled. As the scattering vertex and wavefunction RG flows are distinct for various phases, we expect that the same will hold true for their holographic entanglement entropy scaling relations (eq.\eqref{entropy_bound}). Investigating this point is the content of this subsection.
\pin
The entanglement entropy scaling in the regime (eq.\eqref{condFL}) leading to the Fermi liquid (eq.\eqref{HFL}) has the form (using eq.\eqref{entropyScaling})
\begin{eqnarray}
\Delta S^{1}_{(j)} &=&2\sum_{\alpha_{1},\mu,\nu,\nu'}sgn(\mu,\nu,\nu')\Delta\Gamma^{4,(j)}_{\mu\nu}G^{4,(j)}_{\nu\nu'}C^{(j)}_{\nu'}C^{0,(j)}_{\alpha_{1}}~.\label{EntropyFL}
\end{eqnarray}
Note that in arriving at this RG equation, we have restricted ourselves to the contributions from only 4-point ($\Delta\Gamma^{4,(j)}_{\mu\nu}$) vertex RG flows (eq.\eqref{4and6 particle vertex}). Further, we assume a uniform magntiude for the diagonal and off-diagonal couplings: $\Gamma^{4,(j)}_{\mu\nu}G_{0,(j)}= V$, $\Gamma^{4,(j)}_{\mu\nu}+c=\Gamma^{4,(j)}_{\mu\mu}$~($c$ is a constant) and $G^{-1}_{0,(j)}=\omega-\epsilon_{(j)}$. This leads to the continuum RG equation for the scattering vertex
\begin{eqnarray}
\frac{dV}{d\log\frac{\Lambda}{\Lambda_{0}}}&=&V(\omega G_{0,\Lambda}-1) +\frac{V^{2}}{1-cG_{0,\Lambda}-V}~,~ 
\label{contRG}
\end{eqnarray}
where $G^{-1}_{0,\Lambda}=\omega-\hbar v_{F}\Lambda$, we have assumed a spherical Fermi surface $\epsilon_{\Lambda_{j}\hat{s}}=\hbar v_{F}\Lambda$, and we have replaced the finite difference $\Delta(\log\frac{\Lambda}{\Lambda_{0}})$ by the differential quantity $d\log\frac{\Lambda}{\Lambda_{0}}$. For $cG_{0,\Lambda}>1$, the off-diagonal vertices are RG irrelevant ($dV<0$) as $\Lambda\to 0$
\begin{eqnarray}
\frac{dV}{d\log\Lambda/\Lambda_{0}}=-\frac{\omega V^{2}}{c-\omega}~\Rightarrow ~V(\Lambda)=\frac{V_{0}}{1+\frac{\omega}{c-\omega}V_{0}\log\frac{\Lambda_{0}}{\Lambda}}\approx \frac{c-\omega}{\omega}\frac{1}{\log\frac{\Lambda_{0}}{\Lambda}}~.\label{couplingRGFermiSurface}
\end{eqnarray}
In reaching eq.\eqref{couplingRGFermiSurface} from eq.\eqref{contRG}, we have dropped the first term of eq.\eqref{contRG}, as $(\omega G_{0,\Lambda}-1)\to 0$ as $\Lambda\to 0$. We have also dropped $V$ in the denominator of the second term of eq.\eqref{contRG}, as $V\to 0$ and $G_{0,\Lambda}\to \frac{1}{\omega}$. The RG relation for the entanglement entropy bound of the FL then has the form
\begin{eqnarray}
\frac{dS^{1}(\Lambda)}{d\log\frac{\Lambda}{\Lambda_{0}}}=2\left(\frac{c-\omega}{\omega}\right)^{2}\sum_{\alpha_{1},\mu,\nu,\nu'}\frac{sgn(\mu,\nu,\nu')}{\left(\log\frac{\Lambda_{0}}{\Lambda}\right)^{2}}C^{(j)}_{\nu'}C^{0,(j)}_{\alpha_{1}}~.
\end{eqnarray}
As we approach the Fermi surface, the coefficient tensor $C^{0,(j)}_{\alpha_{1}}$ (corresponding to the ground state configuration $|\alpha\rangle$) scale towards $1$, while all other coefficients scale towards $0$. As a result, the fermion signatures for the RG scaling towards the ground state vanish. Taking these points into account, we find that the entropy scaling relation is given by
\begin{eqnarray}
S^{1}(\Lambda)=\left(\frac{c-\omega}{\omega}\right)^{2}\frac{1}{\log\frac{\Lambda_{0}}{\Lambda}}~.
\end{eqnarray}
Finally, we obtain the holographic entanglement entropy bound obtained by \emph{isolating the Fermi surface (in two dimensions)} from the rest of the system is given by
\begin{eqnarray}
S_{H}(\Lambda)=2\pi k_{F}\left(\frac{c-\omega}{\omega}\right)^{2}\frac{1}{\log\frac{\Lambda_{0}}{\Lambda}}~.\label{EntropyFL1}
\end{eqnarray}
For the MFL and reduced BCS theories, the entanglement entropy RG equations are given respectively by
\begin{eqnarray}
\Delta S^{MFL}&=&2\sum_{\alpha_{1},\mu,\nu,\nu'}sgn(\mu,\nu,\nu')\Delta\Gamma^{6,(j)}_{\mu\nu}G^{6,(j)}_{\nu\nu'}C^{(j)}_{\nu'}C^{0,(j)}_{\alpha_{1}},\label{MFLEntropy}\\
\Delta S^{RBCS}&=&2\sum_{\alpha_{1},\mu,\nu,\nu'}\Delta\Gamma^{4,(j)}_{\mu\nu}G^{4,(j)}_{\nu\nu'}C^{(j)}_{\nu'}C^{0,(j)}_{\alpha_{1}}~.\label{RBCSEntropy}
\end{eqnarray}
In the MFL, the renormalization is carried out primarily by six-point vertices as the quasiparticle degrees of freedom are ill-defined (eq. \eqref{frequency_dependent_self_energy}). On the other hand, for the reduced BCS theory, the dominant two particle vertex RG flow is present in the zero pair-momentum subspace (eq.\eqref{constraint_condense_1}). This leads to condensation of the pairs, and the fermion exchange phases are mitigated in the coefficient RG equations~\cite{anirbanurg1}. As a result, the fermion exchange phases are also absent in the entanglement scaling relation of the RBCS phase (eq.\eqref{RBCSEntropy}). These deviations in the entanglement RG equations for the MFL (eq.\eqref{MFLEntropy}) and the RBCS (eq.\eqref{RBCSEntropy}) phases from that obtained for the FL (eq.\eqref{EntropyFL1}) will likely to lead to a deviation of entropy bound scaling relations of these phases as well. Finally, the Mott liquid phase is described by pseudospins (eq.\eqref{pseudospin_ham_repulsive}) analogous to the RBCS phase, implying similar conclusions for the Mott liquid. We leave a detailed study of this aspect to a future work.
\section{Gauge theories and Topological order for emergent gapped quantum liquids} \label{emergentGaugeTheories}
\pin
In Ref.\cite{hansson2004}, Hansson et al. show that a $U(1)$ symmetry broken superconductor possesses signatures of topological order upon coupling to a dynamical electromagnetic field, i.e., it supports ground state degeneracy on the torus, edge states, charge fractionalization, together with a many-body gap that protects these properties. Importantly, the quantum fluctuations of the combined system restores the broken $U(1)$ phase rotation symmetry of the Cooper pair condensate. In this section, we adopt a different route in unveiling the universal features of topological order for a wide variety of gapped symmetry-preserved quantum liquid ground states arising from electronic correlations. For this, we start with the effective Hamiltonians obtained from RG fixed points which are written in terms of pseudospins. Indeed, we recall that such effective Hamiltonians describe pseudospin dynamics constitute a network of four point vertex tensors $\Gamma^{4,*}_{\alpha\beta}$, i.e., a Hamiltonian tensor network. The idea is to rewrite such a network of pseudospins
in terms of dual nonlocal objects, i.e. Wilson lines, leading to the formulation of a gauge theory. We will show that, for a simple case, such a gauge theory shows well-known signatures of topological order. 
\par\noindent
The condensates we are concerned with arise from parent metallic systems with a connected Fermi surface (FS). The destabilization of the  FS and its neighbourhood due to pseudospin-flip scattering processes (eq.\eqref{constraint_condense_1}) leading to a many-body gap in the 1-particle spectrum, signaling the condensation phenomenon within a momentum-space shell $\Lambda^{*}_{\hat{s}}$ around the erstwhile FS. The pseudospin condensate is now part of an emergent $SU(2)^{\otimes N}$ Hilbert space that originated from the electronic Fock space $\mathcal{F}^{N}$. Below, we consider the XXZ reduced BCS Hamiltonian $H^{*,XXZ}_{RBCS}$ (eq.\eqref{XXZ_BCS_reduced_theory}) as an example in order to demonstrate the origin and signatures of topological order. 
\par
\begin{figure}
\centering
\includegraphics[width=0.85\textwidth]{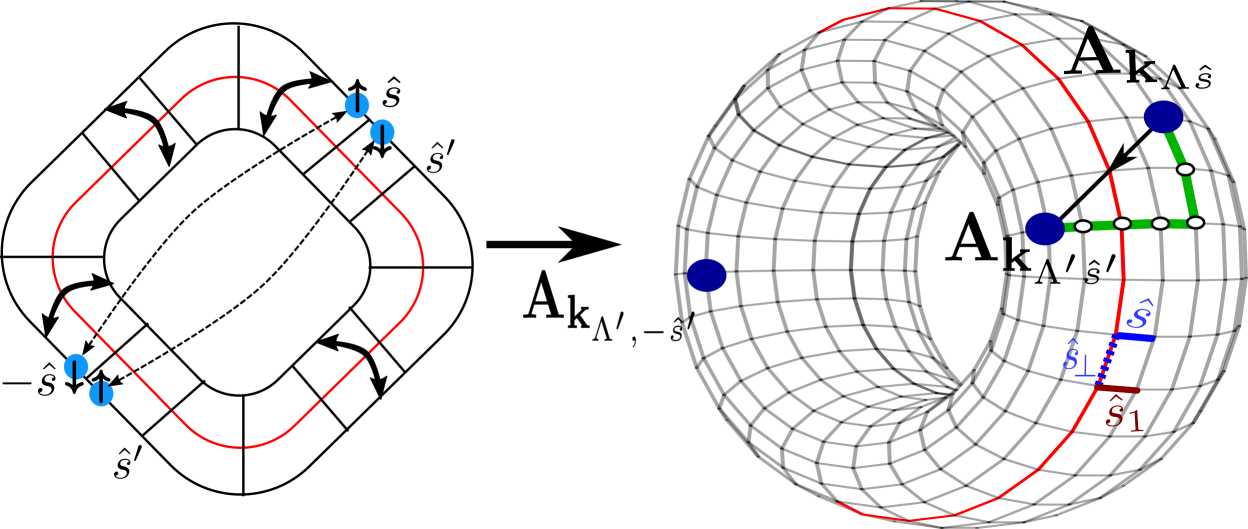}
\caption{(Left) Emergent momentum-space window where pairs of electronic states with opposite-spin and net-momentum $\mathbf{p}=0$ form bound pairs (dashed double-headed arrows). Thick double-headed arrows indicate the imposition of periodic boundary conditions, and the red curve represents the Fermi surface. (Right) Anderson pseudospins $\mathbf{A}_{\mathbf{k}_{\Lambda\hat{s}}}$ living on a torus. A composition of local green lines (a Wilson line) represents the inter- pseudospin interaction. Blue dashed/solid line represents the reference pair of orthonormal Wilson lines located on Fermi surface (and along $\hat{s}_{\perp}$)/normal to the Fermi surface (and along $\hat{s}=\hat{x}$). The brown line represents the rotated normal direction $\hat{s}_{1}$.}\label{pseudospin_emergent_topology}
\end{figure}

\noindent 
As shown in Fig.(\ref{pseudospin_emergent_topology}), we notie that the geometry of the 2D momentum-space shell in the pseudospin basis for the zero-pair momentum ($\mathbf{p}=0$) states with periodic boundary condition (PBC) is topologically equivalent to a torus.
Now, the inter-pseudospin interaction terms in the reduced BCS Hamiltonian $H^{*,XXZ}_{RBCS}$ (eq.\eqref{XXZ_BCS_reduced_theory}) 
can be represented as a Wilson line (dark black line in Fig. (\ref{pseudospin_emergent_topology})) as follows 
\begin{eqnarray}
4A^{o}_{\Lambda,\hat{s}}A^{o}_{\Lambda',\hat{s}'}=\exp(i\pi(A^{o}_{\Lambda,\hat{s}} - A^{o}_{\Lambda',\hat{s}'})),\label{wilson-line}
\end{eqnarray}
where we have rewritten the pseudospin operator $A^{o}_{\mathbf{k}_{\Lambda,\hat{s}}}$ as $A^{o}_{\Lambda,\hat{s}}$ for $o=(x,y,z)$, $\Lambda' =\Lambda +n'\delta\Lambda$, $\delta\Lambda=2\pi L^{-1}$ and $\hat{s}' =R^{m'}\hat{s}$. This Wilson line can in turn be represented as a composition of local Wilson lines (green lines with arrows in Fig.(\ref{pseudospin_emergent_topology})) along the $\hat{s}_{\perp}$ axes (i.e., $\hat{s}_{\perp}$ is perpendicular to $\hat{s}$, and along the direction of the hatched blue line in Fig.(\ref{pseudospin_emergent_topology})) and the $\hat{s}$ axes (along the direction of the solid blue line in Fig.(\ref{pseudospin_emergent_topology}))
\begin{eqnarray}
4A^{o}_{\Lambda,\hat{s}}A^{o}_{\Lambda',\hat{s}'} &=& W^{o}_{\Lambda ,\hat{s}\to R\hat{s}}W^{o}_{\Lambda, R\hat{s}\to R^{2}\hat{s}}\ldots W^{o}_{\Lambda, R^{m'-1}\hat{s}\to\hat{s}'}\nonumber\\
							&\times & W^{o}_{\Lambda\to \Lambda+\delta\Lambda ,\hat{s}'}\ldots W^{o}_{\Lambda +(n'-1)\delta\Lambda\to\Lambda'\hat{s}'}~,~~~~~\label{compositionOfWilsonLines}
\end{eqnarray} 
where $W^{o}_{\Lambda ,\hat{s}\to \hat{s}_{1}}=\exp(i\pi(A^{o}_{\Lambda,\hat{s}}-A^{o}_{\Lambda,\hat{s}_{1}}))$ ($\hat{s}_{1}$ is the rotated normal Fig.\ref{pseudospin_emergent_topology})is the local Wilson line along $\hat{s}_{\perp}$, and $W^{o}_{\Lambda\to \Lambda+\delta\Lambda ,\hat{s}}=\exp(i\pi(A^{o}_{\Lambda+\delta\Lambda,\hat{s}}-A^{o}_{\Lambda,\hat{s}}))$ represents the Wilson line for translation by $\delta\Lambda$ along $\hat{s}$. Here, $R\hat{x}=-\sin\theta\hat{x}+\cos\theta\hat{y}$ is the smallest $\delta$ rotation of the $\hat{x}$ vector normal to the FS. We adopt the gauge choice of first multiplying all Wilson lines along one direction, and then multiply the result obtained with the Wilson lines along the perpendicular direction. All other paths with the end points fixed are equivalent gauge choices, such that the net Wilson line is path independent. 
\par\noindent
We will now define two sets of generalized translation and twist operators for the $\hat{s}$ and $\hat{s}_{\perp}$ directions~\cite{oshikawa2000topological,oshikawa2003insulator} in the center of mass position and momentum spaces in representing the above Wilson lines
\begin{eqnarray}
\mathcal{T}_{\hat{s}_{\perp}}&:&A^{o}_{\Lambda,\hat{s}}\to A^{o}_{\Lambda, R\hat{s}},\label{1-translation}\\
\mathcal{T}_{\hat{s}}&:&A^{o}_{\Lambda,\hat{s}}\to A^{o}_{\Lambda+\delta\Lambda, \hat{s}},\label{2-translation}\\
\hat{O}^{o}_{\Lambda}&=&\exp\left[\frac{\pi}{2} i\sum_{n=0}^{N-1}nA^{o}_{\Lambda, R^{n}\hat{s}}\right],\\
\hat{O}^{o}_{\hat{s}'}&=&\exp\left[\frac{\pi}{2} i\sum_{n=0}^{N-1}nA^{o}_{n\delta\Lambda,\hat{s}'}\right].
\end{eqnarray}
The local Wilson lines along  a reference pair of directions $\hat{s},\hat{s}_{\perp}=\hat{x}$,$-\hat{y}$ (blue hatched/solid line in Fig.\ref{pseudospin_emergent_topology}) can then be translated to any orthogonal pairs of Wilson lines as follows
\begin{eqnarray}
\textit{Wilson line along }\hat{s}_{\perp}~: \hspace*{2cm}&&\nonumber\\
\mathcal{T}^{m}_{\hat{s}_{\perp}}\mathcal{T}^{n}_{\hat{x}}(A^{o}_{\mathbf{k}_{F,\hat{x}}}-A^{o}_{\mathbf{k}_{F,R\hat{x}}})\mathcal{T}^{\dagger n}_{\hat{x}}\mathcal{T}^{\dagger m}_{\hat{s}_{\perp}} &=& A^{o}_{\Lambda ,\hat{s}}-A^{o}_{\Lambda,R\hat{s}}~,\nonumber\\
\mathcal{T}^{m}_{\hat{s}_{\perp}}\mathcal{T}^{n}_{\hat{x}} W^{o}_{F,\hat{x}\to R\hat{x}}\mathcal{T}^{\dagger n}_{\hat{x}}\mathcal{T}^{\dagger m}_{\hat{s}_{\perp}} &=& ~W^{o}_{\Lambda,\hat{s}\to R\hat{s}} ,\label{Translated Wilson line 1}\\
\textit{Wilson line along }\hat{s}~:\hspace*{2cm}&&\nonumber\\
\mathcal{T}^{m}_{\hat{s}_{\perp}}\mathcal{T}^{n}_{\hat{x}}(A^{o}_{\mathbf{k}_{F,\hat{x}}}-A^{o}_{\mathbf{k}_{F+\delta k_{F},\hat{x}}})\mathcal{T}^{\dagger n}_{\hat{x}}\mathcal{T}^{\dagger m}_{\hat{s}_{\perp}} &=& A^{o}_{\Lambda ,\hat{s}}-A^{o}_{\Lambda +\delta\Lambda,\hat{s}}~,\nonumber\\
\mathcal{T}^{m}_{\hat{s}_{\perp},\hat{s}}\mathcal{T}^{n}_{\hat{x},\Lambda} W^{o}_{0\to \delta\Lambda,\hat{x}}\mathcal{T}^{\dagger n}_{\hat{x},\Lambda}\mathcal{T}^{\dagger m}_{\hat{s}_{\perp},\hat{s}} &=& W^{o}_{\Lambda\to \Lambda +\delta\Lambda,\hat{s}},~~~~~\label{Translated Wilson line 2}
\end{eqnarray}
where the number of pseudospins is taken to be $N=2(2k+1)$. The Wilson lines $(W^{o}_{F,\hat{x}\to R\hat{x}}$,$W^{o}_{0\to \delta\Lambda,\hat{x}})$ in eq.\eqref{Translated Wilson line 1},eq\eqref{Translated Wilson line 2} are the momentum-space projections of the Wilson loop defined in the center of mass position-momentum space for the major axis (along $\hat{s}_{\perp}$) and minor axis (along $\hat{s}$) of the torus 
\begin{eqnarray}
\mathcal{T}_{\hat{s}_{\perp}}^{2}\hat{O}^{o}_{\Lambda=0}\mathcal{T}^{2\dagger}_{\hat{s}_{\perp}}\hat{O}^{o\dagger}_{\Lambda=0} &=& 2^{4k+2}W^{o}_{F,\hat{x}\to R\hat{x}}W_{F},\\\label{Wilson loop for major axis}
\mathcal{T}_{\hat{s}}^{2}\hat{O}^{o}_{\hat{s}}\mathcal{T}^{2\dagger}_{\hat{s}}\hat{O}^{o\dagger}_{\hat{s}} &=& 2^{4k+2}W^{o}_{0\to \delta\Lambda,\hat{x}}W_{\hat{x}}~,\label{Wilson loop for minor axis}
\end{eqnarray}
and where $W_{F} = \prod_{m=0}^{N-1}A^{o}_{\mathbf{k}_{FR^{m}\hat{x}}}$ and $W_{\hat{x}}=\prod_{n=0}^{N-1}A^{o}_{n\delta\Lambda,\hat{x}}$ are Wilson loops for the minor and major axis of the torus along the reference directions (solid blue/hatched lines). The interaction terms $A^{o}_{\Lambda,\hat{s}}A^{o}_{\Lambda',\hat{s}'}$ can now be represented as the momentum-space projections of the product of translated Wilson loops as follows
\begin{eqnarray}
4A^{o}_{\Lambda,\hat{s}}A^{o}_{\Lambda',\hat{s}'} &=& \prod_{j_{1}=m,j_{2}=n}^{m',n'}(\mathcal{T}_{\hat{x}})^{j_{1}}(\mathcal{T}_{\hat{s}_{\perp}})^{j_{2}} W^{o}_{F,\hat{x}\to R\hat{x}}W^{o}_{0\to \delta\Lambda,\hat{x}}\nonumber\\
&\times &[(\mathcal{T}_{\hat{x}})^{j_{1}}(\mathcal{T}_{\hat{s}_{\perp}})^{j_{2}}]^{\dagger}~.
\end{eqnarray}  
The reduced BCS Hamiltonian can in turn be written as a $U(1)$ gauge theory in terms of non-local Wilson loops
\begin{eqnarray}
&&H^{*,XXZ}_{RBCS}(\omega)=\nonumber\\
&&\sum_{\mathbf{k}}\epsilon^{(j^{*})}_{\mathbf{k}_{\Lambda\hat{s}}}\ln(W\mathcal{T}_{\hat{s}_{\perp},\hat{s}}\mathcal{T}_{\hat{x},\Lambda}\mathcal{T}_{\hat{s}_{\perp}}\hat{O}^{z}_{\hat{s}_{\perp},0}\mathcal{T}^{\dagger}_{\hat{s}_{\perp}}\hat{O}^{z\dagger}_{\hat{s}_{\perp},0}[\mathcal{T}_{\hat{s}_{\perp},\hat{s}}\mathcal{T}_{\hat{x},\Lambda}]^{\dagger})\nonumber\\&-&\frac{1}{4}\sum_{\mathbf{k}_{\Lambda\hat{s}}} \Gamma^{4,(j^{*})}_{\alpha\beta} [\sum_{o=x,y}\prod_{j_{1}=m,j_{2}=n}^{m',n'}(\mathcal{T}_{\hat{x}})^{j_{1}}(\mathcal{T}_{\hat{s}_{\perp}})^{j_{2}} W^{o}_{F,\hat{x}\to R\hat{x}}W^{o}_{0\to \delta\Lambda,\hat{x}}[(\mathcal{T}_{\hat{x}})^{j_{1}}(\mathcal{T}_{\hat{s}_{\perp}})^{j_{2}}]^{\dagger}]\nonumber\\
&+&\frac{1}{4}\Gamma^{4,(j*),||}_{\alpha\alpha'} [\prod_{j_{1}=m,j_{2}=n}^{m',n'}(\mathcal{T}_{\hat{x}})^{j_{1}}(\mathcal{T}_{\hat{s}_{\perp}})^{j_{2}} W^{z}_{F,\hat{x}\to R\hat{x}}W^{z}_{0\to \delta\Lambda,\hat{x}}[(\mathcal{T}_{\hat{x}})^{j_{1}}(\mathcal{T}_{\hat{s}_{\perp}})^{j_{2}}]^{\dagger}]~.\label{gauge_theory}
\end{eqnarray}
\par\noindent
The above Hamiltonian commutes with the global Wilson loop given by
\begin{eqnarray}
W = \prod_{n=0}^{N-1} W^{z}_{n\delta\Lambda}~,~ [H^{RBCS}_{(j^{*}),XXZ}(\omega),W]=0~,\label{Global_Wilson_loop_emergent_window}
\end{eqnarray}
where $W_{n\delta\Lambda} = \mathcal{T}^{n}_{\hat{x}}W_{F}\mathcal{T}^{\dagger n}_{\hat{x}}$ is a Wilson loop obtained by translating $W_{F}$ by $n$ units. Remarkably, the Wilson loop $W$ (eq.\eqref{Global_Wilson_loop_emergent_window}) is an emergent topological invariant for the \textit{Luttinger zero patch} at the RG fixed point, seen from the Friedel's phase shift  that takes accounts of the total number of bound states in the emergent window
\begin{eqnarray}
\Delta N = -\frac{i}{\pi}\ln W~.
\label{friedelforsc}
\end{eqnarray}
\par\noindent
We will now present a simpler version of the above gauge theory for the case of the effective Hamiltonian $H^{*,XXZ}_{RBCS}$ with the couplings $\Gamma^{4,(j*)}_{\alpha\beta}=J_{\perp}$ and $\Gamma^{4,(j*),||}_{\alpha\alpha'}=J_{||}$. Our goal is to write once more the effective Hamiltonian in terms of nonlocal Wilson loop operators. For this, we first we write $H^{*,XXZ}_{RBCS}$ in terms of collective pseudospin operators
\begin{eqnarray}
H^{*,XXZ}_{RBCS}= -J_{\perp}(A^{2}_{x}+A^{2}_{y})+J_{||}A^{2}_{z}~,\label{collective_model}
\end{eqnarray}
where $A_{o}=\sum_{n,m}A^{o}_{m\delta\Lambda,R^{n}\hat{s}}$ represent the various components of the collective pseudospin vector. We now define nonlocal versions of the twist operators in the space of pseudospins
\begin{eqnarray}
\hat{O}^{o}_{\hat{s}}&=&\prod_{n=0}^{N-1}(\hat{O}^{o}_{R^{n}\hat{s}})^{\frac{4}{N}}=\exp\left[\frac{2\pi}{N} i\sum_{\substack{n=0,\\m=0}}^{N-1}mA^{o}_{m\delta\Lambda,R^{n}\hat{s}}\right].~~~
\end{eqnarray}
The collective pseudospin vectors can be written in terms of nonlocal twist ($\hat{O}^{o}_{\hat{s}}$) and translation ($T_{\hat{s}}$) operators. To show this, we obtain the following identity for the nonlocal Wilson loop composed of twist and translation operators
\begin{eqnarray}
\mathcal{T}_{\hat{s}}\hat{O}^{o}_{\hat{s}}\mathcal{T}^{\dagger}_{\hat{s}}\hat{O}^{o\dagger}_{\hat{s}}&=&\exp\left[\frac{2\pi i}{N} \sum_{\substack{n=0,\\m=0}}^{N-1}A^{o}_{m\Lambda,R^{n}\hat{s}}\right]\exp\left[2\pi i\sum_{n=0}^{N-1}A^{o}_{\Lambda=0,R^{n}\hat{s}}\right]~.
\end{eqnarray}
For every normal vector $\hat{s}$, there exists a opposite normal vector $-\hat{s}$, such that the total number of pseudospins ($N$) is even. Therefore, the overall phase collected from the strip along one of the minor circles of the torus is trivial: \begin{equation}
\exp\left[2\pi i\sum_{n=0}^{N-1}A^{o}_{\Lambda=0,R^{n}\hat{s}}\right]=1~.
\end{equation}
Thus, the collective pseudospin components $A_{o}$ can be represented by a nonlocal Wilson loop 
\begin{eqnarray}
A_{o}=\frac{N}{2\pi i}\log\left[\mathcal{T}_{\hat{s}}\hat{O}^{o}_{\hat{s}}\mathcal{T}^{\dagger}_{\hat{s}}\hat{O}^{o\dagger}_{\hat{s}}\right]~.\label{nonlocal_wilson_loop_pseudospin}
\end{eqnarray}
This enables us to write the Hamiltonian eq.\eqref{collective_model} as
\begin{eqnarray}
H^{*,XXZ}_{RBCS}&=&J_{\perp}\frac{N^{2}}{4\pi^{2}}\left(\log\left[\mathcal{T}_{\hat{s}}\hat{O}^{x}_{\hat{s}}\mathcal{T}^{\dagger}_{\hat{s}}\hat{O}^{x\dagger}_{\hat{s}}\right]\right)^{2}+ J_{\perp}\frac{N^{2}}{4\pi^{2}}\left(\log\left[\mathcal{T}_{\hat{s}}\hat{O}^{y}_{\hat{s}}\mathcal{T}^{\dagger}_{\hat{s}}\hat{O}^{y\dagger}_{\hat{s}}\right]\right)^{2}\nonumber\\
&-&J_{||}\frac{N^{2}}{4\pi^{2}}\left(\log\left[\mathcal{T}_{\hat{s}}\hat{O}^{z}_{\hat{s}}\mathcal{T}^{\dagger}_{\hat{s}}\hat{O}^{z\dagger}_{\hat{s}}\right]\right)^{2}~.\label{XXZ-Ham}
\end{eqnarray}
\par\noindent
We will now display certain features of topological order for this emergent gauge theory, e.g., ground state degeneracy and charge fractionalisation. Note that the transformations carried out by $T_{\hat{s}}$ eq\eqref{2-translation} impart equal and opposite momentum to opposite spin electrons, $\mathbf{k}_{\Lambda,\hat{s}},\uparrow\to \mathbf{k}_{\Lambda+\delta\Lambda,\hat{s}},\uparrow$ and  $\mathbf{k}_{\Lambda,-\hat{s}},\downarrow\to \mathbf{k}_{\Lambda+\delta\Lambda,-\hat{s}},\downarrow$, such that there is no net pair momentum $\mathbf{p}=0$. This observation supports the following representation of $T_{\hat{s}}$ in the position basis
\begin{eqnarray}
\mathcal{T}_{\hat{s}} = \exp\left[\frac{2\pi i}{2N}\sum_{\mathbf{r}}\mathbf{r}\cdot\hat{s}(\hat{n}_{\mathbf{r}\uparrow}-\hat{n}_{\mathbf{r}\downarrow})\right]~,\label{twist-op}
\end{eqnarray}
where the spacing in the momentum along $\hat{s}$ is $\delta\Lambda = \frac{2\pi}{N}$. A degeneracy of the ground state manifold can show up in its nontrivial topology. Below we probe this using spectral flow arguments that originated with the work of Lieb, Schultz and Mattis~\cite{lieb1961two}, and more recently extended to higher dimensions~\cite{yamanaka1997nonperturbative,hastings2004lieb,
altman1998haldane,oshikawa2000topological,
oshikawa2003insulator,pal2019magnetization,
pal2020topological,pal2019}. 
Initially, we compute the action of the twist operator $\mathcal{T}_{\hat{s}}$ on an eigenstate of $H^{*,XXZ}_{RBCS}$ ($|\Psi\rangle$)
\begin{eqnarray}
\langle\Psi|T^{\dagger}_{\hat{x}}\mathcal{T}_{\hat{s}}T_{\hat{x}}|\Psi\rangle &=& \exp\left[\frac{2\pi i}{N}\sum_{\mathbf{r}}S^{z}_{\mathbf{r}}\right]\exp\left[2\pi i\sum_{\mathbf{r}\cdot\hat{s}_{\perp}}S^{z}_{\mathbf{r}\cdot\hat{s}_{\perp},\mathbf{r}\cdot\hat{x}}\right]\nonumber\\
&\times & \langle\Psi|\mathcal{T}_{\hat{s}}|\Psi\rangle~. 
\end{eqnarray}
The Hamiltonian $H^{*,XXZ}_{RBCS}$ commutes with $S^{z}=\sum_{\mathbf{r}}S^{z}_{\mathbf{r}}$, and its low-energy manifold is comprised of states $|S=2k+1,S^{z}=0\rangle$ with net $S^{z}=0$.
Using the fact that the total number of pseudospins  $N= 2(2k+1)=L_{x}L_{y}$ (i.e., $L_{x}$, $L_{y}$ corresponds to the number of pseudospins along the $x$ and $y$ directions of the torus), the second exponential term in the above expression is simple: 
\begin{equation}
e^{\left[2\pi i\sum_{\substack{\mathbf{r}\cdot\hat{s}_{\perp}}}S^{z}_{\mathbf{r}\cdot\hat{s}_{\perp},\mathbf{r}\cdot\hat{x}}\right]}=e^{2\pi(2k+1)\frac{1}{2}}=e^{i\pi}~.
\label{groundStatedegeneracy}
\end{equation}
In this way, we obtain an equivalent of the LSM relation~\cite{lieb1961two} for higher dimensions~\cite{hastings2004lieb}. For this case, $\langle\Psi|T^{\dagger}_{\hat{x}}\mathcal{T}_{\hat{s}}T_{\hat{x}}|\Psi\rangle = -\langle\Psi|\mathcal{T}_{\hat{s}}|\Psi\rangle$, implying that the two states $|\Psi\rangle$ and $T_{\hat{x}}|\Psi\rangle$ are orthogonal. Finally the important relation
\begin{eqnarray}
[H^{*,XXZ}_{RBCS},\mathcal{T}_{\hat{s}}]=0\label{emergentcommutation}
\end{eqnarray}
implies that the eigenstates of the twist operator $\mathcal{T}_{\hat{s}}$ (corresponding to eigenstates of the center of mass momentum $\mathbf{P}_{cm}=0,\pi\hat{s}$) are simultaneously eigenstates of the Hamiltonian. Therefore, the two groundstates $|\mathbf{P}_{cm}=0,S=2k+1,S^{z}=0\rangle$ and $|\mathbf{P}_{cm}=\pi\hat{s},S=2k+1,S^{z}=0\rangle$, both possessing ground state energy $E_{g}=-J_{\perp}(2k+1)(2k+2)$, are degenerate and protected from excitations via a many-body gap $2J_{\perp}(2k+1)$. The adiabatic passage between these degenerate ground states, achieved via the application of the twist operator $\mathcal{T}_{\hat{s}}$, involves the creation of a charge-$1/2$ excitation~\cite{wen1990,oshikawa2006,chen2010local}. Additionally, we note that given the microscopic Hamiltonian $[H_{SFIM},\mathcal{T}_{\hat{s}}]\neq 0$, the the commutation relation eq\eqref{emergentcommutation} for the low energy effective Hamiltonian $H^{*,XXZ}_{RBCS}$ is emergent under RG flow. 
\par\noindent 
Similar gauge theoretic constructions can be attained for the other gapped phases of $H_{SFIM}$ in terms of the appropriate non-local Wilson loop operators (defined in terms of the respective pseudospin Hilbert spaces). On the other hand, for the gapless Fermi liquid (eq.\eqref{HFL}) of $H_{SFIM}$, only the first and last terms in eq.\eqref{gauge_theory} appear. In this case, the equivalent of eq.\eqref{friedelforsc} yields the Luttinger zeroes ~\cite{seki2017topological}. 
\section{RG analysis of the generalized Sachdev-Ye$_{4}$ model}\label{SY}
\pin
In order to understand the interplay between disorder and interactions in a model of correlated electrons~\cite{dagotto2005complexity}, we consider a generalized  electronic Sachdev-Ye(\text{SY}$_{4}$)~\cite{sachdev1993} model with random spin-independent hopping ($t_{ij}$), random on-site potential ($\epsilon_{i}$) and a random four-Fermi interaction ($V^{\sigma\sigma'}_{ijkl}$) 
\begin{eqnarray}
H_{SY_{4}} &=& \sum_{i\neq j,\sigma\sigma'}t_{ij}c^{\dagger}_{i\sigma}c_{j\sigma'} + \sum_{i\sigma}\epsilon_{i}\hat{n}_{i\sigma}+\sum_{ijkl,\sigma\sigma'}V^{\sigma\sigma'}_{ijkl}c^{\dagger}_{i\sigma}c^{\dagger}_{j\sigma'}c_{k\sigma'}c_{l\sigma}~,\label{Sachdev-Ye model}
\end{eqnarray}
and place it on a $D$ spatial-dimensional volume containing $2N$ points and with a specified geometry. Here, the indices ranging $i\in[1,2N]$ ~($2N$ being number of points) correspond to the real-space position vectors $\mathbf{r}_{i}$. The terms $V^{\sigma\sigma'}_{ijkl}=V^{\sigma\sigma'*}_{klij}$, $t_{ji}=t_{ij}$ and $\epsilon_{i\sigma}$ are random 
tensors drawn from separate Gaussian distributions, each with a well-defined mean and standard deviation. The single particle energies $\epsilon_{i\sigma}$ can be sorted as follows
\begin{eqnarray}
\epsilon_{1}\leq\ldots\leq\epsilon_{N}~,\label{ordering energies}
\end{eqnarray}
and employed for implementing the iterative steps of the unitary RG, (here \emph{disorder bandwidth} is defined as $W=\epsilon_{N}-\epsilon_{1}$).
\par\noindent 
The Hamiltonian RG flow equation is given by eq.\eqref{HamRGflow}, where the unitary transformation $U^{\dagger}_{(j)}=U^{\dagger}_{(j\uparrow)}U^{\dagger}_{(j\downarrow)}$. From here, we can extract the hierarchy of 2-, 4- and 6-particle vertex flow equations given in eq.\eqref{vertexRGflows} and shown in Fig.(\ref{vertex_flow_eqn_2,4,6}). 
From the flow equations, we obtain the various parameter regimes belonging to the generalised SY$_{4}$ that lead, under RG, to models with emergent translational invariance (TI, e.g., $H_{SFIM}$ eq.\eqref{SFIM}) as well as non-translationally invariant (NTI) models associated with the physics of \textit{localization}. Having already presented a quantitative verification of the RG flows for various TI phases in the previous section, we will present only those for the NTI phases here.
\begin{figure}
\centering
\includegraphics[scale=0.75]{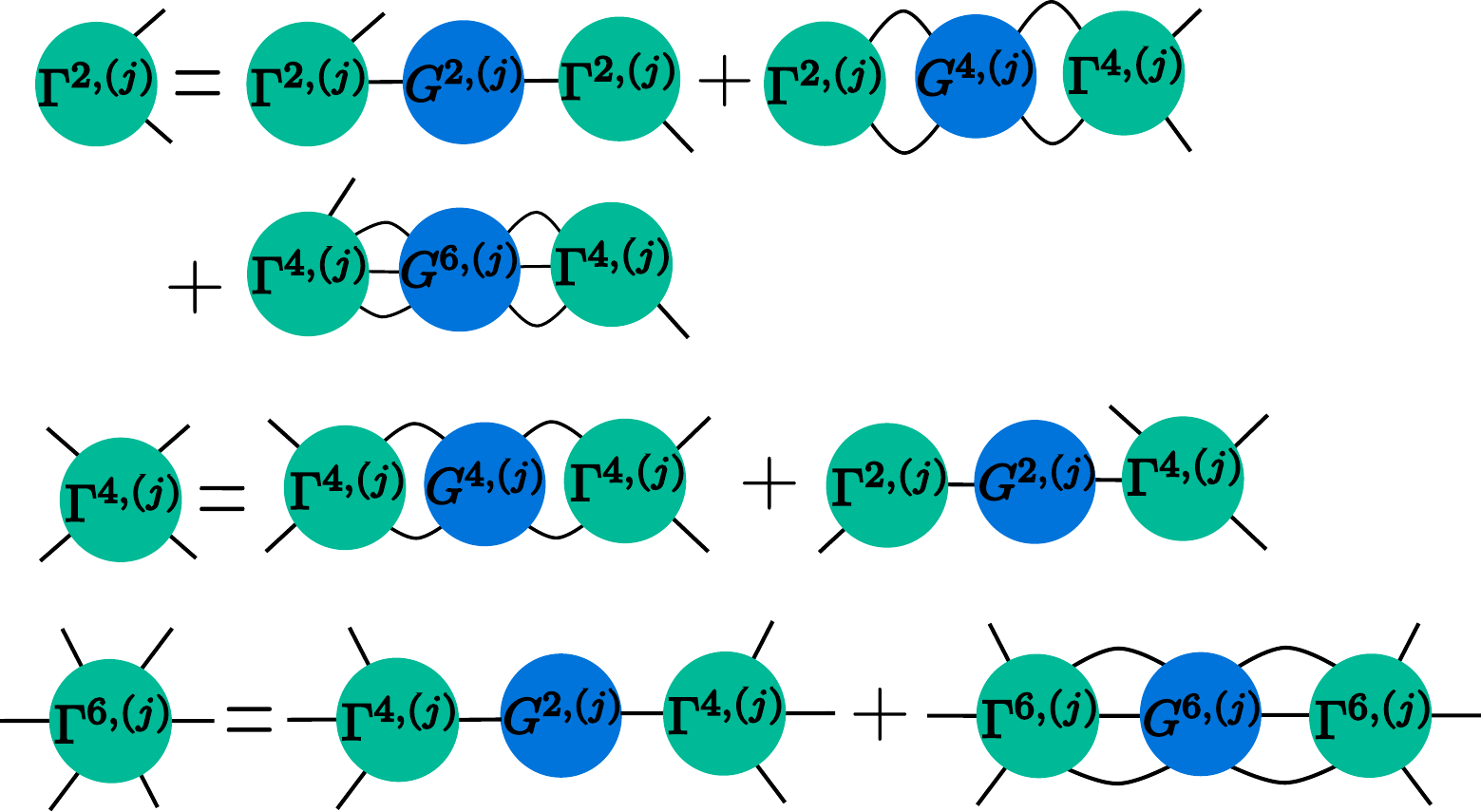} 
\caption{Schematic representation of RG flow equations for $2$-point, $4$-point and $6$-point vertices. See text for discussion.}\label{vertex_flow_eqn_2,4,6}
\end{figure}
\par\noindent
For this, we first write the microscopic parameters as a sum of translational invariant (TI) and non-invariant (NTI) parts
\begin{eqnarray}
t_{ik}&=&t(\mathbf{r}_{i}-\mathbf{r}_{k})+t'_{ik}~,~\epsilon_{i} = \epsilon + \epsilon_{i}',\label{2-point_trans_decom}\\
V^{\sigma\sigma'}_{ijkl} &=& V^{\sigma\sigma'}(\mathbf{r}_{i}-\mathbf{r}_{j},\mathbf{r}_{j}-\mathbf{r_{k}},\mathbf{r}_{k}-\mathbf{r}_{l})+V^{'\sigma\sigma'}_{ijkl},\label{4-point_X_trans_decom}\\
V^{\sigma\sigma'}_{ik} &=& V^{\sigma\sigma'}(\mathbf{r}_{i}-\mathbf{r}_{k})+V^{',\sigma\sigma'}_{ik}~.\label{4-point_D_trans_decom}
\end{eqnarray} 
We note that a similar decomposition for the 2-point Green's function was carried out by Ishikawa and Matsuyama~\cite{ishikawa1987microscopic} for showing the preservation of the momentum-space Ward-Takahashi identity in the integer quantum Hall problem. By analyzing a class of fixed points of the RG flow equations presented in eq.\eqref{Translational invariant hopping term_RG}~--~eq.\eqref{Translational non-invariant current-current interaction term_RG} (in Appendix \ref{SY4}) for the parameter regimes I-VI shown in Tables \ref{tab:parameter_regimes_SY_2,4} and \ref{tab:parameter_regimes_SY_4_thermal}, we obtain fixed point Hamiltonians displayed in Table \ref{tab:fixed_point_Hamiltonians}. We now discuss the physics of each of the 6 regimes in turn. 
\par
\begin{figure}
\centering
\includegraphics[width=0.7\textwidth]{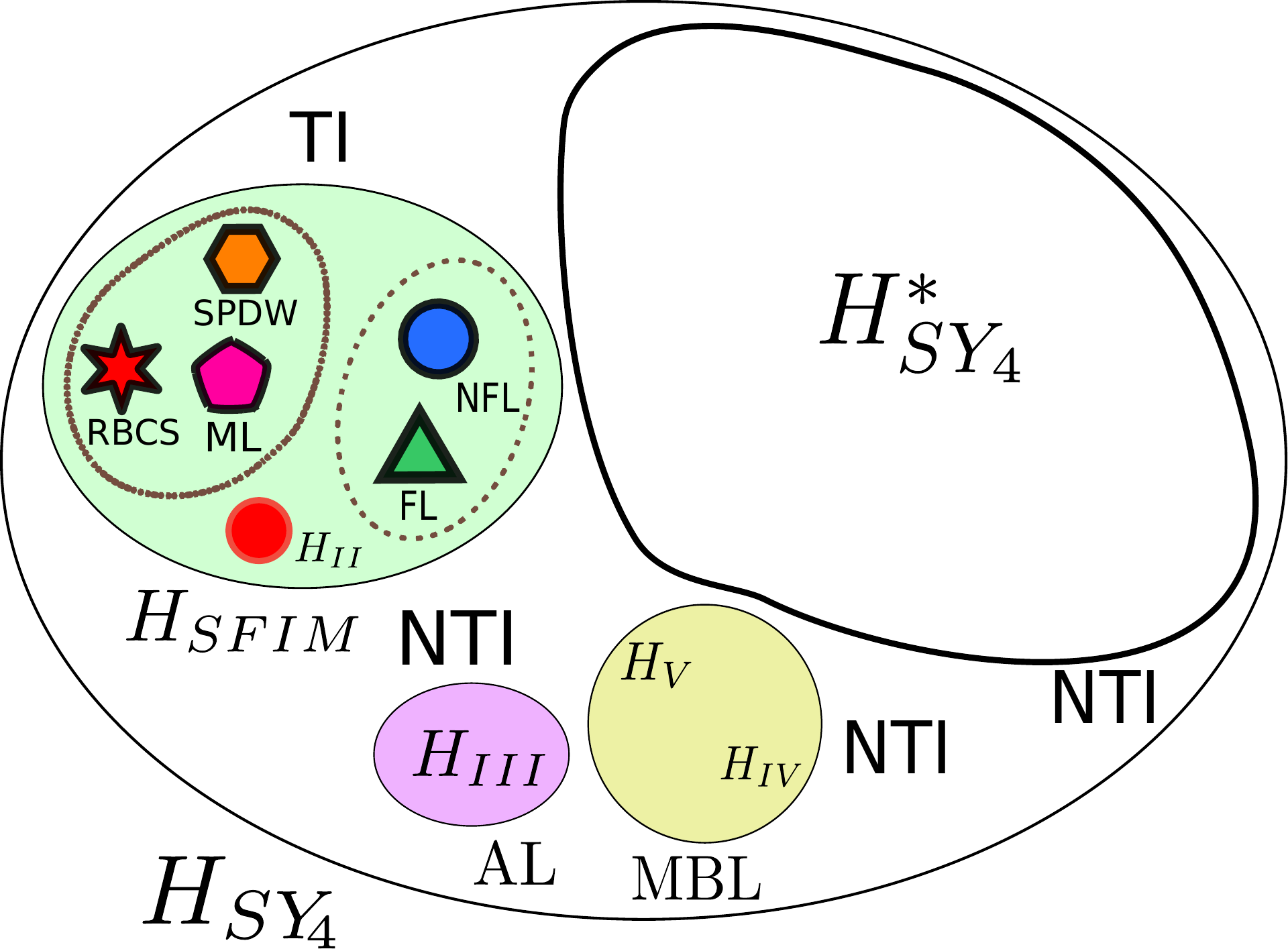} 
\caption{Schematic description of various parameter-space regimes of the SY$_{4}$ model leading to various translationally invariant (TI) and non-translationally invariant (NTI) fixed point Hamiltonians under RG flow. The light green circle comprises various TI phases (arising from an emergent effective Hamiltonian $H_{I}=H_{SFIM}$ of Table \ref{tab:fixed_point_Hamiltonians}): (i) Gapped phases: reds star, orange hexagon, magenta pentagon within the dashed boundary represent the reduced BCS model (RBCS), symmetry unbroken PDW (SPDW) and the Mott liquid Hamiltonians (ML) respectively; (ii) Gapless phases: blue circle and green triangle within the dotted boundary represent the Fermi liquid (FL) and Non Fermi liquid (NFL) respectively; (iii) the red circle represents the Hubbard models with long ranged hopping ($H_{II}$ of Table \ref{tab:fixed_point_Hamiltonians}). The light yellow oval represents many-body localized (MBL) phases, with effective Hamiltonians $H_{IV}$ and $H_{V}$ are $2$-electron and $3$-electron interacting number-diagonal NTI models respectively (see Table \ref{tab:fixed_point_Hamiltonians}). The pink oval is the Anderson disorder localization regime, with the NTI model $H_{III}$ of Table \ref{tab:fixed_point_Hamiltonians}. The large white region with black border represents the themalised phase, and corresponds to a generalized $SY_{4}$ model reproduced under RG ($H_{VI}$ of Table \ref{tab:fixed_point_Hamiltonians}).}\label{SY-to-others}
\end{figure}

\noindent
Regime I (green oval in Fig.\ref{SY-to-others}, Table \ref{tab:parameter_regimes_SY_2,4}) leads to a general TI Hamiltonian $H_{I}$ (see Table \ref{tab:fixed_point_Hamiltonians}) which, in the single band limit, is equivalent to the Hamiltonian $H_{SFIM}$ (eq.\eqref{SFIM}) considered in the previous section. To see this, we replace the real-space creation/annhilation operators in the fixed point theory by their Fourier transforms $c^{\dagger}_{\mathbf{k}\sigma}= \sqrt{\text{Vol}^{-1}}\sum_{\mathbf{r}}e^{i\mathbf{k}\cdot\mathbf{r}}c^{\dagger}_{\mathbf{r}\sigma}$. This generic TI model of interacting electrons is obtained in the regime of low randomness, i.e., the magnitude of the NTI parameters (with opposite signs for the hopping, and 4-fermi interactions) compared to their TI counterparts. Following our detailed RG analysis of $H_{SFIM}$ in the previous section, we know that a non-Fermi liquid phase (blue circle in Fig.\ref{SY-to-others}) is obtained via a second level of the RG on $H_{SFIM}$, i.e., $H_{\text{SY}_{4}}\xrightarrow[RG]{}H_{SFIM}\xrightarrow[RG]{}$ (eq\eqref{3-numberdiag}). This non-Fermi liquid is characterized by a logarithmically dependent self-energy (eq.\eqref{frequency_dependent_self_energy}), a $T$-linear resistivity (eq.\eqref{T-linear}), a vanishing quasiparticle residue $Z$, a finite temperature geometric entanglement content (eq.\eqref{entanglement_spectral_weight_content}) etc. We recall that non-Fermi liquid phases were also obtained from large $N$ analyses of the  spin-$S$ Heisenberg Sachdev-Ye model~\cite{sachdev1993}, and as well as in electronic Sachdev-Ye-Kitaev (SYK) model~\cite{song2017}. We also recall that the $H_{SFIM}$ possesses an emergent gapless FL phases, as well as several gapped phases that emerge from instabilities of the non-Fermi and Fermi liquids -- reduced BCS, symmetry unbroken PDW's, Mott liquid etc.(shown within the green oval in Fig.\ref{SY-to-others}). Similar pairing instabilities of the non-Fermi liquid phase in the SYK model have also been reported recently~\cite{bi2017instability}. Regime II in Table ~\ref{tab:parameter_regimes_SY_2,4} is one where attractive extended interactions are RG irrelevant, whereas on-site repulsion is RG relevant. This leads under RG flow to the Hubbard model with long-range hopping (shown as a red circle within green oval in Fig.\ref{SY-to-others}, see Table \ref{tab:fixed_point_Hamiltonians}). 
\begin{figure}
\hspace*{-2cm}
\includegraphics[width=0.63\textwidth]{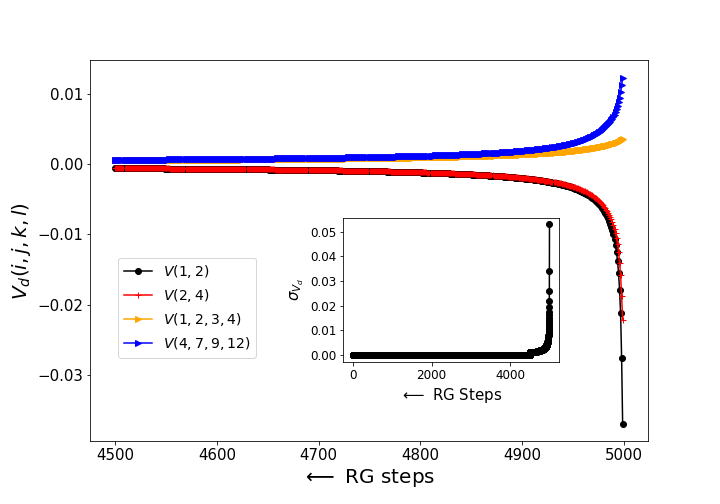}
\includegraphics[width=0.63\textwidth]{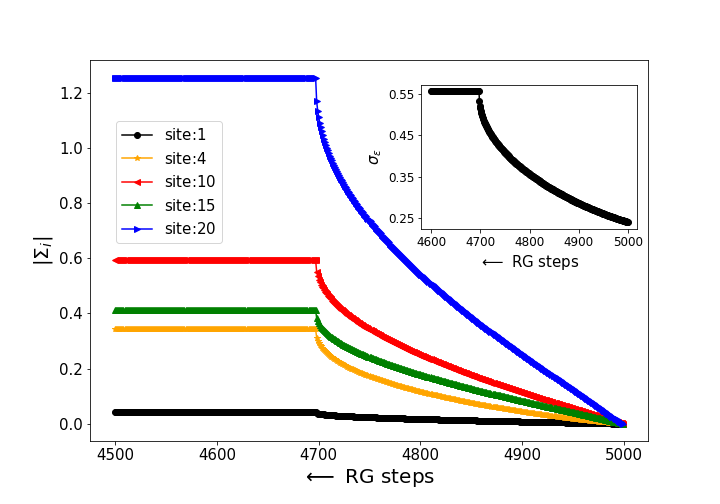}
\caption{(Left Panel) Figure represents RG flows for number diagonal/off-diagonal interaction couplings $V(i,j)$/$V(i,j,k,l)$ that are sampled from a guassian with mean ($\overline{V(i,j,k,l)}=0.1$) and standard deviation ($\sigma_{V_{d}}=0.05$), $\omega=0.5\epsilon'_{N}$, system size $N=5000$, disorder bandwidth ($W=2$), $\epsilon'_{N}=-2W$. Inset plot in left panel tracks the renormalization of standard deviation $\sigma_{V_{d}}$.(Right Panel) Figure represents RG flows for self energy $\Sigma_{i}$ of the onsite disordered energies $\epsilon'_{i}$  sampled from a gaussian distribution with $\overline{\epsilon'_{i}}=-W$ and $\sigma_{\epsilon'_{i}}=0.25$. Inset plot in right panel tracks the RG flow for $\sigma_{\epsilon'}$ of the renormalized onsite energies $\epsilon'$.}\label{AndLoc1}
\end{figure}
\par
\begin{figure}
\includegraphics[width=\textwidth]{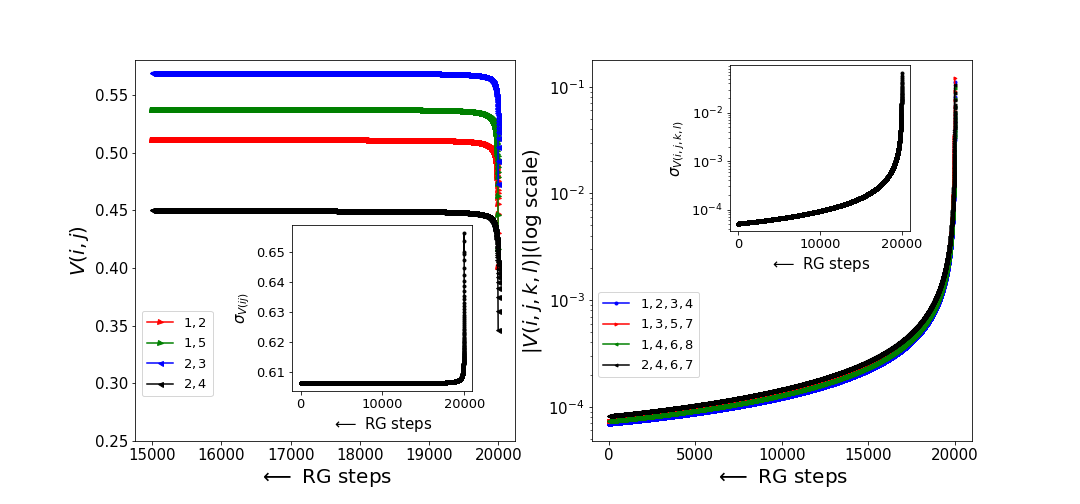}
\caption{Left panel: RG flow for the number-diagonal interactions $V(i,j)$ for $\omega=\epsilon_{N}+0.1$. The $V(i,j)$ are initially sampled from a gaussian distribution with mean $\overline{V(a,b)}=0.5$ and $\sigma_{V(a,b)}=0.1$. Inset in left panel tracks the RG flow for $\sigma_{V(a,b)}$. Right panel: RG flow for the number off-diagonal interactions $V(i,j,k,l)$, whose bare values are sampled from a gaussian distribution with mean $\overline{V(i,j,k,l)}=-0.1$ and $\sigma_{V(i,j,k,l)}=0.01$. Inset in right panel tracks the RG flow for $\sigma_{V(i,j,k,l)}$.}\label{GlassI_1}
\end{figure}

\begin{figure}
\includegraphics[width=\textwidth]{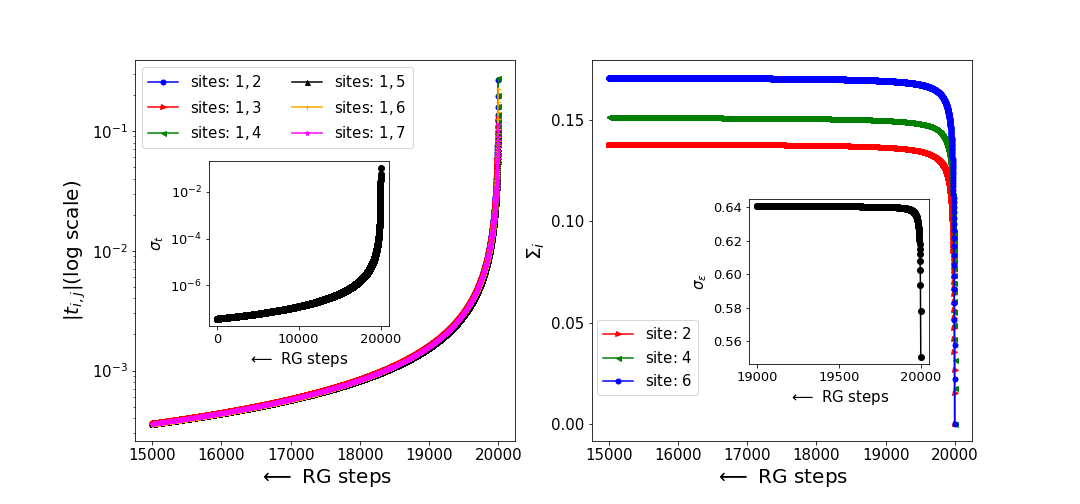}
\caption{(Left Panel) Figure represents RG flow for the hopping strengths $t'_{i,j}$. Bare values are sampled from a gaussian distribution with mean $\overline{t'_{i,j}}=-0.2$ and $\sigma_{t_{d}}=0.05$. Inset in left panel tracks the RG flow for $\sigma_{t_{d}}$. (Right Panel) Figure represents RG flow for the disordered onsite potentials $\epsilon_{i}$'s, the initial values are sampled from a gaussian distribution with mean $\overline{\epsilon_{i}'}=1.0$ and $\sigma_{\epsilon_{i}'}=0.56$. Inset in right panel tracks the RG flow for $\sigma_{\epsilon_{i}'}$.}\label{GlassI_2}
\end{figure}

\begin{figure}
\includegraphics[width=\textwidth]{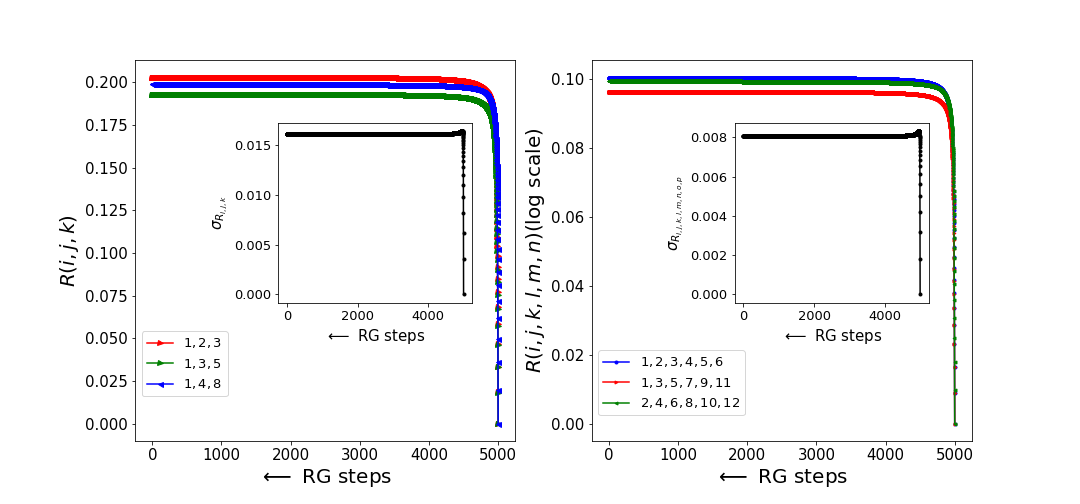}
\caption{Left panel: RG flow for the emergent three-particle number diagonal interactions $R(i,j,k)$. Inset in left panel diplays the RG flow for $\sigma_{R(i,j,k)}$. Right panel: RG flow for the number off-diagonal interactions $R(i,j,k,l,m,n)$. Inset in right panel shows the RG flow for $\sigma_{R(i,j,k,l,m,n)}$.}\label{GlassII_2}
\end{figure}

\begin{figure}
\includegraphics[width=\textwidth]{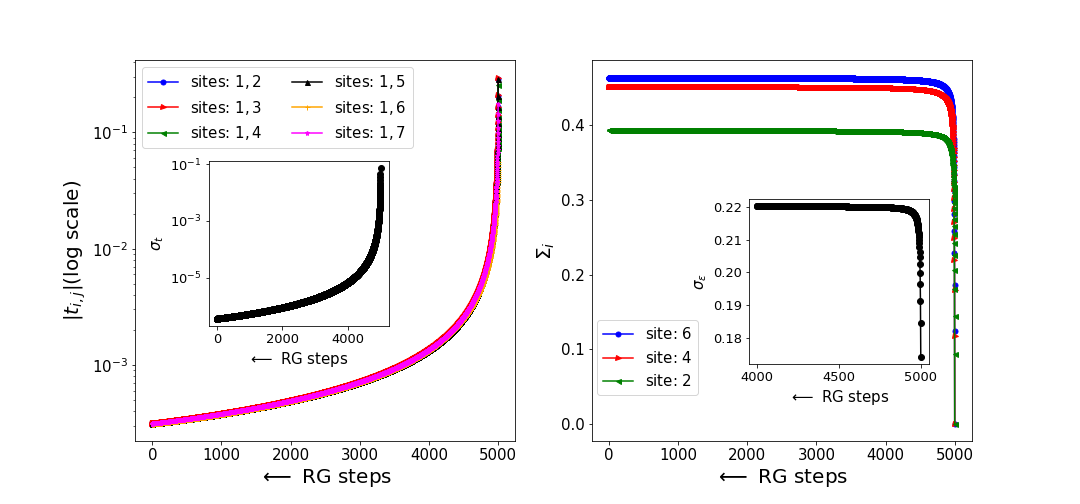}
\caption{Left panel: RG flow for the hopping parameters $t'_{ij}$, whose bare values are sampled from a gaussian distribution with mean $\overline{t'_{i,j}}=-0.2$ and $\sigma_{t'_{ij}}=-0.1$. Inset in right panel displays the RG flow for $\sigma_{t_{ij}}$. Right panel: RG flow for the onsite energies $\epsilon_{i}$, whose bare values are sampled from a gaussian distribution with mean $\overline{\epsilon_{i}}=1$ and $\sigma_{\epsilon_{d}}=0.15$. Inset in left panel shows the RG flow for $\sigma_{\epsilon_{i}}$.}\label{GlassII_3}
\end{figure}

\begin{figure}[!ht]
\includegraphics[width=\textwidth]{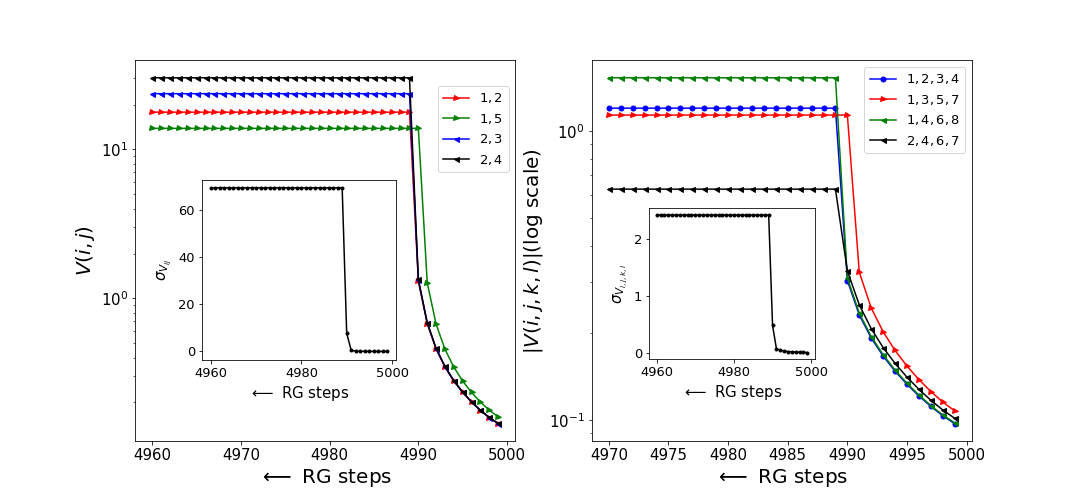}
\caption{Left panel: RG flow for the number diagonal interactions $V(i,j)$, whose initial values are sampled from a gaussian distribution with mean $\overline{V(a,b)}=0.15$ and $\sigma_{V(a,b)}=0.01$. Inset in left panel displays the RG flow for $\sigma_{V(a,b)}$. Right panel: RG flow for the number off-diagonal interactions $V(i,j,k,l)$, whose bare values are sampled from a gaussian distribution with mean $\overline{V(i,j,k,l)}=-0.1$ and $\sigma_{V(i,j,k,l)}=0.01$. Inset in right panel shows the RG flow for $\sigma_{V(i,j,k,l)}$.}\label{SY4_2}
\end{figure}

\begin{figure}[!ht]
\centering
\includegraphics[scale=0.45]{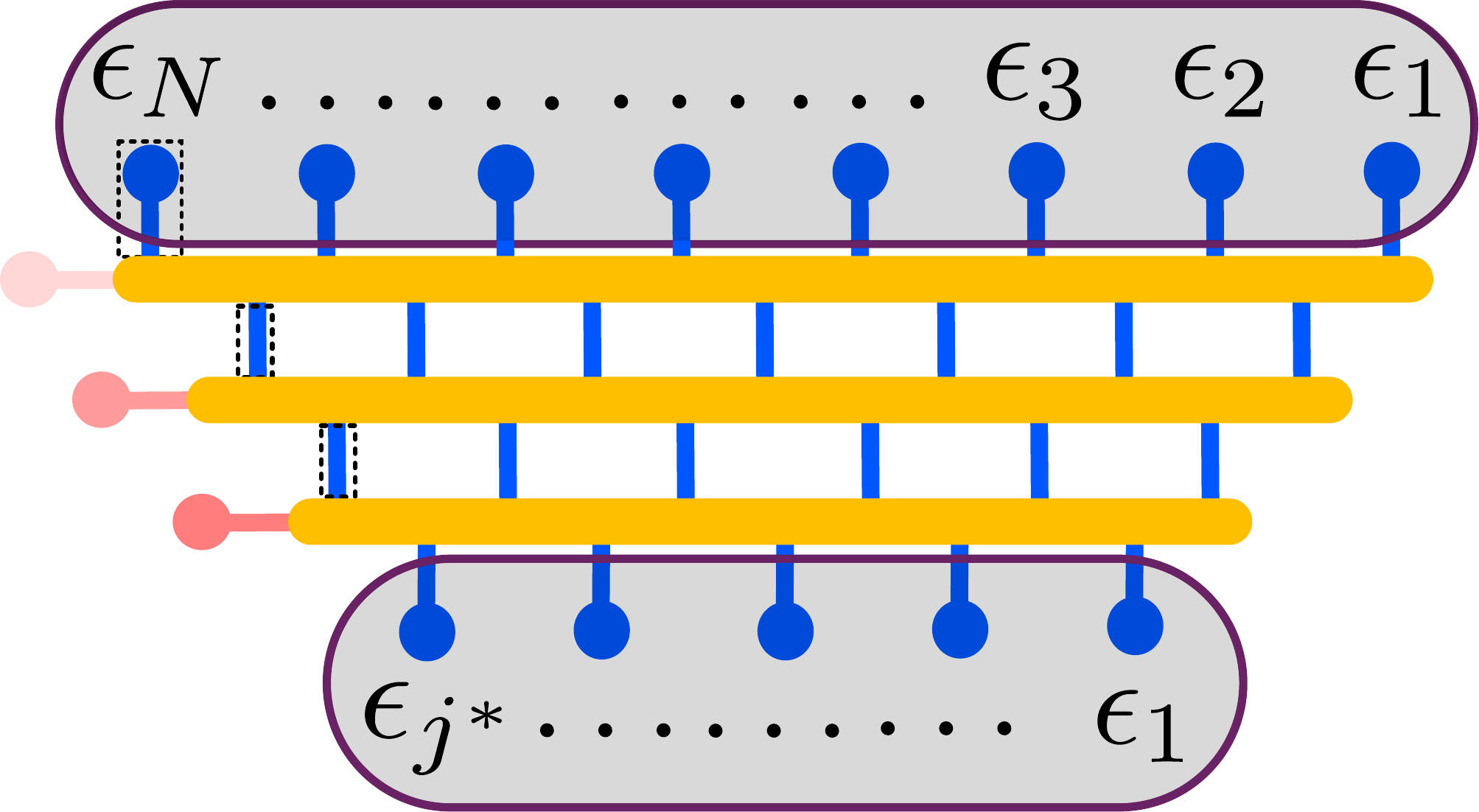}
\caption{EHM tensor network representation of RG flow towards the thermalised phase (regime VI of Tables \ref{tab:parameter_regimes_SY_4_thermal} and \ref{tab:fixed_point_Hamiltonians}) of the generalized $SY_{4}$ model. The color of the blue legs do not change, indicating a similarity in the nature of the intermediate theories leading upto a fixed point $SY_{4}$ model. The rounded square boxes indicate the interaction comprising all the degrees of freedom in the initial and final points of the RG flow.}\label{TN-SY}
\end{figure}

\noindent
In regime-III of Table\ref{tab:parameter_regimes_SY_2,4} , we obtain a model displaying the phenomenon of Anderson localisation (AL): disordered noninteracting electrons with long range hopping ($H_{III}$ in Table~\ref{tab:fixed_point_Hamiltonians}). This is obtained from a relevant RG flow for random on-site potential, together with an irrelevant RG flow for all random hopping processes as well as all four-fermionic interactions. The left panel of Fig.\ref{AndLoc1} represents the numerical evaluation of the RG flow of the disordered interaction strengths $V(i,j)\hat{n}_{i}\hat{n}_{j}$~($V(i,j)=V_{i,j,j,i}$) and $V(i,j,k,l)c^{\dagger}_{i}c^{\dagger}_{j}c_{k}c_{l}$ ~($V(i,j,k,l)=V_{i,j,k,l}$). We find that both $V(i,j)$ and $V(i,j,k,l)$ are RG irrelevant. The inset plot in the right panel of Fig.\ref{AndLoc1} shows the vanishing of the standard deviation $\sigma_{V_{d}}^{(j)}$ of the disordered $V(i,j,k,l)$ coupling under RG flow
\begin{eqnarray}
\sigma_{V_{d}}^{(j)}=\sqrt{\frac{\sum\limits_{(a,b,c,d)<j}(V(a,b,c,d)^{(j)})^{2}}{N^{4}}-\left(\frac{\sum\limits_{(a,b,c,d)<j}V(a,b,c,d)}{N^{4}}\right)^{2}}~.\label{stdDevInteraction}
\end{eqnarray}
On the other hand, the RG flows for the single-particle self-energy $\Sigma^{(j)}_{i}$ $\epsilon_{i}^{'(j)}=\epsilon_{i}+\Sigma^{(j)}_{i}$) are observed to grow in Fig.\ref{AndLoc1} (right panel), finally saturating at an IR fixed point. The inset plot in the right panel of Fig.\ref{AndLoc1} shows the growth and saturation of the standard deviation $\sigma_{\epsilon}^{(j)}$ of the renormalised onsite energies $\epsilon_{i}^{'(j)}$ under the RG flow
\begin{eqnarray}
\sigma_{\epsilon}^{(j)}=\sqrt{\frac{1}{N}\sum_{i}(\epsilon_{i}^{'(j)})^{2}- (\frac{1}{N}\sum_{i}\epsilon_{i}^{'(j)})^{2}}~.\label{stdDevonsite}
\end{eqnarray}
\begin{figure}
\includegraphics[width=\textwidth]{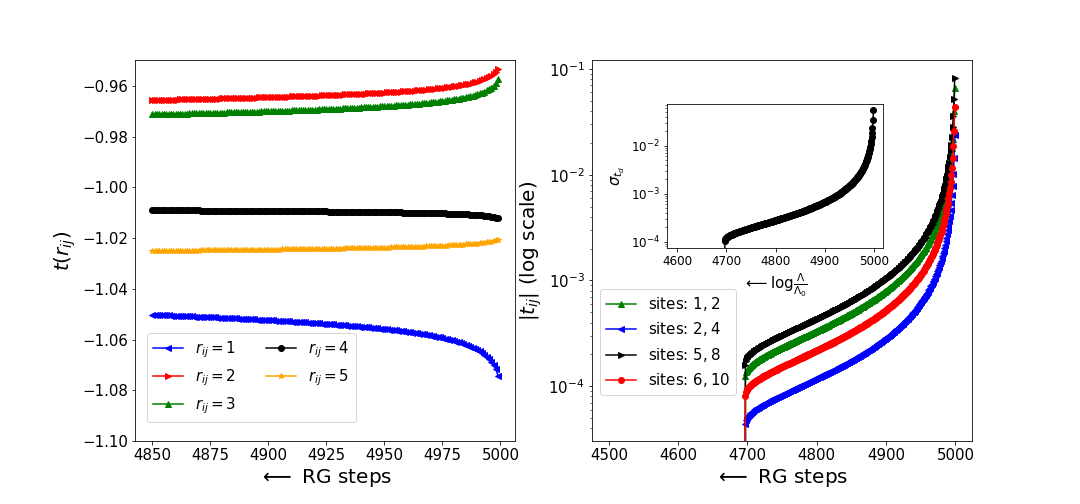}
\caption{Left panel: RG flows for translationally invariant hopping strengths $t(r_{ij})$ for various distances $r_{ij}$. Right panel: RG flows of the disordered hopping strengths $t'_{ab}$ sampled from a guassian with mean ($\overline{t'_{ab}}=0.2$) and standard deviation ($\sigma_{t_{d}}=0.05$). Inset plot in right panel displays the RG flow for the standard deviation ($\sigma_{t_{d}}$) of disordered hopping strengths.}\label{AndLoc2}
\end{figure}
Finally, the left panel of Fig.\ref{AndLoc2} shows the saturation under RG flow of translationally invariant variable-range hopping strengths $t(r_{ij})$ at the IR fixed point. On the other hand, in the right panel of Fig.\ref{AndLoc2}, we find that the disordered hopping $t'_{ik}$  are found to be RG irrelevant and vanish at an IR fixed point. Similarly, the inset plot in the right panel tracks the reduction under RG flow of the standard deviation $\sigma_{t_{d}}$ in the random hopping
\begin{eqnarray}
\sigma_{t_{d}}=\sqrt{\frac{1}{N^{2}}\sum_{a,b}t^{(j),2}_{a,b}-(\frac{\sum_{i,j}t_{i,j}}{N^{2}})^{2}}\label{stdDevHop}~.
\end{eqnarray}
Taken together, Figs.\ref{AndLoc1} and \ref{AndLoc2} establish that in Regime III, the IR fixed point effective Hamiltonian is that for disordered noninteracting electrons with long range hopping ($H_{III}$ in Table~\ref{tab:fixed_point_Hamiltonians}). 
\pin
Regime IV of Table~\ref{tab:parameter_regimes_SY_4_thermal}, with an effective IR fixed point Hamiltonian $H_{IV}$ (see Table~\ref{tab:fixed_point_Hamiltonians}), corresponds to a phase that is a glassy variant of the Fermi liquid (known as the interacting Fermi insulator), and involves the phenomenon of many-body localization (MBL) in Fock space~\cite{pal2010many,ros2015, parameswaran2017non}. Fig.\ref{GlassI_1} (left panel) represents the growth under RG flow and saturation of the number-diagonal interactions at the low-energy fixed point. The inset shows that the standard deviation $\sigma_{V(a,b)}$ of the renormalized couplings $V_{ij}$ reduces in magnitude under RG and saturates to a finite value at the fixed point. On the other hand, in right panel of Fig.\ref{GlassI_1}, the off-diagonal interaction couplings $V_{i,j,kl}$ are found to be RG irrelevant. In the inset, we observe that the standard deviation of the off-diagonal scattering vertices $\sigma_{V_{i,j,k,l}}$ diminishes under RG flow, eventually vanishing at the low-energy fixed point.
Fig.\ref{GlassI_2} (left panel) represents the RG irrelevant flows for the hopping strength $t_{ij}$. The inset in the left panel shows the reduction in $\sigma_{t_{d}}$ under RG flow, eventually vanishing at the IR fixed point. On the other hand, the right panel of Fig.\ref{GlassI_2} represents the relevant RG flow for the onsite self-energies $\Sigma_{i}$, displaying a growth and saturation at the low-energy fixed point. The inset plot in right panel of Fig.\ref{GlassI_2} shows that the standard deviation of the renormalized energies $\sigma_{\epsilon'_{i}}$ also grows under RG and saturates at low energies. Together, Fig.\ref{GlassI_1} and \ref{GlassI_2} indicate the onset of many-body localization with a Hamiltonian $H_{IV}$ Table \ref{tab:fixed_point_Hamiltonians} describing the effective low energy theory.   
\begin{figure}
\includegraphics[width=\textwidth]{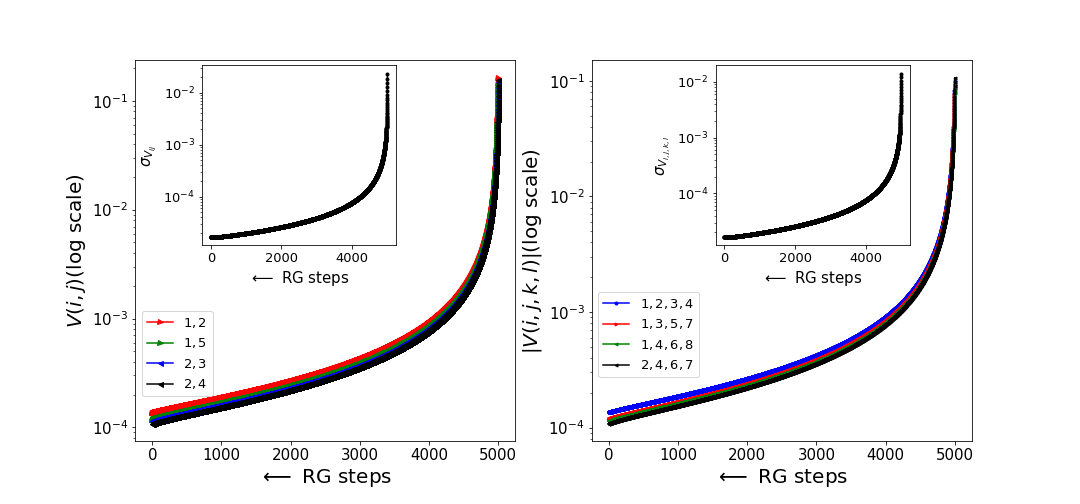}
\caption{Left panel: RG flow for the number diagonal interactions $V(i,j)$ for $\omega=0.5\epsilon_{N}+0.1$. The bare $V(i,j)$ are sampled from a gaussian distribution with mean $\overline{V(i,j)}=0.15$ and $\sigma_{V(i,j)}=0.03$. Inset in left panel displays the RG flow for $\sigma_{V(i,j)}$. Right panel: RG flow for the number off-diagonal interactions $V(i,j,k,l)$, whose bare values are sampled from a gaussian distribution with mean $\overline{V(i,j,k,l)}=-0.1$ and $\sigma_{V(i,j,k,l)}=0.01$. Inset in right panel shows the RG flow for $\sigma_{V(i,j,k,l)}$.}\label{GlassII_1}
\end{figure}
\pin
Similarly, regime V in Table~\ref{tab:parameter_regimes_SY_2,4} corresponds to a many-body localised (MBL) phase that is the glassy variant of a non-Fermi liquid Hamiltonian ($H_{V}$ in Table~\ref{tab:fixed_point_Hamiltonians})~\cite{pal2010many, parameswaran2017non} and we call it the marginal Fermi insulator. The left and right panels of Fig.\ref{GlassII_1} represent the RG flows for the number diagonal and off-diagonal interactions respectively. Both are found to be RG irrelevant in this regime. The inset plot in the left and right panels show that both $\sigma_{V_{i,j,k,l}}$ and $\sigma_{V_{i,j}}$ vanish at low-energies. Nevertheless, even as the two-particle interactions are found to be RG irrelevant, they lead to the generation of RG relevant three-particle off-diagonal scattering terms $R(i,j,k,l,m,n)c^{\dagger}_{i}c^{\dagger}_{j}c^{\dagger}_{k}c_{l}c_{m}c_{n}$ (right panel of Fig.\ref{GlassII_2}) and two electron-one hole number diagonal interactions $R(i,j,k)\hat{n}_{i}\hat{n}_{j}(1-\hat{n}_{k})$ (left panel of Fig.\ref{GlassII_2}) that are observed to reach finite values at low-energies. The inset plots in both left/right panels of Fig.\ref{GlassII_2} show that the standard deviation of both the three-particle off-diagonal interaction ($\sigma_{R_{i,j,k,l,m,n}}$) and two electron-one hole diagonal interaction ($\sigma_{R_{i,j,k}}$) grow under RG, and finally saturate to a finite value at the RG fixed point. 
\pin
Importantly, the hopping strengths $t_{i,j}$'s (left panel in Fig.\ref{GlassII_3} are also found to be RG irrelevant, such that they reduce in magnitude and vanish at the IR fixed point. On the other hand, the onsite self-energies $\Sigma_{i}$ are RG relevant (right panel in Fig.\ref{GlassII_3}), as they grow and saturate at a IR fixed point. The inset of the right panel of Fig.\ref{GlassII_3} shows the growth of the renormalized onsite disoredered potential $\sigma_{\epsilon'_{d}}$ under RG flow, and its saturation at an IR fixed point.
\begin{figure}
\includegraphics[width=\textwidth]{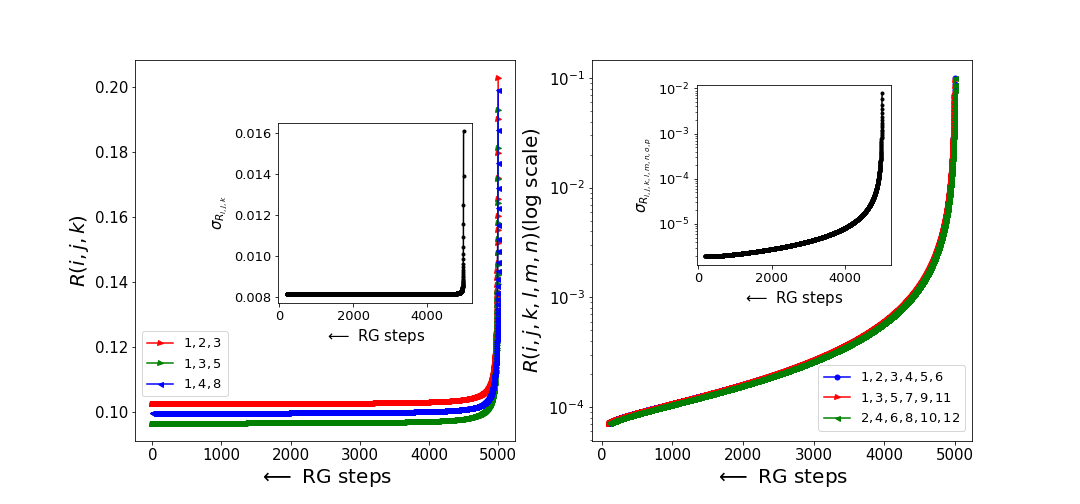}
\caption{Left panel: RG flow for the three-particle off-diagonal interactions $R(i,j,k,l,m,n)$. Inset in left panel displays the RG flow for the standard deviation of these couplings ($\sigma_{R(i,j,k,l,m,n)}$). Right panel: RG flow for the two electron-one hole number diagonal interactions $R(i,j,k)$. Inset in right panel shows the RG flow for $\sigma_{R(i,j,k)}$.}\label{GlassII_4}
\end{figure}
In order to study the effect of three-particle number off-diagonal terms $R_{i,j,k,l,m,n}$ at low-energies, we perform a second level of the URG analysis. The left and right panels of Fig.\ref{GlassII_4} represent the RG flows of the two-electron 1-hole $R(i,j,k)$ and three particle off-diagonal couplings $R(i,j,k,l,m,n)$ respectively. The number off-diagonal couplings in the right panel are found to be RG irrelevant. The inset in right panel of Fig.\ref{GlassII_4} shows that the $\sigma_{R_{i,j,k,l,m,n}}$ is also RG irrelevant, and diminishes at low-energies. The number diagonal interactions in the left panel of Fig.\ref{GlassII_4} are RG relevant, and saturate to a finite value.  The inset of left panel in Fig.\ref{GlassII_4} also shows a similar saturation to a finite value at low-energies. Altogether, the plots Fig.\ref{GlassII_1}-\ref{GlassII_4} provide a numerical verification of the effective Hamiltonian $H_{V}$ in Table~\ref{tab:fixed_point_Hamiltonians}. As is expected for many-body localised phases of matter, the effective fixed point Hamiltonians obtained for phases IV and V are obtained at higher values of the quantum fluctuation energyscale $\omega >0$ than those for all other phases (where $\omega <0$, see Tables~\ref{tab:parameter_regimes_SY_2,4} and \ref{tab:parameter_regimes_SY_4_thermal}). The RG flows to these phases also confirm that an extensive number of single-particle occupation numbers ($n_{i}$) are transformed into integrals of motion under the RG flow~\cite{ros2015}. We note that effective Hamiltonians describing many-body localization similar to $H_{IV}$ and $H_{V}$ have been proposed recently in Refs.\cite{ros2015,rademaker2017many,you2016entanglement}. 
\begin{figure}[!ht]
\includegraphics[width=\textwidth]{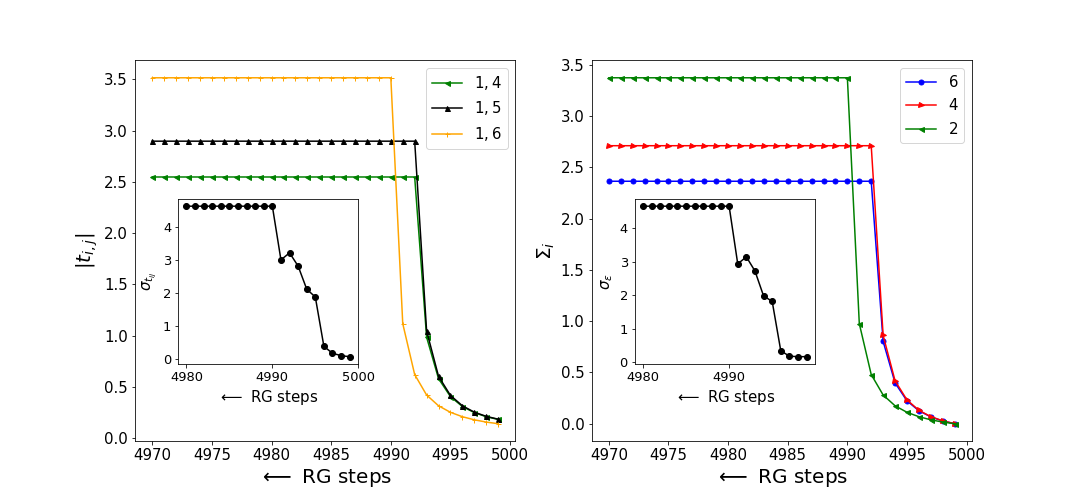}
\caption{Left panel: RG flow for the hopping strengths ($t_{i,j}$), whose initial values are sampled from a gaussian distribution with mean $\overline{t_{i,j}}=-0.2$ and $\sigma_{t_{i,j}}=0.07$ and $\omega=-0.15$. Inset in left panel displays the RG flow for $\sigma_{t_{d}}$. Right panel: RG flow for the onsite self-energies, whose bare values are sampled from a gaussian distribution with mean $\overline{\epsilon_{i}}=1$ and $\sigma_{\epsilon_{d}}=0.1$. Inset in right panel shows the RG flow for $\sigma_{\epsilon_{d}}$.}\label{SY4_1}
\end{figure}
\pin
Last but not least, in regime VI, a numerical evaluation of the RG equations for all the couplings $t_{i,j}$, $\epsilon_{i}'$, $V_{i,j}'$, $V_{i,j,k,l}'$ is shown in Fig.\ref{SY4_1} and Fig.\ref{SY4_2}. Importantly, we find that \textit{all} the couplings are found to be RG relevant, with a growth and eventual saturation at an IR fixed point. Further, the standard deviation of all of these couplings is also found to grow under RG and saturate at the IR fixed point. Thus, in this phase, it is safe to say that \textit{none} of the disordered couplings vanish under RG, thereby preserving the form of the bare Hamiltonian given in (eq.\eqref{Sachdev-Ye model}) but with renormalized couplings. The IR fixed point effective Hamiltonian is shown as $H_{VI}$ (Table \ref{tab:fixed_point_Hamiltonians}) in Table~\ref{tab:parameter_regimes_SY_4_thermal}, and corresponds to the generalized Sachdev-Ye model itself as the stable fixed point theory. Indeed, $H_{VI}$ possesses the greatest parameter space, and corresponds to a thermalized regime: the many-particle entanglement content of the eigenstates of this phase possess the greatest complexity. This is reflected in the marginality of all off-diagonal scattering vertices in $H_{SY_{4}}$, as well as in the fact that very few (i.e., of $\mathcal{O}(10)$ out of $5000$ in the numerical simulations) occupation numbers ($n_{i}$) are transformed into integrals of motion under the RG flow in Regime VI (as can be seen in Figs.\ref{SY4_1} and \ref{SY4_2}). Finally, Regimes IV and V possess tensor network representations similar to the Fermi liquid (Fig.\ref{TN-FL}) and marginal Fermi liquid (Fig.\ref{TN-MFL}) respectively. The tensor network representation of regime VI is shown in Fig.\ref{TN-SY}.
\begin{table}
\hspace*{-1.5cm}
\centering
\resizebox{1.2\textwidth}{!}{  
\begin{tabular}{|c|c|c|c|}
\hline
\textbf{Parameters}&\textbf{Regime-I}&\textbf{Regime-II}&\textbf{Regime-III}\\\hline
$\omega$ & $\omega<\frac{1}{2}(\epsilon'_{l}-\epsilon'_{j})$, $\epsilon'_{l}<\epsilon'_{j}$ & same as \textbf{I} & same as \textbf{I} \\[0.2cm]

$t(\mathbf{r}_{jl})$ & $t(\mathbf{r}_{jl})<0$ & same as \textbf{I} & same as \textbf{I} \\[0.2cm]

$V^{\sigma\sigma'}(\mathbf{r}_{jl})$ & $V^{\sigma\sigma'}(\mathbf{r}_{jl})>0$ & $<\frac{1}{2}(\epsilon'_{l}-\epsilon'_{j})-\omega$& $V^{\sigma -\sigma}(\mathbf{r}_{jl})<0$\\[0.2cm]

$V^{'\sigma\sigma'}_{jl}$ & $V^{'\sigma\sigma'}_{jl}<0$ &same as \textbf{I} & same as \textbf{I}\\[0.2cm]

 $t_{il}'$  & $0<t_{il}'<|t(\mathbf{r}_{jl})|$&same as \textbf{I} & same as \textbf{I} \\[0.2cm]
 
 $\epsilon$ & $\epsilon >2|\omega|$&same as \textbf{I}& same as \textbf{I}\\[0.2cm]
 
 $\epsilon'_{j}$ & $0<\epsilon^{'}_{j}<\epsilon$&same as \textbf{I}& $\epsilon'_{j}<0$ \\[0.2cm]
 
 $V^{'\sigma\sigma'}_{ijkl}$ & $V^{'\sigma\sigma'}_{ijkl}>0$ & same as \textbf{I}& same as \textbf{I}\\[0.2cm]
 
 $V^{\sigma\sigma'}(\mathbf{r}_{ij},\mathbf{r}_{ik},\mathbf{r}_{il})$ & $V^{\sigma\sigma'}(\mathbf{r}_{ij},\mathbf{r}_{ik},\mathbf{r}_{il})>|V^{\sigma\sigma'}_{ijkl}|$& $|V^{\sigma\sigma'}_{ijkl}|<V^{\sigma\sigma'}(\mathbf{r}_{ij},\mathbf{r}_{ik},\mathbf{r}_{il})<-2\omega$ & same as \textbf{I}\\[0.2cm]
 \hline
\end{tabular}
}
\caption{Parameter space regimes for RG flows of $H_{SY_{4}}$. Regime-I corresponds to a general $4$-fermion translational invariant fixed point theory on a $D$-dimensional lattice. Regime-II corresponds to the Hubbard model with long-ranged hopping under RG flow. Regime-III leads to the Anderson disorder model with long-ranged hoppings.}\label{SYcond}
\label{tab:parameter_regimes_SY_2,4}
\end{table}
\begin{table}
\hspace*{-1.5cm}
\begin{tabular}{|c|c|c|c|}
\hline
\textbf{Parameters}&\textbf{Regime-IV}&\textbf{Regime-V}&\textbf{Regime-VI}\\\hline
$\omega$ & $\omega>\frac{1}{2}(\epsilon_{l}+\epsilon_{j})$, $\epsilon_{l}>0$ & $\frac{1}{2}(\epsilon_{j}+\epsilon_{l})>\omega>\frac{1}{2}\epsilon_{j}$ & $\omega<0$ \\[0.2cm]

$t(\mathbf{r}_{jl})$ &  same as \textbf{I} & same as \textbf{I} & no condition \\[0.2cm]

$V^{\sigma\sigma'}(\mathbf{r}_{jl})$ & $V^{\sigma\sigma'}(\mathbf{r}_{jl})<0$ & $V^{\sigma\sigma'}(\mathbf{r}_{jl})>\omega - \frac{1}{2}(\epsilon_{j}+\epsilon_{l})$ & $V^{\sigma\sigma'}(\mathbf{r}_{jl})>0$\\[0.2cm]

$V^{'\sigma\sigma'}_{jl}$ & $V^{'\sigma\sigma'}_{jl}>|V^{'\sigma\sigma'}_{ijkl}|$ & same as \textbf{IV} & $V^{'\sigma\sigma'}_{jl}>0$ \\[0.2cm]

 $t_{il}'$  & $t_{il}'<0$ & same as \textbf{IV} & $sgn(t(\mathbf{r}_{jl}))$ \\[0.2cm]
 
 $\epsilon$ &  same as \textbf{I} & same as \textbf{I} & $\epsilon>0$ \\[0.2cm]
 
 $\epsilon'_{j}$ & no condition & same as \textbf{I}&$\epsilon'_{j}>0$ \\[0.2cm]
 
$V^{'\sigma\sigma'}_{ijkl}$ & $V^{'\sigma\sigma'}_{ijkl}<0$ & same as \textbf{IV} & no condition\\[0.2cm]
 
 $V^{\sigma\sigma'}(\mathbf{r}_{ij},\mathbf{r}_{ik},\mathbf{r}_{il})$ & $0<V^{\sigma\sigma'}(\mathbf{r}_{ij},\mathbf{r}_{ik},\mathbf{r}_{il})<V^{\sigma\sigma'}(\mathbf{r}_{jl})$ & $V^{\sigma\sigma'}(\mathbf{r}_{ij},\mathbf{r}_{ik},\mathbf{r}_{il})>0$ & $sgn(V^{'\sigma\sigma'}_{ijkl})$ \\[0.2cm]
 \hline
\end{tabular}
\caption{Parameter space regimes for RG flows of $H_{SY_{4}}$. Regimes-IV and V corresponds to many-body localized phases. Parameter regime VI leads to the thermalised phase, corresponding to a generalized SY$_{4}$ model.}
\label{tab:parameter_regimes_SY_4_thermal}
\end{table}
\begin{table}
\centering
\begin{tabular}{|c|c|}
\hline
\textbf{Regime}&\textbf{Fixed point Hamiltonian}\\\hline
\textbf{I} & \hspace*{-1.2cm}$H_{I} = \sum_{ij}t^{*}(\mathbf{r}_{ij})c^{\dagger}_{i\sigma}c_{j\sigma} +\epsilon^{*}\sum_{i}\hat{n}_{i\sigma}$\\
  & \hspace*{-1.5cm}$+\sum_{ij}V^{\sigma\sigma',*}(\mathbf{r}_{ij})\hat{n}_{i\sigma}n_{j\sigma'}$\\
  & \hspace*{-0.7cm}$~~~~~~~~~~~+\sum_{ijkl}V^{\sigma\sigma',*}(\mathbf{r}_{ij},\mathbf{r}_{ik},\mathbf{r}_{il})c^{\dagger}_{i\sigma}c^{\dagger}_{j\sigma'}c_{k\sigma'}c_{l\sigma}$\\[0.3cm]
\textbf{II} & \hspace*{-0.2cm}$H_{II} = \sum_{ij}t^{*}(\mathbf{r}_{ij})c^{\dagger}_{i\sigma}c_{j\sigma} +\sum_{i}V^*\hat{n}_{i\sigma}n_{i-\sigma}$\\[0.3cm]
\textbf{III} & \hspace*{-1.0cm}$H_{III} = \sum_{ij}t^{*}(\mathbf{r}_{ij})c^{\dagger}_{i\sigma}c_{j\sigma} +\sum_{i}\epsilon^{*}_{i}\hat{n}_{i\sigma}$ \\[0.3cm]
\textbf{IV} & \hspace*{-1.0cm}$H_{IV} = \sum_{i}\epsilon^{*}_{i}\hat{n}_{i\sigma}+\sum_{i}V^{\sigma\sigma'*}_{ij}\hat{n}_{i\sigma}\hat{n}_{j\sigma'}$\\[0.3cm]
\textbf{V} & \hspace*{0.1cm}$H_{V} = \sum_{i}\epsilon^{*}_{i}\hat{n}_{i\sigma}+\sum_{ijk}V^{\sigma\sigma'*}_{ijk}\hat{n}_{i\sigma}\hat{n}_{j\sigma'}(1-\hat{n}_{k\sigma})$\\[0.3cm]
\textbf{VI} & $H_{VI} = H^{*}_{SY_{4}}$\\
\hline
\end{tabular}
\caption{Fixed point Hamiltonians obtained via RG from $H_{SY_{4}}$ in regimes I-VI.}
\label{tab:fixed_point_Hamiltonians}
\end{table}
\section{Conclusions}\label{conclusions}
\pin
In this work, we have applied the URG formalism~\cite{anirbanmotti,anirbanmott2,anirbanurg1} to certain prototypical models of strongly correlated electrons. The model of a single band of tight binding electrons with momentum-dependent interactions ($H_{SFIM}$) leads to a diverse family of IR fixed point Hamiltonians including gapless phases in the Fermi and non-Fermi liquids, as well as various insulating liquid phases arising from large momentum transfer/back-scattering across the Fermi surface. In a companion work~\cite{anirbanurg1}, we have shown how the $2n$-point vertex RG flow equations can be interpreted as a tensor network. The nodes of this vertex tensor network is composed of the $2n$-point vertices, while the edges represent the electronic states. At each RG step, the vertex tensor network transforms via disentanglement of electronic states, and   
the simultaneous renormalization of the vertex tensors. 
\pin
Here, we have restricted our attention to the study of the RG flows of two-, four- and six-point vertex tensors. We represent the diagonal and off-diagonal vertex tensors in a tree diagram (see Fig.\ref{246vertices-tree}). Each node of the tree represents a subclass of scattering processes. The parameter subspaces are classified in terms of the relative magnitude of the off-diagonal and diagonal vertex tensors. This assists in identifying different IR fixed points reached under RG flow. A numerical evaluation of various RG equations shows that certain vertex scattering processes vanish at these stable fixed points, while certain others become dominant. As a result, the different emergent phases are classified in terms of their distinct tree representations and vertex tensor network diagrams. For the Fermi and non-Fermi liquid phases, the four-point and six-point off-diagonal vertices vanish. This results in their vertex tensor networks being completely disentangled. The fixed point theories differ nevertheless: in the Fermi liquid, each output leg describes a electronic degree of freedom, while in the non-Fermi liquid, each composite degree of freedom is described as a composition of three output legs (two in electron-like and one in hole-like configuration). On the other hand, the vertex tensor networks for gapped phases such as the reduced BCS and Mott liquid Hamiltonians display greater complexity: pairs of electronic legs (indicating bound states that have condensed in the IR) are coupled to each other via four-point scattering vertices. Further, the IR fixed point theories describing such gapped phases can be rewritten in terms of nonlocal Wilson loops, leading to a Hamiltonian gauge theory construction. The zero mode of the Hamiltonian gauge theory reveals interesting topological features, e.g., ground state degeneracy, charge fractionalization etc. 
\pin
In the generalised Sachdev-Ye ($SY_{4}$) model, the non-translationally invariant (NTI) phases can be separated into two subclasses: one with remnant electronic interactions and the other without. The first NTI subclass contains many-body localised (MBL) and thermalized phases. Our study reveals the MBL phases to be glassy variants of the Fermi and non-Fermi liquid phases. On the other hand, we find that the thermalized phase is described by a theory that involves only a marginal deformation of the parent Hamiltonian ($H_{SY_{4}}$), and involves a strong interplay between fermion exchange signatures and electronic correlation. The second NTI subclass contains a phase corresponding to the Anderson model of disordered electrons. Further, there is also a finite parameter-space window obtained from the $SY_{4}$ for translationally invariant (TI) models such as $H_{SFIM}$, which upon further renormalization lead to the various metallic and insulating phases described earlier for $H_{SFIM}$. 
\pin  
We have also shown that the entanglement renormalization towards gapless and gapped IR phases is distinct: while gapless phases are characterised by the presence of fermion exchange phases along the RG flow, the passage to gapped phases displays the mitigation of the effects arising from fermion signs. This is due to the fact that dominant RG flow in the latter case occurs in a reduced pseudospin subspace where the elementary degrees of freedoms are pairwise electronic states. Furthermore, we obtained the RG scaling form for the holographic entropy bound of the Fermi liquid phase, and argued with regards to its distinction from that for the marginal Fermi liquid and reduced BCS phases. We also showed separately that the effective IR theories for gapped models support a gauge-theoretic description. In this way, the URG offers an ab-initio formulation of the gauge-gravity duality: the passage from UV to IR involves
the holographic generation of spacetime via entanglement renormalization~\cite{mukherjee2020,anirbanurg1} as well as an effective gauge theory from vertex renormalization. Among several exciting future directions, this paves the way for further investigations on the nature the many-particle entanglement of strongly interacting quantum liquids.
\pin\\
\textbf{Acknowledgments}\\
The authors thank R. K. Singh, A. Dasgupta, S. Patra, A. Taraphder, N. S. Vidhyadhiraja, S. Pal and P. Majumdar for several discussions and feedback. We thank an anonymous referee for valuable suggestions that have helped improve the analysis substantially. A. M. thanks the CSIR, Govt. of India for funding through a junior and senior research fellowship. S. L. thanks the DST, Govt. of India for funding through a Ramanujan Fellowship during which a part of this work was carried out.
\appendix
\renewcommand{\thesection}{\Alph{section}}
\section{1-particle self-energy RG for three-particle MFL Hamiltonian}\label{Appendix-1pselfenergy}
\pin
The net correlation energy for the five electronic states $E^{j}_{5,1}$  close to the Fermi surface in eqs.\eqref{tot energy 1,3} and \eqref{NetFiveCorrelationEnergy} attains a simplified expression via the following approximation
\begin{eqnarray}
\epsilon^{(j)}_{\mathbf{k}_{\Lambda_{j}\hat{s}}} &=& \epsilon^{(j)}_{\mathbf{p}-\mathbf{k}_{\Lambda_{j}\hat{s}}} = \epsilon^{(j)}_{\mathbf{p}'-\mathbf{k}_{\Lambda_{j}\hat{s}}}  \Rightarrow E^{(j)}_{1} = \frac{1}{2}\epsilon^{(j)}_{\mathbf{k}_{\Lambda_{j}\hat{s}}}~\mathrm{and}~
\epsilon^{(j)}_{\mathbf{k}_{\Lambda_{j}\hat{s}}}=\epsilon^{(j)}_{\mathbf{k}_{\Lambda_{1}\hat{s}}},
\label{energy_space_constraint}
\end{eqnarray}
such that
\begin{eqnarray}
E^{(j)}_{5,1}=E^{(j)}_{1}+\epsilon^{(j)}_{\mathbf{k}_{\Lambda_{j}\hat{s}}}=\frac{3}{2}\epsilon_{\mathbf{k}_{\Lambda_{j}\hat{s}}}~.\label{1-particle_effective_tot}
\end{eqnarray}
Similarly, we also approximate the net 3-particle correlation energy coming from six-particle terms: $E^{(j)}_{5,3} = -\frac{3}{4}\Gamma^{3,(j)}_{(k\sigma,\mu)}$. Putting these two correlation energies into the 1-particle self-energy RG (eq.\eqref{1-particle_self_energy_RG}) near the fixed point eq.\eqref{3-number_diag_fp}, we find that close to the Fermi surface
\begin{eqnarray}
&&\hspace*{-0.9cm}\Delta\Sigma^{(l^{*})}_{\mathbf{k}_{\Lambda_{l^{*}}\hat{s}}}(\omega) = \sum_{\epsilon^{(l^{*})}_{\mathbf{k}_{\Lambda_{j}\hat{s}}}<\epsilon^{(l^{*})}_{\mathbf{k}_{\Lambda_{l^{*}}\hat{s}}},\hat{s}'}\frac{(\Gamma^{6,(l^{*})}_{\alpha\gamma}(\omega))^{2}}{\omega -\frac{1}{2}\epsilon^{(j)}_{\mathbf{k}_{\Lambda_{j\hat{s}}}}+\frac{3}{4}\Gamma^{6,(l^{*})}_{\gamma\gamma'}(\omega)}~,\nonumber\\
&&\hspace*{-0.9cm}=\frac{N(0)}{\Gamma^{6,(l^{*})}_{\gamma\gamma'}(\omega)}\int_{\Lambda_{j}\to 0,F}^{\Lambda_{l^{*}}} d\Lambda  \frac{(\Gamma^{6,(l^{*})}_{\alpha\gamma}(\omega))^{2}}{\omega -\frac{1}{2}\epsilon^{(j)}_{\mathbf{k}_{\Lambda_{\hat{s}}}}+\frac{3}{4}\Gamma^{6,(l^{*})}_{\gamma\gamma'}(\omega)}~.~~~~~
\end{eqnarray}
In the above expression, the set $\alpha$
\begin{eqnarray}
\alpha &=& \bigg\lbrace (\mathbf{k}_{\Lambda_{j}\hat{s}}\sigma ,1)~,~(\mathbf{k}'\sigma',1)~,~(\mathbf{k}_{\Lambda_{1}\hat{s}}\sigma'',1)~,~
(\mathbf{p}-\mathbf{k}_{\Lambda_{1}\hat{s}}\sigma''',1)~,~(\mathbf{k}''\sigma',0)\bigg\rbrace
\end{eqnarray}
comprises the 4-electron 1-hole intermediate configuration whose energy appears in the 5-partcle Green's function. The index $\nu =(\mathbf{k}_{\Lambda_{j^{*}}\hat{s}}\sigma ,1)$ labels the electronic state whose self-energy is being renormalized. Integrating the RG equation results in the expression
\begin{eqnarray}
&&\hspace*{-0.9cm}\Delta\Sigma^{(l^{*})}_{\mathbf{k}_{\Lambda_{l^{*}}\hat{s}}}(\omega)=N(0)\frac{(\Gamma^{6,(l^{*})}_{X,\mathbf{k}_{\Lambda_{l^{*}}\hat{s}}}(\omega))^{2}}{\Gamma^{6,(l^{*})}_{D,\mathbf{k}_{\Lambda_{l^{*}}\hat{s}}}(\omega)}\ln\bigg\vert\frac{\omega -\frac{3}{2}\epsilon^{(l^{*})}_{\mathbf{k}_{\Lambda_{l^{*}}\hat{s}}}+\frac{1}{8}\Gamma^{6,(l^{*})}_{D,\mathbf{k}_{\Lambda_{l^{*}}\hat{s}}}}{\omega}\bigg\vert ~,\nonumber\\
\label{self_energy_flow_final_simp}
\end{eqnarray}
where $N(0)$ is the density of states on the Fermi surface and 
\begin{equation}
\Gamma^{6,(l^{*})}_{\mathbf{k}_{\Lambda_{l^{*}}\hat{s}}\sigma,\alpha\nu}(\omega)=\Gamma^{6,(l^{*})}_{X,\mathbf{k}_{\Lambda_{l^{*}}\hat{s}}}(\omega)~,~ \Gamma^{6,(l^{*})}_{\gamma\gamma'}(\omega) = \Gamma^{6,(l^{*})}_{D,\mathbf{k}_{\Lambda_{l^{*}}\hat{s}}}(\omega)~.
\end{equation}
Using the fixed point relation eq.\eqref{3-number_diag_fp}, the self-energy flow relation (eq.\eqref{self_energy_flow_final_simp}) can be further simplified to
\begin{eqnarray}
\Delta\Sigma^{(l^{*})}_{\mathbf{k}_{\Lambda_{l^{*}}\hat{s}}}(\omega) &=& N(0)\frac{(\Gamma^{6,(l^{*})}_{X,\mathbf{k}_{\Lambda_{l^{*}}\hat{s}}}(\omega))^{2}}{\Gamma^{6,(l^{*})}_{D,\mathbf{k}_{\Lambda_{l^{*}}\hat{s}}}(\omega)}\ln\bigg\vert\frac{\epsilon_{\mathbf{k}_{\Lambda^{*}\hat{s}}}}{\omega}\bigg\vert ~.~~~~~~~~~
\end{eqnarray}
As the Fermi energy is reached by taking the limits of $\omega \to 0$ and $\Lambda_{j}\to 0$, the change in self-energy $\Delta\Sigma^{(j)}_{\mathbf{k}_{\Lambda\hat{s}}}(\omega)$ has a branch-cut log singularity. Thus, the self-energy attains the familiar logarithmic form that was proposed on phenomenological grounds for the Marginal Fermi liquid~\cite{varma-physrevlett.63.1996}
\setlength{\abovedisplayskip}{3pt}
\begin{eqnarray}
\Sigma^{(l^{*})}_{\mathbf{k}_{\Lambda\hat{s}}}(\omega)&\approx &\Delta\Sigma^{(j)}_{\mathbf{k}_{\Lambda\hat{s}}}(\omega)+O(\omega)\nonumber\\
&\approx &N(0)\frac{(\Gamma^{6,(l^{*})}_{X,\mathbf{k}_{\Lambda_{l^{*}}\hat{s}}}(\omega))^{2}}{\Gamma^{6,(l^{*})}_{D,\mathbf{k}_{\Lambda_{l^{*}}\hat{s}}}(\omega)}\ln\bigg\vert\frac{\epsilon_{\mathbf{k}_{\Lambda_{(l^{*}}\hat{s}}}}{\omega}\bigg\vert ~.~~~~~~~~
\end{eqnarray}
\section{RG equations for SY$_{4}$ model}\label{SY4}
\pin
Using the diagrammatic contributions of the RG flow hierarchy eq.\eqref{vertexRGflows} (see also Fig.\ref{vertex_flow_eqn_2,4,6}), the RG equations for random hopping amplitudes $t_{ik}$, on-site potentials $\epsilon_{i}$ and random four-fermion interaction amplitudes $V^{\sigma\sigma'}_{ijkl}$ for the generalized Sachdev-Ye model (eq.\eqref{Sachdev-Ye model}) is given by
\begin{eqnarray}
\hspace*{-0.7cm}&&\Delta t^{(j)}_{ik} = t^{(j)}_{ij}G^{2,(j)}_{j}t^{(j)}_{jk}+ \sum_{m<j}t^{(j)}_{lj}G^{4,\sigma\sigma,(j)}_{jl}V^{\sigma\sigma}_{ijlk}
+\sum_{i,l,m<j}V^{\sigma\sigma',(j)}_{ijlm}G^{6,(j)}_{jlm}V^{\sigma\sigma',(j)}_{jlmk},\\\label{hopping-RG}
\hspace*{-0.7cm}&&\Delta \epsilon^{(j)}_{i} = t^{(j)}_{ij}G^{2,(j)}_{j}t^{(j)}_{ji}+\sum_{l,k<j}V^{\sigma\sigma',(j)}_{ijlk}G^{6,(j)}_{jlk}V^{\sigma\sigma',(j)}_{klji},~~~~~\\\label{on-site_pot-RG}
\hspace*{-0.7cm}&&\Delta V^{\sigma\sigma',(j)}_{ik} = -\sum_{l}V^{\sigma\sigma',(j)}_{ijkl}G^{4,\sigma\sigma',(j)}_{jl}V^{\sigma\sigma',(j)}_{lkji},\label{four_fermi-D-RG}\\
\hspace*{-0.7cm}&&\Delta V^{(j)}_{iklm} = \sum_{s}V^{\sigma\sigma',(j)}_{ijsm}G^{4,(j)}_{js}V^{\sigma\sigma',(j)}_{ksjl}+t^{(j)}_{jm}G_{j}^{2,(j)}V^{\sigma\sigma',(j)}_{iklj}.\label{four_fermi-X-RG}
\end{eqnarray}
By decomposing the $2$-point vertex parameters into translational invariant ($\epsilon^{(j)}$~,~$t^{(j)}(\mathbf{r})$) and non-translation invariant ($\epsilon_{i}^{(j)}$~,~$t_{ij}^{'(j)}$) parts (eq.\eqref{2-point_trans_decom}), and denoting $\mathbf{r}_{ik}=\mathbf{r}$, $\mathbf{r}_{ij}=\mathbf{r}'$, $\mathbf{r}_{il}=\mathbf{r}''$, $N_{j} = \sum_{\mathbf{r}'<\max\limits_{i<j}(|\mathbf{r}_{ij}|)} 1$, we can use eqs.\eqref{hopping-RG} and \eqref{on-site_pot-RG} to write their separate RG flow equations in the e-h configuration $\hat{n}_{j\sigma}=0,\hat{n}_{i\sigma'}=1$ as follows:\\
\pin
\textit{1. Translational invariant hopping term ---}
\begin{eqnarray}
\hspace*{-0.5cm}&&\Delta t^{(j)}(\mathbf{r}) = \frac{1}{N_{j}}\sum_{\mathbf{r}_{ij}<\max\limits_{i<j}(|\mathbf{r}_{ij}|)}\frac{t^{(j)}(\mathbf{r}_{jk})t^{(j)}(\mathbf{r}_{ij})}{\omega + \frac{1}{2}\epsilon^{(j)}+\frac{1}{2}\epsilon'^{(j)}_{j}}\nonumber\\ &+&\sum_{\mathbf{r}_{ij},\mathbf{r}_{il}}\frac{\frac{1}{N_{j}}t^{(j)}(\mathbf{r}_{jl})V^{\sigma\sigma,(j)}(\mathbf{r},\mathbf{r}_{ij},\mathbf{r}_{il})}{\omega +\frac{1}{2}\epsilon^{'(j)}_{j}-\frac{1}{2}\epsilon^{'(j)}_{l}+\frac{1}{4}V^{\sigma\sigma,(j)}(\mathbf{r}_{jl})+\frac{1}{4}V^{'\sigma\sigma,(j)}_{jl}}~,~~~~~~\label{Translational invariant hopping term_RG}
\end{eqnarray}
\textit{2. Translational non-invariant hopping term ---}
\begin{eqnarray}
\hspace*{-0.5cm}&&\Delta t^{'(j)}_{ik} = \frac{t^{(j)}(\mathbf{r}_{ij})t^{'(j)}_{jk}+t^{(j)}(\mathbf{r}_{jk})t^{'(j)}_{ij}+t^{'(j)}_{ij}t^{'(j)}_{jk}}{\omega +\frac{1}{2}\epsilon'^{(j)}_{j}+\frac{1}{2}\epsilon^{(j)}}\nonumber\\ &+& \sum_{\mathbf{r}''}\frac{t^{(j)}(\mathbf{r}'')V^{'\sigma\sigma,(j)}_{ijkl}+t^{'(j)}_{jl}V^{\sigma\sigma,(j)}(\mathbf{r},\mathbf{r}',\mathbf{r}'')}{\omega +\frac{1}{2}\epsilon^{'(j)}_{j}-\frac{1}{2}\epsilon'^{(j)}_{l}+\frac{1}{4}V^{\sigma\sigma,(j)}(\mathbf{r}_{jl})+\frac{1}{4}V^{'\sigma\sigma,(j)}_{jl}}~,~~~~~~\label{Translational non-invariant hopping term_RG}
\end{eqnarray}
\textit{3. Translational invariant chemical potential term ---}
\begin{eqnarray}
&&\Delta \epsilon^{(j)} = \frac{1}{N_{j}}\sum_{\mathbf{r}<\max\limits_{i<j}(|\mathbf{r}_{ij}|)}\frac{(t^{(j)}(\mathbf{r}))^{2}}{\omega +\frac{1}{2}\epsilon'^{(j)}_{j}+\frac{1}{2}\epsilon^{(j)}}\nonumber\\
\hspace*{-0.5cm}
&& +\frac{1}{N_{j}}\sum_{l,k<j,\mathbf{r}'}(V^{\sigma\sigma,(j)}(\mathbf{r},\mathbf{r}_{ij},\mathbf{r}_{il}))^{2}G^{6,\sigma\sigma',(j)}_{jlk}~,\label{Translational invariant chemical potential term_RG}
\end{eqnarray}
\textit{4. Translational non-invariant on-site potential term ---}
\begin{eqnarray}
\hspace*{-0.5cm}&&\Delta \epsilon'^{(j)}_{i} = \frac{1}{N_{j}}\sum_{\mathbf{r}<\max\limits_{i<j}(|\mathbf{r}_{ij}|)}\frac{t^{(j)}(\mathbf{r}')t^{'(j)}_{ji}+t^{'(j)}_{ij}t^{'(j)}_{ji}}{\omega + \frac{1}{2}\epsilon'^{(j)}_{j}+\frac{1}{2}\epsilon^{(j)}}\nonumber\\
&& +\frac{1}{N_{j}}\sum_{l,k<j,\mathbf{r}'}(V^{\sigma\sigma,(j)}(\mathbf{r},\mathbf{r}',\mathbf{r}''))^{2}G^{6,\sigma\sigma',(j)}_{jlk}~.\label{Translational non-invariant on-site potential term_RG}
\end{eqnarray}
Similarly, by decomposing the $4$-point vertexes $V^{\sigma\sigma'}_{ijkl}$ diagonal and off-diagonal parts into translational invariant  (eq.\eqref{4-point_X_trans_decom}) and non-invariant parts (eq.\eqref{4-point_D_trans_decom}), their RG equations are obtained from eqs.\eqref{four_fermi-D-RG} and \eqref{four_fermi-X-RG} as follows:\\

\pin 
\textit{5. Translational invariant density-density interaction term ---}
\begin{eqnarray}
\hspace*{-0.4cm}&&\Delta V^{\sigma\sigma',(j)}(\mathbf{r}) = -\sum_{\mathbf{r}_{ij},\mathbf{r}_{il}} \frac{\frac{1}{N_{j}}(V^{\sigma\sigma',(j)}(\mathbf{r},\mathbf{r}_{ij},\mathbf{r}_{il}))^{2}}{\omega
 +\frac{1}{2}\epsilon'^{(j)}_{j}-\frac{1}{2}\epsilon^{'(j)}_{l}+\frac{1}{4}V^{\sigma\sigma'(j)}(\mathbf{r}_{jl})+\frac{1}{4}V_{jl}^{'\sigma\sigma'(j)}}~,\label{Translational invariant density-density(dd) interaction term_RG}
 \hspace*{-0.7cm}
 \end{eqnarray}
 \textit{6. Translational non-invariant density-density interaction term ---}
 \begin{eqnarray}
 \hspace*{-0.6cm}&&\Delta V^{'\sigma\sigma',(j)}_{ik}= -\sum_{\mathbf{r}_{ij},\mathbf{r}_{il}} \frac{\frac{1}{N_{j}}V^{\sigma\sigma',(j)}(\mathbf{r}_{ik},\mathbf{r}_{ij},\mathbf{r}_{il})V^{'\sigma\sigma',(j)}_{ijkl}+V^{'\sigma\sigma',(j)}_{ijkl}V^{'\sigma\sigma',(j)}_{ijkl}}{\omega+\frac{1}{2}\epsilon'^{(j)}_{j}-\frac{1}{2}\epsilon'^{(j)}_{l}+\frac{1}{4}V^{(j)}(\mathbf{r}_{jl})+\frac{1}{4}V_{jl}^{'\sigma\sigma'(j)}}~,\label{Translational non-invariant density-density interaction term_RG}
  \hspace*{-0.7cm}
\end{eqnarray}
\textit{7. Translational invariant current-current interaction term ---}
\begin{eqnarray}
&&\Delta V^{\sigma\sigma',(j)}(\mathbf{r}_{is},\mathbf{r},\mathbf{r}_{im}) =\sum_{\mathbf{r}_{ij},\mathbf{r}_{il}} \frac{\frac{1}{N_{j}}V^{\sigma\sigma',(j)}(\mathbf{r}_{ji},\mathbf{r}_{js},\mathbf{r}_{jl})V^{\sigma\sigma',(j)}(\mathbf{r}_{lk},\mathbf{r}_{lm},\mathbf{r}_{lj})}{\omega
 +\frac{1}{2}\epsilon'^{(j)}_{j}-\frac{1}{2}\epsilon'^{(j)}_{l}+\frac{1}{4}V^{(j)}(\mathbf{r}_{jl})+\frac{1}{4}V_{jl}^{'\sigma\sigma'(j)}}~,~~~~~~~\label{Translational invariant current-current interaction term_RG}
 \end{eqnarray}
 \textit{8. Translational non-invariant current-current interaction term ---}
 \begin{eqnarray}
 \hspace*{-0.6cm}&&\Delta V^{'\sigma\sigma',(j)}_{iskm}= \sum_{\mathbf{r}_{ij},\mathbf{r}_{il}} \frac{\frac{1}{N_{j}}V^{\sigma\sigma',(j)}(\mathbf{r}_{ji},\mathbf{r}_{js},\mathbf{r}_{jl})V^{'\sigma\sigma',(j)}_{lkmj}+V^{'\sigma\sigma',(j)}_{jisl}V^{'\sigma\sigma',(j)}_{lkmj}}{\omega+\frac{1}{2}\epsilon'^{(j)}_{j}-\frac{1}{2}\epsilon'^{(j)}_{l}+\frac{1}{4}V^{\sigma\sigma',(j)}(\mathbf{r}_{jl})+\frac{1}{4}V_{jl}^{'\sigma\sigma'(j)}}~.\label{Translational non-invariant current-current interaction term_RG}
  \hspace*{-0.7cm}
\end{eqnarray}
\pin\\
\textbf{\Large{References}}
\bibliographystyle{elsarticle-num}
\bibliography{netbib}
\end{document}